\documentclass[graybox]{svmult}
\usepackage{makeidx}         
\usepackage{multicol}        
\usepackage[bottom]{footmisc}
\usepackage[utf8]{inputenc}
\usepackage{mathtools}
\usepackage{bm}
\usepackage{cite}
\usepackage{url}
\PassOptionsToPackage{hyphens}{url}
\usepackage[colorlinks]{hyperref}

\hypersetup{
    colorlinks=true,
    linkcolor=teal,      
    urlcolor=blue,
    citecolor=magenta
}
\usepackage{comment}
\usepackage{enumitem}
\usepackage{amsmath}
\usepackage{amssymb}
\usepackage{graphicx}
\usepackage{caption}
\usepackage{subcaption}
\usepackage{hyperref}
\usepackage{lscape}
\usepackage{color}

\usepackage[margin=2.5cm]{geometry}
\usepackage{multirow}
\usepackage{algorithm}
\usepackage{algorithmic}
\usepackage{authblk}
\newcommand{\CapRightArrow}[2]{\underrightarrow{ \text{ \hspace{#2} \small{#1} \hspace{#2} } } }

\DeclareMathOperator{\Diag}{\mathbf{diag}}

\DeclareMathOperator*{\argmin}{\arg\!\min} 

\begin{document}
\setcounter{chapter}{4}

\title{Noninvasive Fetal Electrocardiography: Models, Technologies and Algorithms}
\titlerunning{Fetal Cardiac Signal Processing Techniques}
\author{Reza Sameni, PhD}
\authorrunning{R.~Sameni}
\institute{Department of Biomedical Informatics, Emory University School of Medicine, Atlanta, GA, USA\\ \email{rsameni@dbmi.emory.edu}, Web: \url{http://www.sameni.info}\\
\textbf{Author's preprint published as:} R. Sameni. Noninvasive Fetal Electrocardiography: Models, Technologies, and Algorithms, Chapter 5, pages 99–146.
Springer International Publishing, Cham, 2021. ISBN 978-3-030-54403-4. DOI: \url{http://doi.org/10.1007/978-3-030-54403-4_5}%
}

\maketitle
\abstract{
The fetal electrocardiogram (fECG) was first recorded from the maternal abdominal surface in the early 1900s. During the past fifty years, the most advanced electronics technologies and signal processing algorithms have been used to convert noninvasive fetal electrocardiography into a reliable technology for fetal cardiac monitoring. In this chapter, the major signal processing techniques, which have been developed for the modeling, extraction and analysis of the fECG from noninvasive maternal abdominal recordings are reviewed and compared with one another in detail. The major topics of the chapter include: 1) the electrophysiology of the fECG from the signal processing viewpoint, 2) the mathematical model of the maternal volume conduction media and the waveform models of the fECG acquired from body surface leads, 3) the signal acquisition requirements, 4) model-based techniques for fECG noise and interference cancellation, including adaptive filters and semi-blind source separation techniques, and 5) recent algorithmic advances for fetal motion tracking and online fECG extraction from few number of channels.
}


\section{Introduction}
The early assessment of fetal well-being is the major objective of fetal monitoring during pregnancy and labor. The latter is specifically useful for identifying fetuses at risk of hypoxia (oxygen deficiency) during labor. In this context, fetal electrocardiography is one of the emerging technologies, which dates back to 1906 \cite{Cremer1906}, but has gained much more attention during the past two decades. The technology has significantly evolved throughout the past fifty years, from na\"ive visual inspection to multichannel automatic methods of noninvasive fetal electrocardiogram (fECG) extraction, using advanced signal processing methods \cite{SameniClifford2010,jamshidian2018fetal}. The method has become more popular in recent years, due to its relatively low cost and advances in the required signal acquisition and signal processing techniques. In this context,  both invasive methods used after amniotic sac rupture during labor, and noninvasive methods using maternal abdominal leads throughout pregnancy (especially during the third trimester) have been used. Although invasive fECG recording using fetal scalp leads have a higher signal-to-noise ratio (SNR) and require less processing as compared with noninvasive signals captured from the maternal abdomen, due to the potential risks of invasive methods for both the mother and the fetus(es), it is not so popular. On the other hand, despite its advantages, noninvasive fECG extraction is hampered by many practical challenges including: 1) the significantly lower SNR of the fECG as compared with the maternal ECG (mECG), which superposes over the abdominal leads; 2) device and measurement issues related to noninvasive fECG acquisition using single or multiple maternal abdominal sensors; 3) the indirect access to the fetal heart through multiple maternal body layers, which act as a \textit{volume conductor}; 4) artifacts and variations in fECG shape due to fetal movements; 5) baseline wanders of the data due to maternal respiration; 6) measurement and environmental noises such as maternal muscle and uterine contractions, power-line noise and artifacts due to other bedside monitors and devices such as the infusion pumps. Most of these noises overlap with the fECG in time, frequency and space (leads), making fECG extraction a nontrivial challenge, which requires advanced signal processing.

To date, various methods have been developed for fECG extraction with various degrees of success, including adaptive filtering \cite{Farvet1968,Widrow75,Park92,Outram95,Shao2004,Martens2007,swarnalatha2010novel,joachimbehar2014,Ma2015},  Kalman filtering \cite{SSJC06, SSJ08,niknazar2013fetal}, singular value decomposition \cite{Kanjilal1997}, blind and semi-blind source separation using independent and periodic component analysis \cite{Lathauwer2000,Zarzoso01,Sameni2008,Sameni2008a} and wavelet transforms \cite{Khamene2000,Vigneron2003,Liu2015}. Some of these techniques, such as Kalman filters, singular value decomposition, wavelets and adaptive filters (used in line-enhancement mode) have been applied to both single and multichannel abdominal ECG recordings. In contrast, other techniques such as independent component analysis or adaptive noise cancellation using an external reference require two or more channels of measurements. Multichannel techniques based on blind and semi-blind source separation have proved to be very effective to overcome the aforementioned challenges. Nevertheless, various aspects of noninvasive fECG extraction are still open problems and require further studies. For example, issues related to long-time online fECG monitoring (required for fetal Holter monitoring), problems due to fetal movements during signal acquisition, variations in fECG morphology (again due to fetal motion and fetal positioning with respect to the body surface leads), fECG extraction in low SNR using few numbers of channels. There are also several post fECG extraction issues including fetal R-peak detection, heart-rate (HR) calculation, fECG morphology extraction and clinical parameter extraction (QT interval, ST-level calculation, etc.) from noisy fECG signals. From the clinical and industrial perspective, the size and cost of the device, the technology and the number of maternal abdominal leads (preferably only a few leads placed close together in a patch of electrodes) are also of great importance.

In this chapter, the major signal processing techniques, which have been developed for the modeling, extraction and analysis of the fECG from noninvasive maternal abdominal recordings over the past fifty years are reviewed and compared with one another in detail.

\section{Noninvasive fetal electrocardiography data model}
\label{sec:datamodel}
\subsection{Volume conductor model}
\label{sec:volumeconductor}
The physics of the problem of noninvasive fECG measurement from the maternal abdomen follows the general principles of volume conduction theory \cite{webster2009medical}. The properties of the propagation media from the fetal heart to the maternal abdomen have been studied in previous studies \cite{Oostendorp1989,Sameni2008}. The major aspects of the problem, which influence the fECG data model and extraction techniques can be summarized as follows \cite{jamshidian2018fetal}:
\begin{enumerate}
\item \textit{Negligible electric displacement current}: The electromagnetics of the problem is quasi-static. Therefore, the electric and magnetic fields are decoupled, the electric field is proportional to the gradient of the electric scalar potential and the divergence of the current density is zero.
\item \textit{Linear propagation media}: Superposition holds for the electrical potentials due to the maternal heart, fetal heart and other sources of biopotentials.
\item \textit{Negligible capacitive component of the body tissues' electrical impedance}: Due to the relatively low frequency range of interest (below 10~kHz), the tissues are to a very good approximation resistive.
\item \textit{Spatial distribution of the heart}: The source signals are \textit{non-punctual} and different lead configurations provide different views of the heart, conveying different--- although rather redundant and correlated--- information. Therefore, the cardiac source may only be approximated by a current dipole in the far-field.
\item \textit{Non-homogeneous volume conductor}: Low-conductivity layers such as the \textit{vernix caseosa}, which form throughout pregnancy (mainly between weeks 28 and 32 of gestation \cite{SameniClifford2010}) can change the preferred electrical propagation pathways, resulting in morphological variations on the maternal body surface \cite{Oostendorp1989,Stinstra2001}. 
\item \textit{Morphological variability}: During a signal recording session, although the fECG morphology is consistent with respect to the fetal body (as in adult ECG), due to fetal motions such as rotations, movements of extremities and hiccups, the extracted fECG morphology can change with respect to the maternal body coordinate system and the maternal body surface sensors. Moreover, minor fetal and maternal movements, such as maternal respiration, somehow \textit{modulate} the fetal cardiac signals acquired from the maternal abdomen.
\end{enumerate}
These properties imply that temporal parameters such as the R-peak locations, heart-rate, PT and QT intervals, etc. can be very accurate, but parameters, such as the R-wave amplitudes and T-to-R ratios, which rely on amplitudes and ratios of amplitudes are totally unreliable, since they can easily change with fetal positioning, gestation age or a change of lead configurations. Nevertheless, relative variations of amplitude-based parameters can still be accurate between successive fetal heart beats and during real-time monitoring. For example, phenomena such as \textit{T-wave alternates} (TWA), which requires the comparison of the T-wave amplitudes between successive beats is still reliable (up to the signal quality).

Note that items 1 to 4 listed above are also applicable to adult ECG and the mECG that superposes over the abdominal leads. Based on these properties, the problem of noninvasive fECG acquisition from an array of maternal abdominal sensors can be mathematically formulated as follows:
\begin{equation}
\mathbf{x}(t) = \mathbf{H}_m \mathbf{s}_m(t) + \mathbf{H}_f \mathbf{s}_f(t) + \mathbf{H}_v \mathbf{v}(t) + \mathbf{n}(t)
\label{eq:datamodel}
\end{equation}where $\mathbf{x}(t) \in \mathbb{R}^{n}$ are  $n$ channels of maternal body surface measurements acquired differentially with respect to one or more reference channels, $\mathbf{s}_m(t) \in \mathbb{R}^{m}$ are the mECG source components, $\mathbf{s}_f(t) \in \mathbb{R}^{l}$ are the fECG source components, $\mathbf{v}(t) \in \mathbb{R}^{k}$ represent structured (correlated or low-rank) noise corresponding to other biopotential sources (such as maternal muscle contractions) or device noises, and $\mathbf{n}(t) \in \mathbb{R}^{n}$ are unstructured (full-rank) measurement noise, which correspond to sensor-wise noise that are uncorrelated from the other signals and structured noises. In the data model (\ref{eq:datamodel}), $\mathbf{H}_m \in \mathbb{R}^{n\times m}$, $\mathbf{H}_f \in \mathbb{R}^{n\times l}$ and $\mathbf{H}_v \in \mathbb{R}^{n\times k}$ are the \textit{lead-field} matrices, which map the source components to the body surface electrode recordings. The model may be further extended to consider minor maternal body motions (e.g., due to respiration) and fetal movements, by considering $\mathbf{H}_m$, $\mathbf{H}_v$ and $\mathbf{H}_f$ to be functions of time. Also in multiple pregnancies, similar terms can be added to (\ref{eq:datamodel}) for the other fetuses \cite{SCJS06}.

The spatial distribution of the cardiac source implies that in (\ref{eq:datamodel}), $m$ and $l$ theoretically tend to infinity. However, as we get farther from the cardiac sources, far-field approximations are applicable and the cardiac sources behave more like a dipoles \cite{MP95}. Therefore, in practice, each of the cardiac sources can be approximated up to finite \textit{effective number of dimensions} \cite{scher1960factor}. In \cite{SJS06}, it was quantitatively shown that for adult ECG, taking $m$ between 5 to 6 and for fetal ECG, assuming $l$ between 1 to 3 is sufficient to retrieve the major energy fraction of the maternal and fetal ECG components (from the maternal abdominal lead recordings). Apparently, the effective number of dimensions also depends on the sensor position with respect to the maternal and fetal hearts. For example, if the maternal abdominal leads are placed rather distanced from the maternal chest, or if the fetal position is such that the shortest conductive path between the differential sensor pairs do not pass through the fetal heart (i.e., the fetal cardiac electrical fields do not result in significant potential differences between the recording differential pair leads), the effective number of dimensions reduces and in some cases the fECG is not retrievable from the abdominal leads, even by using the most advanced signal processing techniques. It is later shown that the effective number of dimensions and the number of maternal body sensors are specifically important for multichannel fECG extraction algorithms. Some general guidelines for selecting the sensor locations for better fECG retrieval are presented in Section \ref{sec:sensorpositions}.

\subsection{Morphological model}
\label{sec:morphologicmodel}
\subsubsection{Template-based models}
Mathematical modeling of the ECG waveform has vast applications in ECG device test instruments and for educational purposes. To date, the beat-wise ECG morphology has been modeled by various mathematical functions including Bessel functions \cite{sornmo1981method}, Hermite polynomials \cite{laguna1996adaptive}, and Gaussian functions \cite{McSharry2003,jafarnia2007modified}. The latter has an intrinsic dynamic mechanism for generating continuous ECG waveforms, which will be later discussed in details. Other wave-based models can generate a continuous ECG by replicating a fixed waveform that resembles the beat-wise ECG morphology. Accordingly, a single-channel ECG can be modeled as follows
\begin{equation}
\text{ecg}(t) = \sum_n{h(t - T_n;\gamma_n)}, \quad T_n = n T + \eta_n
\label{eq:ECGmodel}
\end{equation}
where $T_n$ denotes the R-peak locations, $T$ is the average RR-interval, $\eta_n$ is the RR-interval deviation, $h(t; \cdot)$ is the ECG morphology and $\gamma_n$ denotes the beat-wise variations of the ECG morphology considered as a model parameter. It is shown in the sequel that this simple pseudo-periodic model can be used for removing mECG interferences from the fECG. The limitation of this model is that the natural beat-wise variations of the heart-rate, which result in the shortening or prolongation of certain segments of the ECG are not explicitly considered in this model. In fact, a more accurate model should permit the compression and expansion of the ECG morphology, as the heart-rate evolves over time. Based on this requirement, the notion of \textit{cardiac phase} has been introduced for modeling and development of ECG filtering and later used for mECG cancellation and fECG extraction from multichannel abdominal recordings.

\subsubsection{The notion of cardiac phase}
\label{sec:cardiacphase}
\begin{figure}[b]\centering
\includegraphics[trim=0in 1.7in 0in 1.5in,clip,width=\columnwidth]{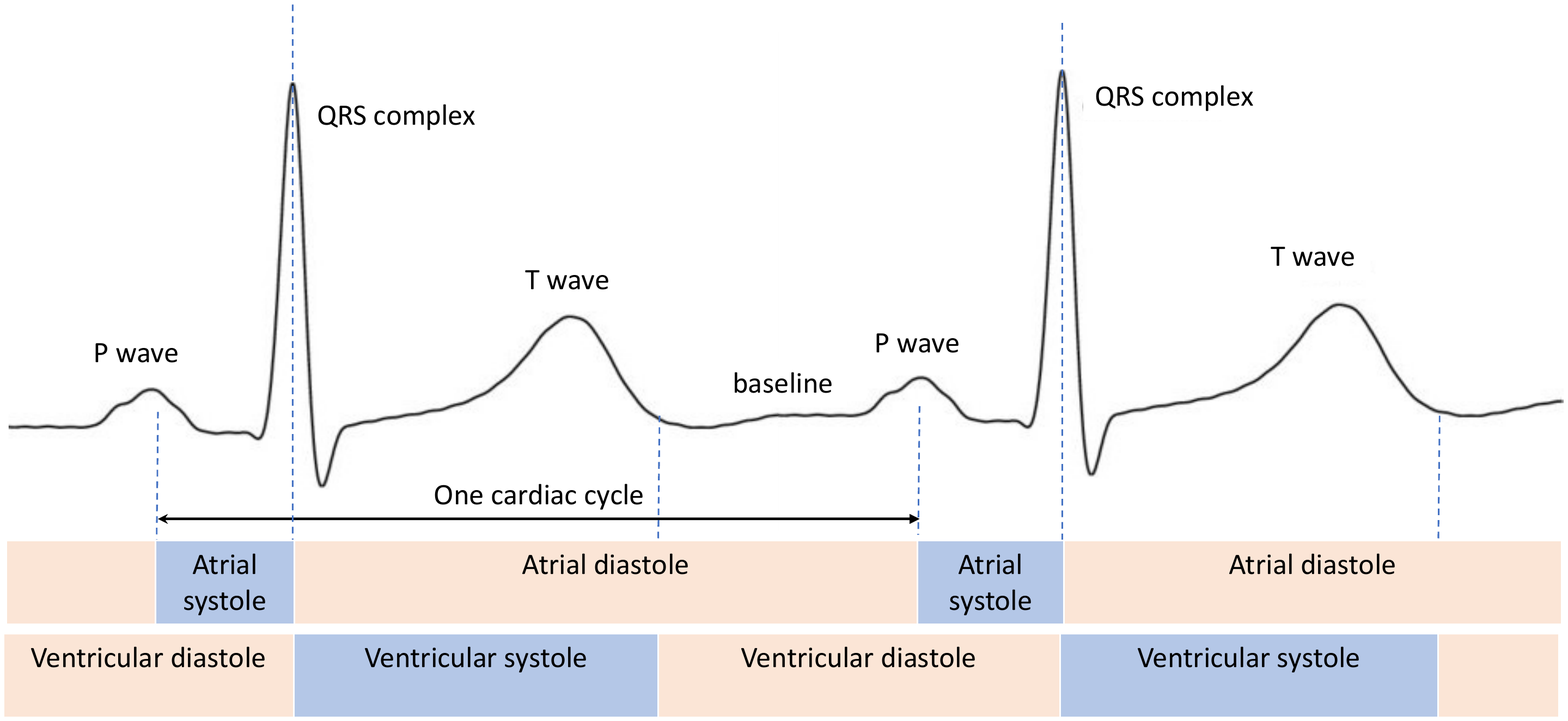}
\caption{The cardiac states across successive beats versus the ECG}
\label{fig:ECG}
\end{figure}

\begin{figure}[tb]
\centering
\includegraphics[width=5cm]{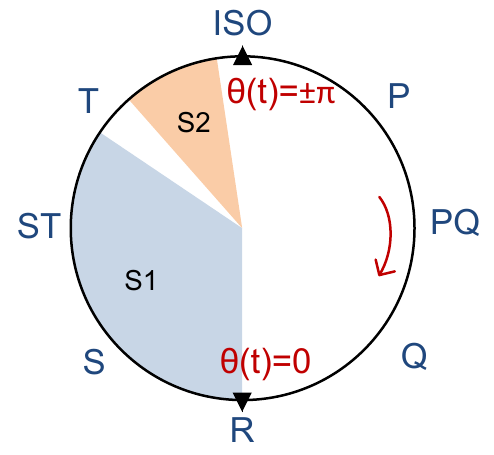}
\caption{The cardiac cycle phase-wrapped on the unit circle using the phase signal. The heart sounds S1 and S2 are also demonstrated for reference to the mechanical activity of the heart}
\label{fig:statemachine}
\end{figure}

\begin{figure}[tb]\centering
\includegraphics[width=.6\columnwidth]{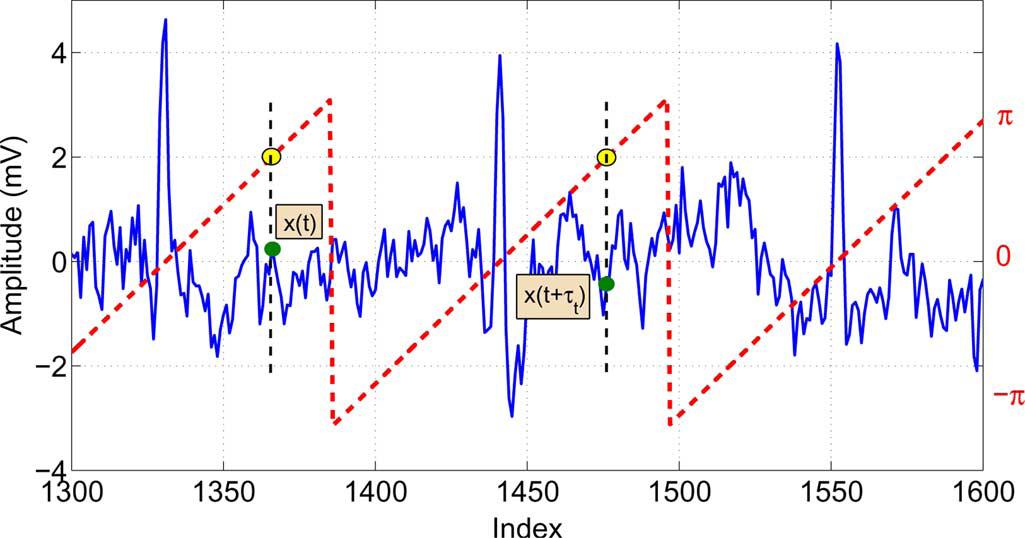}
\caption{The cardiac phase using a linear phase; adopted from \cite{Sameni2008a}}
\label{fig:Linearphase}
\end{figure}

\begin{figure}[tb]
\centering
\includegraphics[trim=0in 0in 0in 0in,clip,width=3.5in]{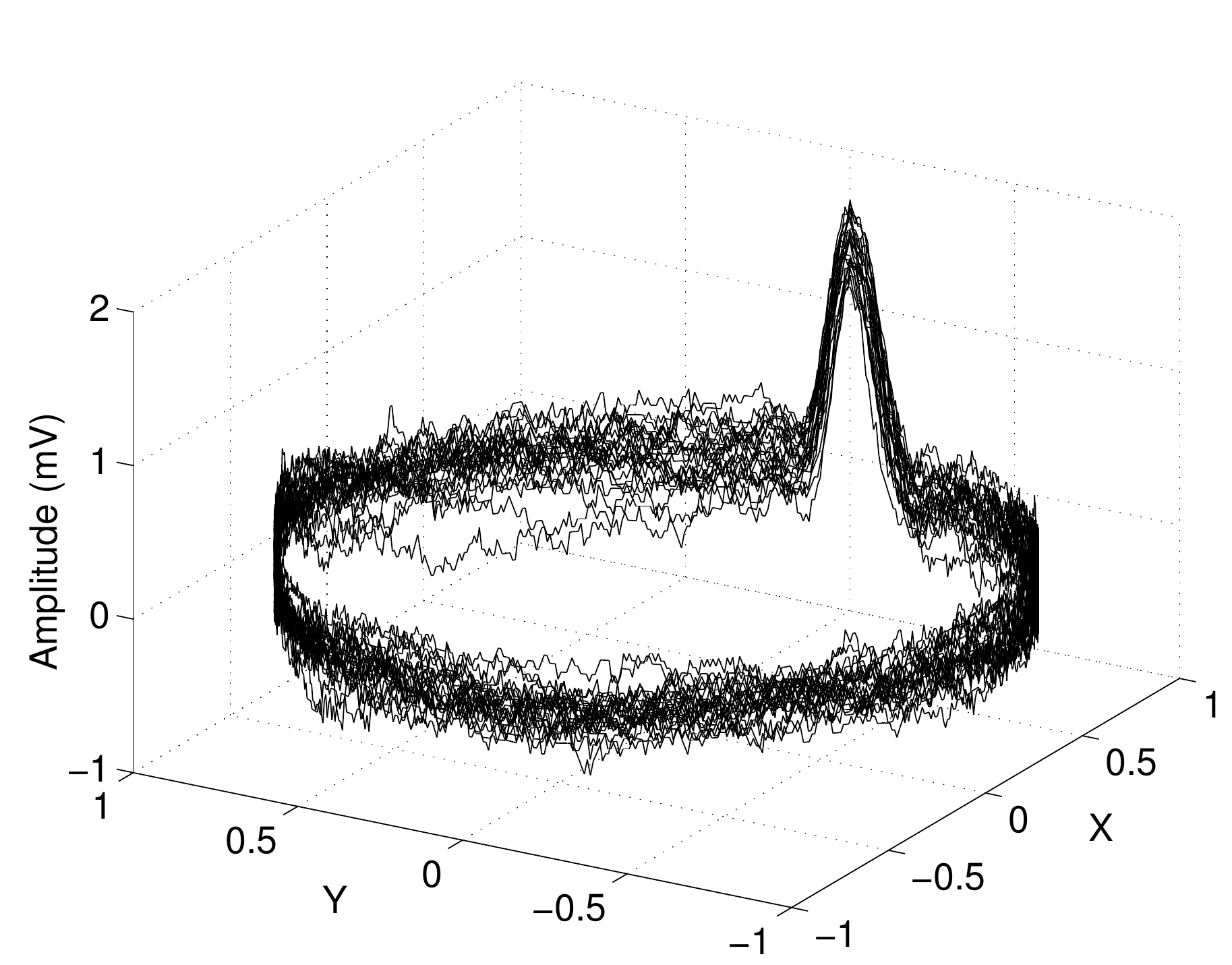}
\caption{Polar representation of a noisy ECG using the cardiac phase signal $\theta(t)$ \cite{Sameni2008}}
\label{fig:PhaseFig}
\end{figure}

The cardiac cycle, or the period from one sinoatrial (SA) node activation to the next, consists of a period of relaxation (diastole), during which the heart is filled with blood, followed by a period of contraction (systole), as shown in Fig.\ref{fig:ECG}. For a normal heart, the contraction and relaxation phases are subject to continuous change, controlled by the autonomic nervous system, and these changes do not necessarily take place ``linearly'' along the beats. In other words, when the heart-rate changes, the different segments of the ECG are not scaled to the same extent. Specifically, it is believed that when the heart-rate increases, e.g., due to physical activity, tachycardia, bradycardia, etc., the duration of the action potentials and the period of the systolic phase also decrease, but not as much as the variations of the diastolic phase of the ECG \cite{hall2006}.
Alternation in the cardiac cycle duration depends on various physiological factors, which can be modeled using the notion of \textit{cardiac phase}. As proposed in \cite{Sameni2008a}, the cardiac phase $\theta(t) \in [-\pi, \pi]$ (or alternatively $[0  , 2\pi]$) can be used as a variable for the mathematical representation of the pseudo-periodic behavior of the heart over different beats. As illustrated in Figs.~\ref{fig:statemachine} and  \ref{fig:Linearphase}, each electrophysiological state of the heart over a full cardiac cycle can be mapped to a unique value between $[-\pi, \pi]$. In other words, the linear phase $\theta(t)$, provides a means of phase-wrapping the RR-interval onto the $[-\pi,\pi]$ interval. Therefore, the ECG--- regardless of its RR-interval deviations--- is converted to a polar representation, in which the ECG components in different beats, such as the P, Q, R, S, and T-waves, are more or less phase-aligned with each other, especially over the QRS segment (Fig. \ref{fig:PhaseFig}). As a result, identical contraction or relaxation states of the heart are mapped to identical values of $\theta(t)$. For example, by convention, the peak of the systole (the R-peak), can be fixed to $\theta(t)=0$. This convention maps the \textit{ventricular diastolic} state of the heart to negative phases and the \textit{ventricular systolic} state to positive phases. In this case, the phase-wrapping from $-\pi$ to $\pi$ takes place just after the T-wave offset, and at the beginning of the relaxation period of the heart, where the ECG level is at its isoelectric or baseline (cf. Figs.~\ref{fig:ECG} and \ref{fig:statemachine}).

From the cardiac phase signal, some other quantities can be calculated, which have been extensively used in the literature, for modeling and denoising adult and fetal ECG signals:

\begin{itemize}
    \item \textit{Cardiac angular frequency and instantaneous heart-rate:}
The \textit{cardiac angular velocity} $\omega(t)$, in $\text{rad}/s$ and the \textit{instantaneous heart-rate} in Hz are defined as follows:
\begin{equation}
\omega(t) = 2\pi f(t) = \frac{d\theta(t)}{dt}
\label{eq:cardiacangularfrequency}
\end{equation}
Therefore, the conventional RR-interval can be considered as the average of the reciprocal of $f(t)$, over one beat. Note that both $f(t)$ and $\omega(t)$ are rather abstract quantities for conventional ECG analysis, in the sense that only the RR-interval is known as a clinical index (the duration between the onsets of successive ventricular systoles). Nevertheless--- again in an abstract sense--- $f(t)$ and $\omega(t)$ can be considered as the speed of cardiac dipole rotation in the myocardium.
\item \textit{Time-varying cardiac period:}
In each ECG cycle, the sample at the time instant $t$ has a \textit{dual sample} in other beats, which have the same phase value. We define the distance between sample $t$ and its dual sample in the previous beat, as the \textit{time-varying period}, denoted by $\tau_t$ and mathematically defined as:
\begin{equation}
\tau_t = \argmin_{\tau > 0}\{\theta(t-\tau)=\theta(t)\}
\label{eq:tau}
\end{equation}

\end{itemize}

\subsubsection{Dipolar models}
According to dipolar models of the heart \cite{MP95,Malmivuo2000}, the signals acquired from different body surface leads are projections of the cardiac dipole vector onto the recording electrode axes. Due to the properties of the fetal and maternal body volume conductors, detailed in Section \ref{sec:volumeconductor}, the signals acquired by all body surface leads are quasi-periodically time synchronous with the cardiac phase. These properties have been used in the literature to develop synthetic models for generating maternal and fetal cardiac waveforms. The first modeling framework, explicitly focused on the fECG, was developed in \cite{Oostendorp1989}. This study, was based on maternal body surface potentials modeling using finite elements methods and assuming a dipolar model for the fetal heart. Another popular model is based on the single-channel ECG model proposed by McSharry and Clifford \cite{McSharry2003,cliffordSPIE04,SSJB05,GDC06}, which was later extended to the fECG in \cite{SCJS06}. Accordingly, the following dynamic model has been proposed for simulating the three dipole coordinates of the \textit{vectorcardiogram} (VCG), denoted by $\mathbf{s}(t)=[x(t), y(t), z(t)]^T$:
\begin{equation}
\begin{array}{l}
\displaystyle\dot{\theta}=\omega\\
\displaystyle\dot{x}=-\sum_{i}\frac{\alpha^x_i\omega\Delta\theta^x_i }{(b^x_i)^2}\exp[-\frac{(\Delta\theta^x_i)^2}{2(b^x_i)^2}]\\
\displaystyle\dot{y}=-\sum_{i}\frac{\alpha^y_i\omega\Delta\theta^y_i}{(b^y_i)^2}\exp[-\frac{(\Delta\theta^y_i)^2}{2(b^y_i)^2}]\\
\displaystyle\dot{z}=-\sum_{i}\frac{\alpha^z_i\omega\Delta\theta^z_i}{(b^z_i)^2}\exp[-\frac{(\Delta\theta^z_i)^2}{2(b^z_i)^2}]
\end{array}
\label{eq:SyntheticDipole}
\end{equation}
where $\Delta\theta^x_i=(\theta-\theta^x_i)\mod(2\pi)$, $\Delta\theta^y_i=(\theta-\theta^y_i)\mod(2\pi)$, $\Delta\theta^z_i=(\theta-\theta^z_i)\mod(2\pi)$, $\omega=2\pi f$ is the cardiac angular velocity and  $f$ is the instantaneous heart-rate, as defined in (\ref{eq:cardiacangularfrequency}). Mathematically, the first equation in (\ref{eq:SyntheticDipole}) generates a circular trajectory, which rotates with the frequency of the heart-rate. In other words each cycle of $\theta$ sweeping from 0 to $2\pi$ corresponds to one cardiac cycle, and the other equations model the dynamics of the three coordinates of the source vector $\mathbf{s}(t)$ as a summation of Gaussian functions with amplitudes $\alpha^x_i$, $\alpha^y_i$, and $\alpha^z_i$, widths $b^x_i$, $b^y_i$, and $b^z_i$, each located at rotational angles $\theta^x_i$, $\theta^y_i$, and $\theta^z_i$. The intuition behind this set of equations is that the baseline of each of the dipole coordinates is pushed up and down, as the trajectory approaches the centers of the Gaussians, resulting in a moving vector in the $(x,y,z)$ coordinate space. In practice, by adding some deviations to the parameters of (\ref{eq:SyntheticDipole}), for example by considering them as random variables rather than deterministic constants, more realistic ECG with inter-beat variations can be generated.

The above model of the rotating dipole vector is rather general, since due to the \textit{universal approximation} property of Gaussian mixtures, any continuous function such as the dipole vector coordinates can be modeled with a sufficient number of Gaussian functions, up-to an arbitrarily close approximation \cite{bb16291}. Moreover, the model is a very good choice for ECG signals of both adults and fetuses, for which the Gaussian kernels can be eventually related to clinical parameters of the ECG. Equation (\ref{eq:SyntheticDipole}) can also be thought as a model for the orthogonal lead VCG coordinates, with an appropriate scaling factor for the attenuations of the volume conductor. This analogy between the orthogonal VCG and the dipole vector was used in \cite{SCJS06} to estimate the parameters of (\ref{eq:SyntheticDipole}) from the three \textit{Frank-lead} VCG recordings.

\begin{figure}[tb]
\centering
\includegraphics[trim={0 0.6in 0 0},clip, width=0.7\columnwidth]{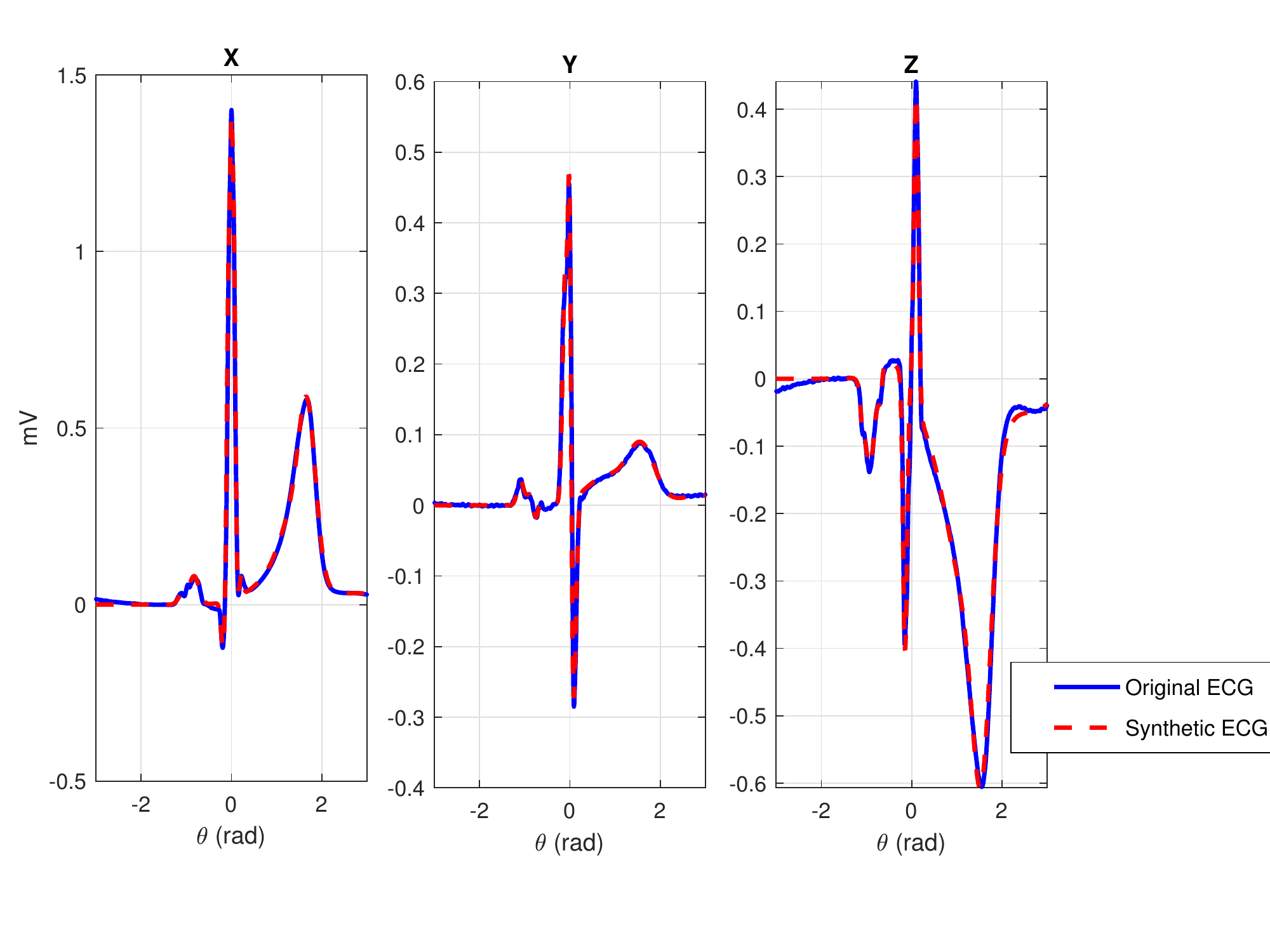}
\caption{Synthetic ECG signals generated by the VCG model in (\ref{eq:SyntheticDipole})}
\label{fig:SynECG}
\end{figure}

\begin{figure}[tb]
\centering
\includegraphics[width=0.5\columnwidth]{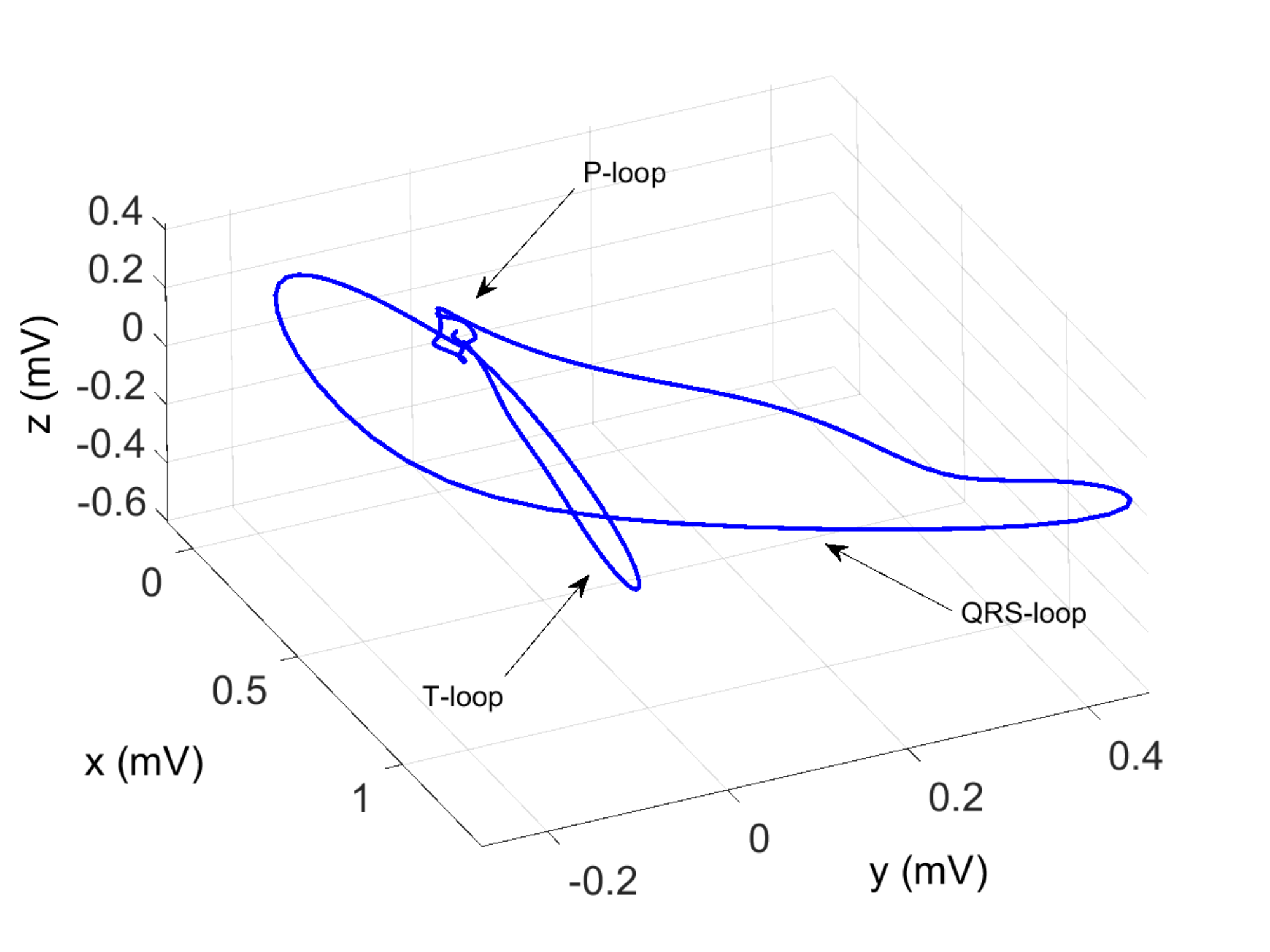}
\caption{Typical synthetic VCG loop. Each clinical lead
is produced by mapping this trajectory onto a 1-dimensional vector in this 3-dimensional space.}
\label{fig:SynVCG}
\end{figure}

\begin{figure}[tb]
\centering
\begin{subfigure}{0.32\columnwidth}
\includegraphics[width=0.95\columnwidth]{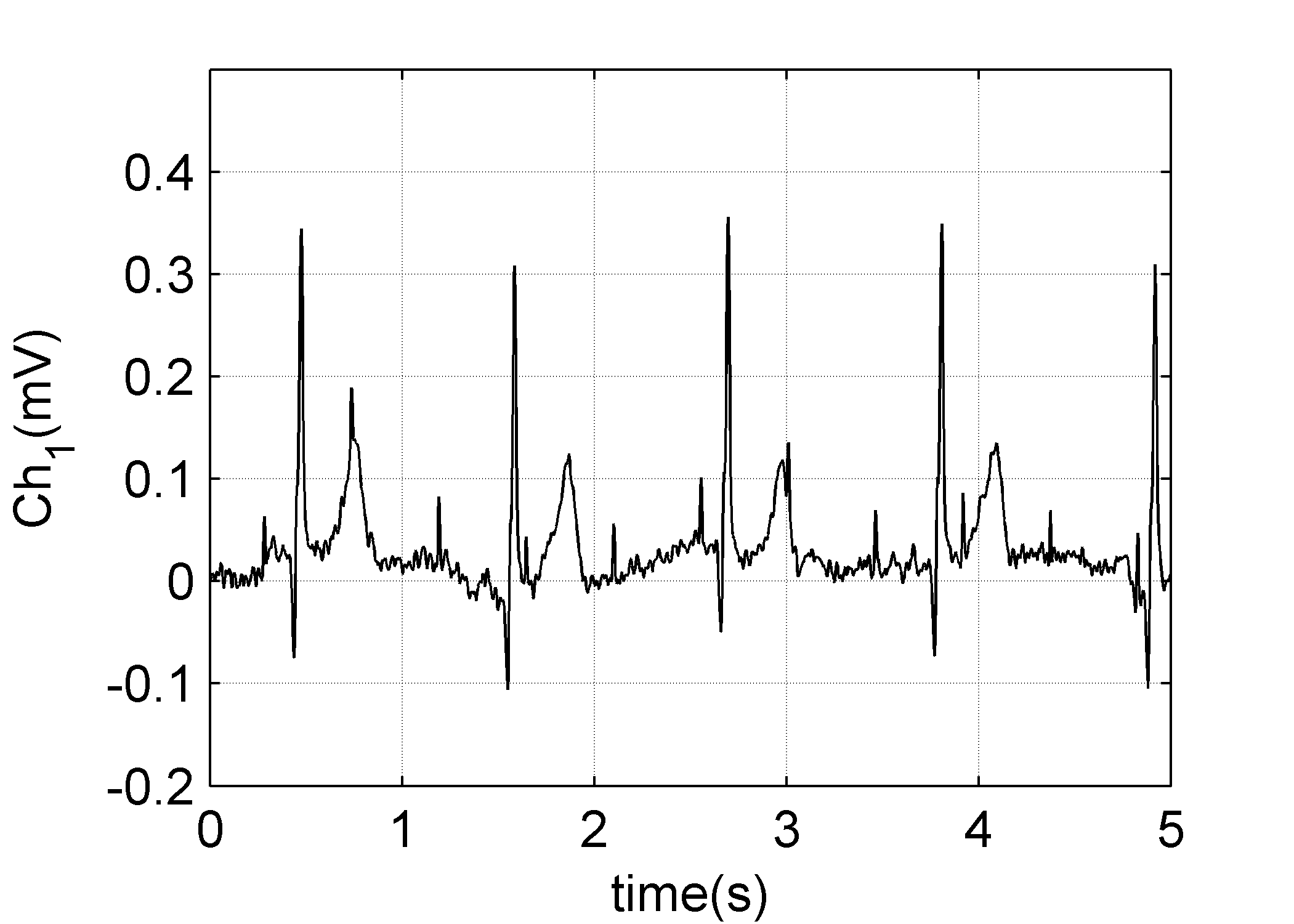}    
\end{subfigure}
\begin{subfigure}{0.32\columnwidth}
\includegraphics[width=0.95\columnwidth]{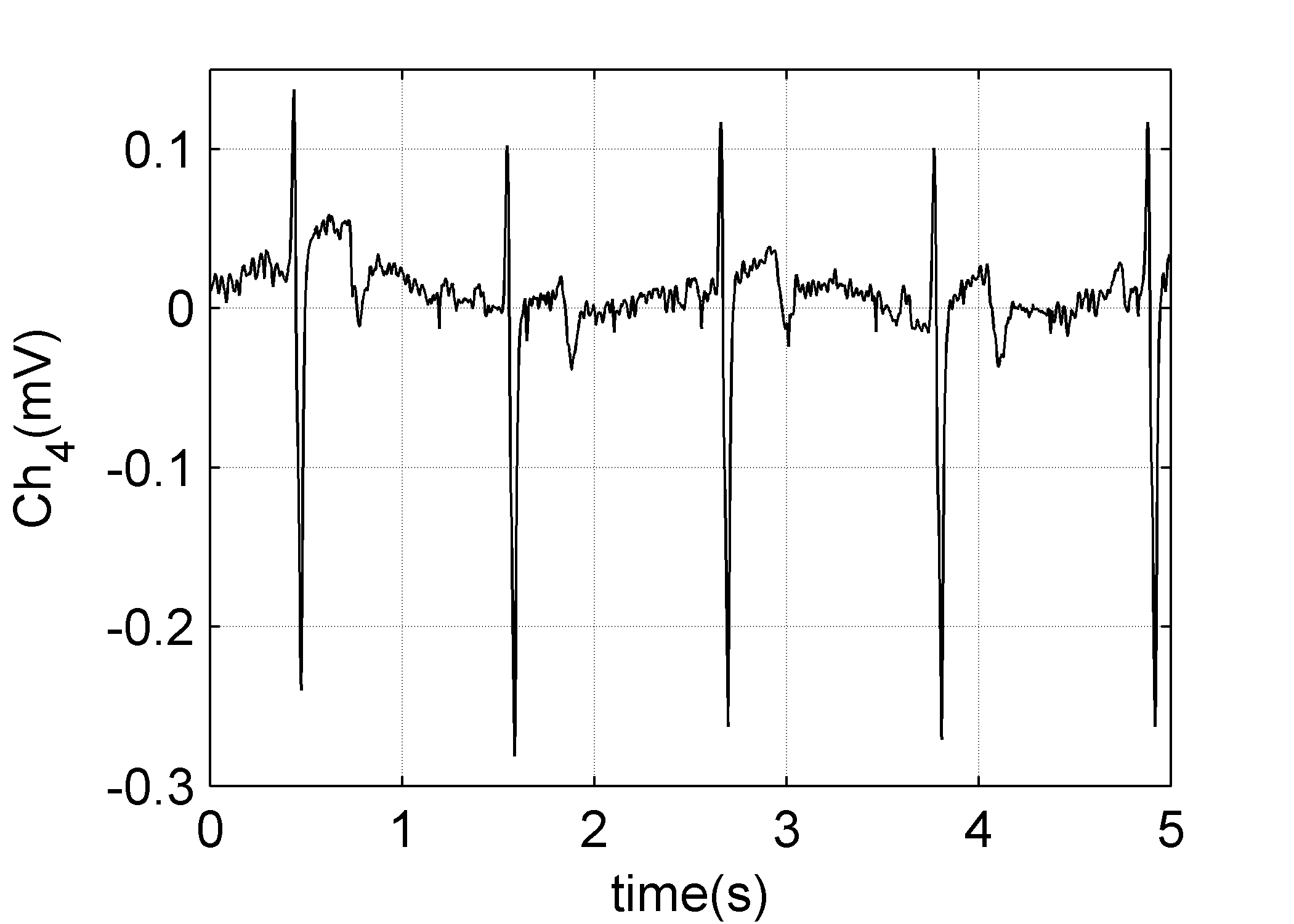}    
\end{subfigure}
\begin{subfigure}{0.32\columnwidth}
\includegraphics[width=0.95\columnwidth]{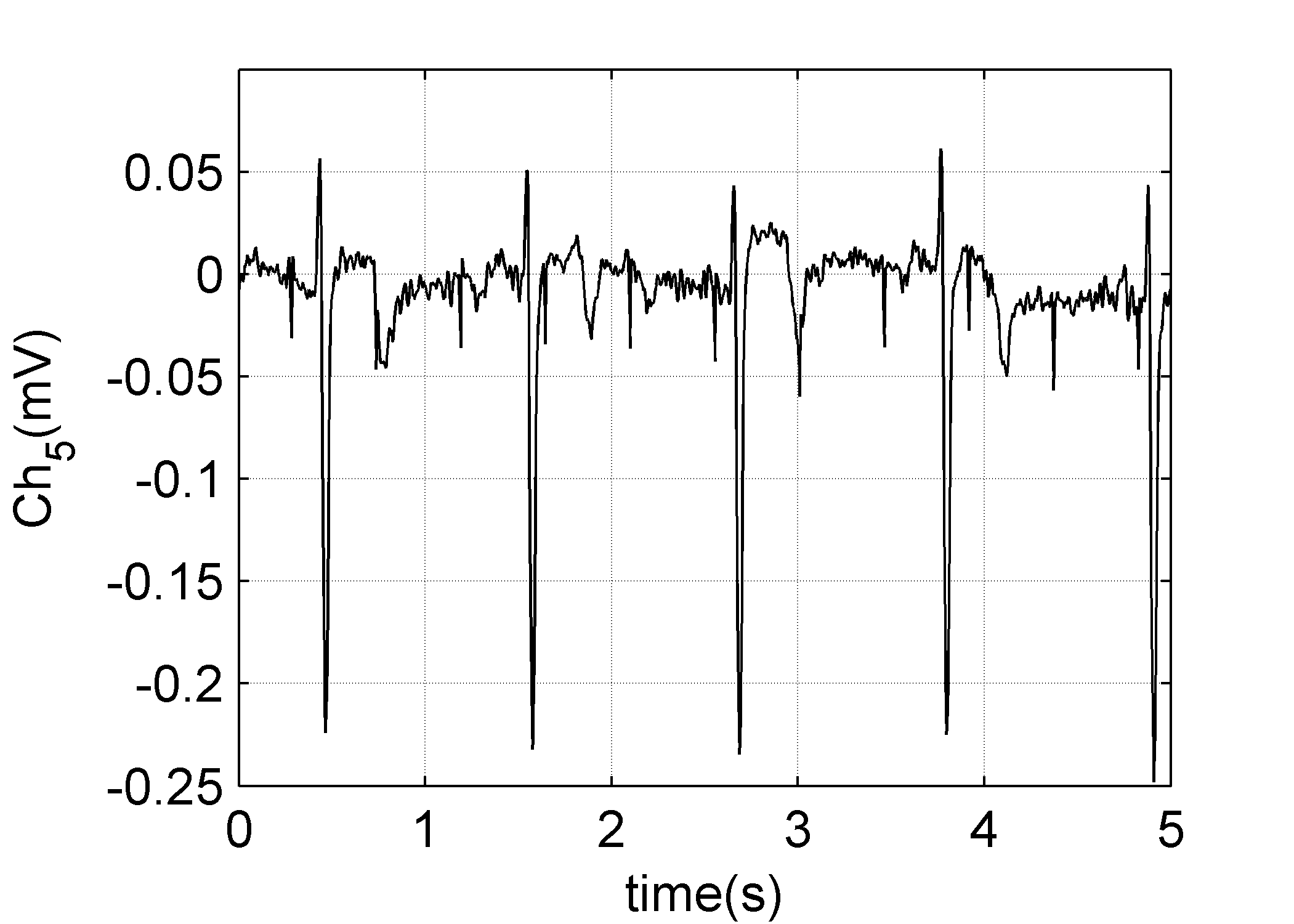}    
\end{subfigure}
\caption{Typical multichannel ECG generated by a synthetic maternal-fetal ECG generator}
\label{fig:syntheticabdominal}
\end{figure}

By placing the resulting cardiac source models of the maternal and fetal cardiac dipoles in (\ref{eq:datamodel}), realistic mixtures of maternal abdominal signals are obtained. In Fig.~\ref{fig:SynECG} and Fig.~\ref{fig:SynVCG}, a sample signal corresponding to the cardiac dipole coordinates and the resulting three-dimensional vectorcardiogram loop are shown for illustration. A multichannel signal generated by this technique plus synthetic noise is also shown in Fig.~\ref{fig:syntheticabdominal}. The functions required for generating synthetic maternal abdominal signals are online available at \cite{OSET3.14}, with the parameter set listed in \cite{SCJS06}. Accordingly, the number of the Gaussian functions used for modeling the maternal and fetal ECG are not necessarily the same for the different channels and they can be selected according to the shape of the desired channel. Databases of synthetic maternal and fetal cardiac signals generated by this method are online available for algorithm evaluation \cite{OSET3.14,Andreotti2016}.


\section{Digital noninvasive fetal ECG acquisition}
\label{sec:acquisition}
\subsection{Acquisition front-end requirements}
\begin{figure}[tb]
\centering\includegraphics[trim=1in 2.8in 2in 2.5in,clip,width=0.8\columnwidth]{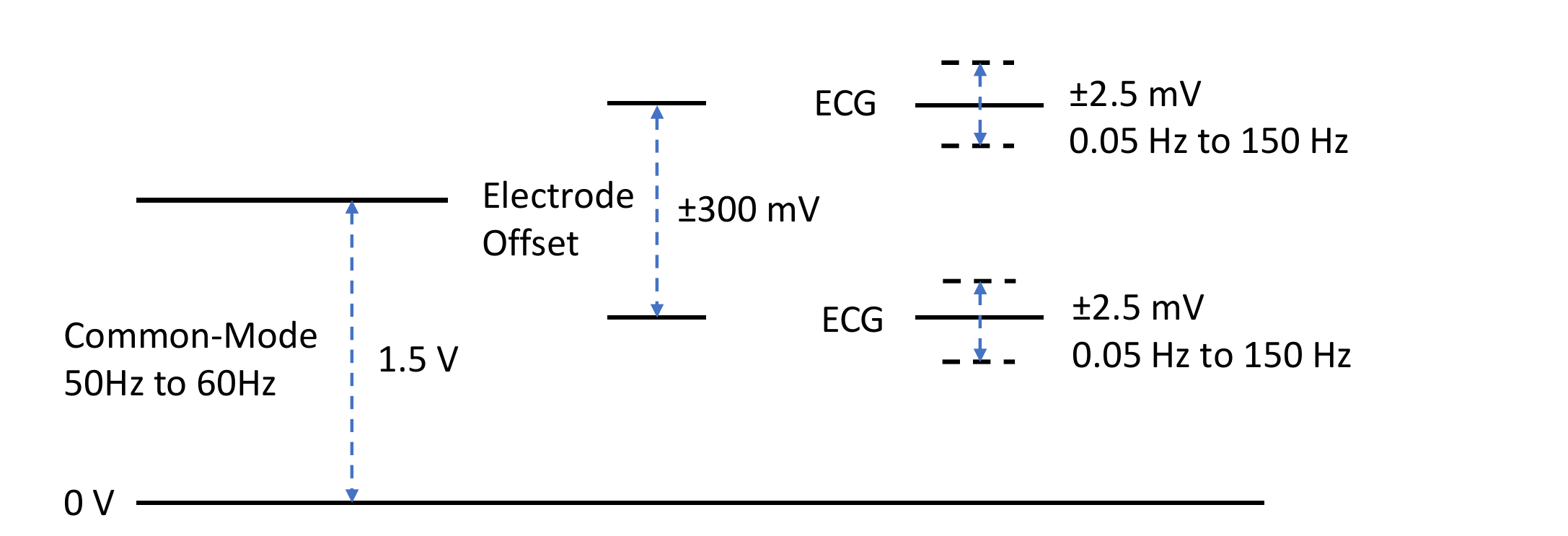}
\caption{The dynamic range of analog ECG frontends; adapted from \cite{TI:SBAA160A}.}
\label{fig:ecgrange}
\end{figure}

To date, there are no standards or widely accepted protocols for fECG acquisition. Nevertheless, the common properties of the fetal and adult ECG and the existing open-access fECG databases can be used to set some baselines. It is known that the effective bandwidth of adult ECG is between 0.05~Hz to 150~Hz, with a maximum span of $\pm$5~mV in magnitude, besides the common-mode and electrode offset voltages, as shown in Fig.~\ref{fig:ecgrange}. It is recommended that the front-end noise of adult ECG devices be bellow 30$\mu$V in root mean square (RMS) \cite{IEC60601-2-25:2011}. On the other hand, in the currently available maternal abdominal datasets, the fECG can be ten to twenty times smaller than the mECG. At the same time, due to the sharper QRS and higher heart-rate of the fetus as compared with the adult ECG, the fECG is wider in bandwidth. As a baseline, a bandwidth between 0.05~Hz to 250~Hz covers the dominant bandwidth of the fECG. In this range, the most informative band is from 10~Hz to 70~Hz, which is used for fetal heart rate detection, while the full bandwidth is recommended for fECG morphological analysis.
    
According to the sampling theorem, the sampling frequency of a signal should be above twice the maximum frequency of the input signal (known as the \textit{Nyquist rate}) to avoid \textit{aliasing} and to guarantee information retrieval. But for biomedical applications, signal visualization is an integral aspect of the analysis and sampling at the minimal Nyquist rate does not result in visually agreeable signals. Therefore, biomedical signals are commonly over-sampled above the Nyquist rate for better visualization and possible SNR improvement during post-processing.
 
\begin{figure}[tb]
\centering\includegraphics[width=\columnwidth]{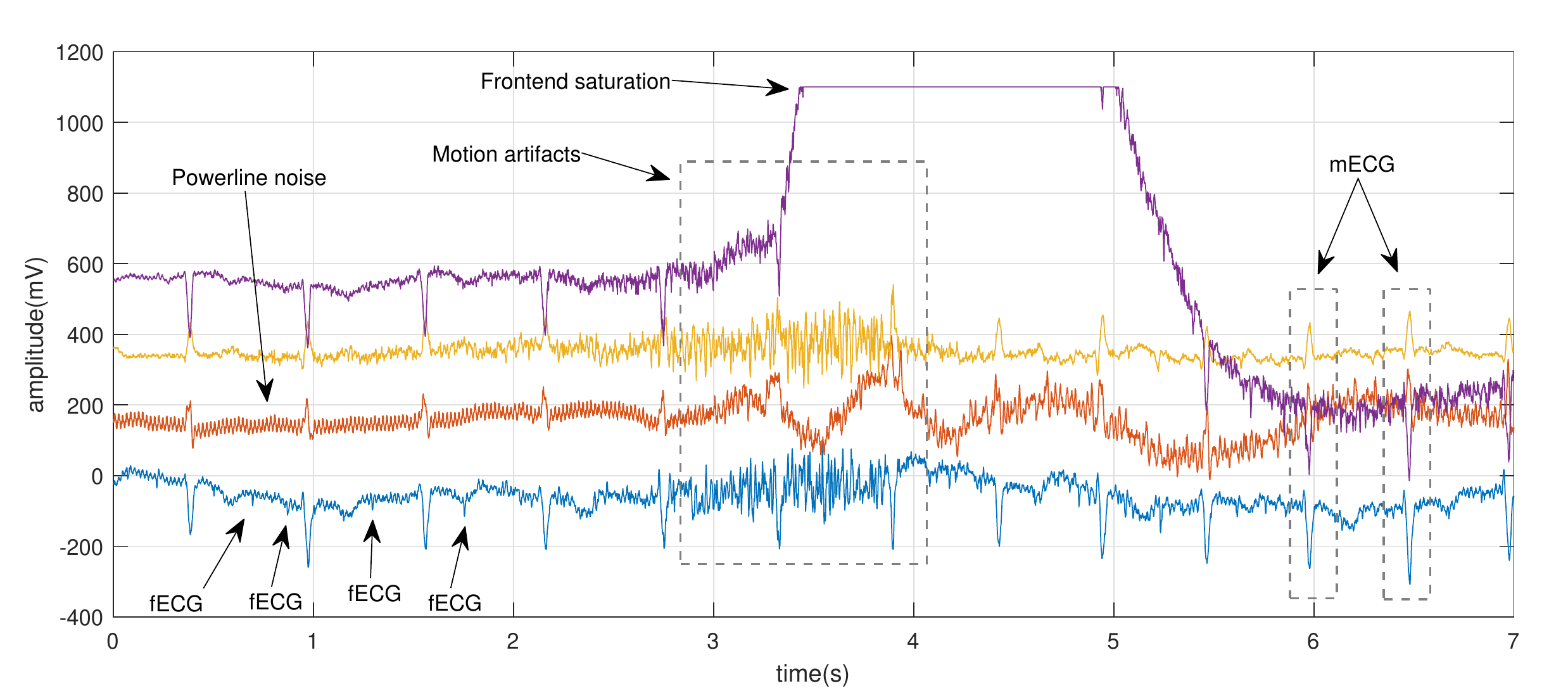}
\caption{A typical segment of maternal abdominal recordings containing various signals and noises. The dynamic range of the digital front-end should be such that the acquired signals would not overflow due to interfering signals such as the maternal ECG. Refer to the text for further details.}
\label{fig:mecgfecgrange}
\end{figure}

\begin{figure}[tb]\sidecaption
\centering\includegraphics[trim=0.1in 0in 0in 0in,clip,width=3.8in]{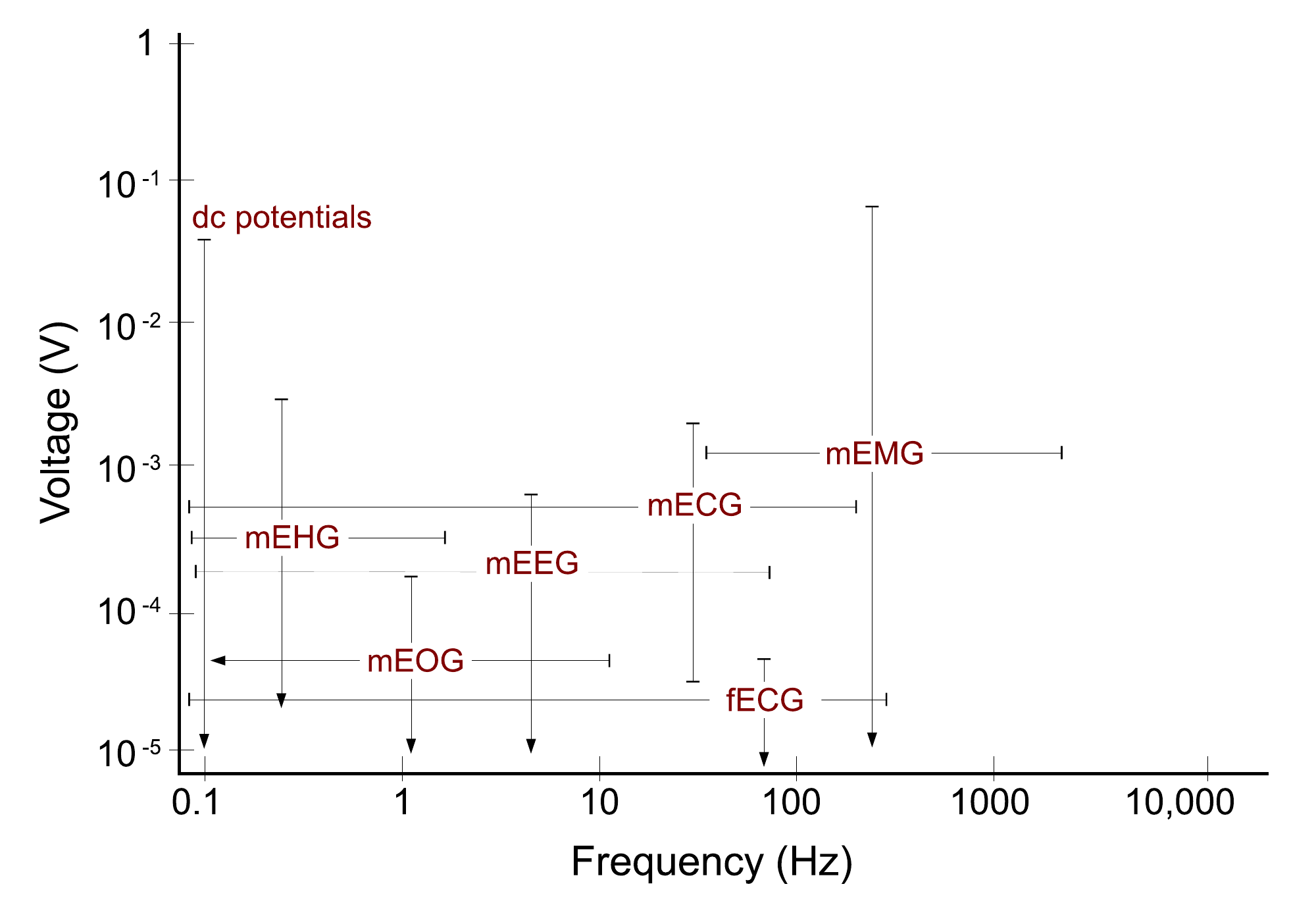}
\caption{The amplitude and frequency range of the maternal electrocardiogram (mECG), electroencephalogram (mEEG), electrooculogram (mEOG), electromyogram (mEMG), electrohystrogram (mEHG), and the fetal ECG (fECG). Accordingly, different biosignals interfere with the fetal ECG  \cite{Webster1998,Devedeux1993,Snowden2001,SCJS06}. Note that the fECG amplitude depends on the sensor position, fetal positioning and age.}
\label{fig:Signals}
\end{figure}

As for the amplitude, fECG acquisition systems should have a broad dynamic range to permit fECG acquisition without overflow or saturation due to interfering signals such as the mECG and power-line noise, as demonstrated in Fig.~\ref{fig:mecgfecgrange}. In Fig. \ref{fig:Signals} the amplitude and frequency range of the fECG is compared with other biosignals and artifacts. Accordingly, the fECG spectrally overlaps with the interfering biosignals and is significantly weaker in amplitude. Therefore, classical frequency domain filtering is ineffective, especially for the mECG, which is the dominant biomedical interfering signal for the fECG.

\subsection{Analog to digital conversion requirements}
The procedure of analog-to-digital signal conversion, inevitably adds quantization noise to the signal and reduces the signal-to-noise ratio (SNR). It is therefore important to keep the quantization noise below or at the same level as the analog signal noise level, to avoid significant signal quality degradation. The SNR due to the quantization procedure can be calculated from the standard equation:
\begin{equation}
    \text{SNR(dB)} = 6.02 b + 1.76 + 10\log_2(\text{OSR})
    \label{eq:snrformula}
\end{equation}
where $b$ is the number of analog-to-digital converter (ADC) bits and $\text{OSR} = f_s/\text{BW}$ is the \textit{over-sampling ratio}, which is the ratio of the sampling frequency $f_s$ and the bandwidth (BW) of the input signal. The SNR improvement due to the OSR term in (\ref{eq:snrformula}) is only obtained by post-filtering, if the signal is sampled above the minimal Nyquist rate. Note that the standard SNR equation (\ref{eq:snrformula}) is based on the assumption of a sinusoidal input signal with close to full-scale amplitude range (typically 1~dB below the ADC full-scale level) applied to a symmetric voltage referenced ADC with uniform quantization levels, and assuming that the quantization noise is uniformly distributed over the entire Nyquist bandwidth \cite{kester2005data}. This standard procedure enables the manufacturers and circuit designers to have a unified comparison between different ADC devices.

It should also be noted that in digital electronics circuits design, the maximum SNR expected from the nominal number of ADC bits is not achievable. In fact, depending on the ADC technology, sampling frequency and the printed circuit board (PCB) design and quality, the \textit{effective number of bits} (ENOB) is what is obtained in practice:
\begin{equation}
    \tilde{b} = \frac{\text{SNR}_{\text{real}} - 1.76\text{dB}}{6.02}
\end{equation}
where $\text{SNR}_{\text{real}}$ is the SNR that is obtained in practice and $\tilde{b}$ is the ENOB, which is not necessarily an integer value. For example, an ADC with 16 nominal bits may practically have 13.5 to 14 ENOBs. The ENOB is one of the standard properties of all ADC, which is documented in the datasheets of ADC devices by the manufacturers. Considering that beyond the ADC chip technology, the ENOB also depends on the circuit design quality, it is measured in practice by sweeping close to full-scale sinusoidal signals within the Nyquist band of the manufactured circuit front-end (by applying a signal generator to the ADC front-end), and by logging the samples acquired by the ADC. The real SNR ($\text{SNR}_{\text{real}}$) can be eventually calculated by analyzing the sampled signals in software. This is a standard procedure that is performed during the design and quality control of all (including medical) equipment.  
The overall recommended front-end specifications for noninvasive fECG acquisition are summarized in Table~\ref{tab:fecgrecommendations}.
\begin{table}[tb]
\caption{The recommended front-end specifications for fetal ECG acquisition}
    \centering
    \begin{tabular}{|p{1.9in}|p{4.1in}|}
    \hline
        \textbf{Property} & \textbf{Range}\\\hline
        Bandwidth (-3dB cutoff frequency) & \multirow{1.8}{4in}{Acceptable: 0.05Hz to 250Hz\\
        Preferred: 0.05Hz to 1kHz (for better fECG-noise separability)}\\&\\\hline
        Amplified analog voltage range & 
        3--5V (preferably differential pairs)\\\hline
        Analog-to-digital resolution &
        \multirow{1.8}{4in}{Low resolution: 16~bits\\
        High resolution: 24~bits}\\&\\\hline
        Sampling frequency & 
        \multirow{1.8}{4in}{Minimum: 500~Hz\\
        Acceptable: 1~kHz\\
        High-resolution: 5--10~kHz}\\&\\&\\\hline
        Sampling sequence & 
        \multirow{1.8}{4in}{Preferred: Simultaneous\\
        Acceptable: Sequential (multiplexed); only at high sampling frequencies}\\&\\&\\\hline
        Number of channels & Between 8 to 32 with dedicated mECG channels used as reference\\\hline        
    \end{tabular}
    \label{tab:fecgrecommendations}
\end{table}

\subsection{Sensor placement}
\label{sec:sensorpositions}
In order to maximize the chance of retrieving the fECG from maternal abdominal leads, it is common to use multiple leads spread over the abdomen, lower back and the two sides of the maternal body. The sensors should ideally be close to the fetus and the referencing of the leads should be such that the electrical fields due to the fetal heart pass through the differential pairs used for acquisition. To date, the number of abdominal channels used for research and clinical usage are very diverse, ranging from as few as one and as many as 144 abdominal channels. From the electronic and manufacturing perspective, using a few leads placed close together in a patch of disposable or reusable electrodes is very advantageous, as compared with using numerous electrodes distributed all over the maternal abdomen and back. However, as explained throughout this chapter, a group of sensors placed close to each other are prone to becoming highly dependent and result in mathematically low-rank and non-invertible mixture of signals, which is inappropriate for multichannel fECG extraction. Therefore, there is a compromise between the simplicity of the acquisition system and the robustness to fetal positioning. The major fECG acquisition technologies use between 8 to 32 channels, including one or more reference leads for the mECG acquired from maternal chest leads.

\section{Single-channel fetal electrocardiogram extraction}
\label{sec:singlechannel}
Single-channel fECG extraction algorithms refer to the category of methods that use a single maternal abdominal channel and possibly a set of reference electrodes for acquiring the mECG from the maternal chest. An interesting comparative survey on the advantages and limitations of these methods was conducted in \cite{behar2014comparison}. In this section, some of the major algorithms of this class of techniques is reviewed in further detail.
\subsection{Na\"ive fetal electrocardiogram detection and extraction}
\label{sec:naive}
Before the advances in digital signal processing in recent decades, fECG detection was performed over raw paper prints of abdominal recordings, without any processing. For instance in \cite{Larks1962}, by visual inspection, several cases were reported in which due to the \textit{vertex} presentation of the fetus, the fetal R-peaks appeared as positive peaks while the maternal R-peaks had negative peaks. It is evident that such studies remained discrete and subjective, since due to the low SNR, fECG detection by visual inspection is not always applicable and highly depend on the fetal presentation and gestational age. Nevertheless, visual inspection remains as the first intuitive test for machine-based fECG extraction algorithms.

\subsection{Template subtraction and cyclostationary random process theory}
\label{sec:templatesubtraction}
Template subtraction is the most basic method for mECG cancellation from maternal abdominal recordings \cite{andreotti2014robust, li2017efficient}. Despite its simplicity, it was shown in \cite{jamshidian2018fetal} that using the theory of cyclostationarity, this technique can be the \textit{optimal cyclostationary Wiener filter}, when applied properly by compensating the inter-beat variations of the mECG. The proof was inspired by the problem of pulse amplitude demodulation, a well-known method in the context of telecommunications \cite[Ch. 4]{gardCyclo}.

Let us consider the signal $x(t) = \sum_n{c_n g(t-nT)}$, where $g(\cdot)$ is an arbitrary \textit{known} function and $c_n$ is a stationary time-sequence. It can be shown that the problem of optimal filtering of $x(t)$, which is a \textit{wide-sense cyclostationary} random process, from the additive mixture $z(t) = x(t) + \eta(t)$ (where $\eta(t)$ is a stationary noise) reduces to the problem of estimating the minimum mean square estimate of $c_n$ and repeating $g(\cdot)$ at multiples of $T$, using the estimated amplitude \cite[p. 253]{gardCyclo}, \cite{jamshidian2018fetal}.

\begin{figure}[tb]
\centering\includegraphics[trim=1in 3in 0.8in 2in,clip,width=\columnwidth]{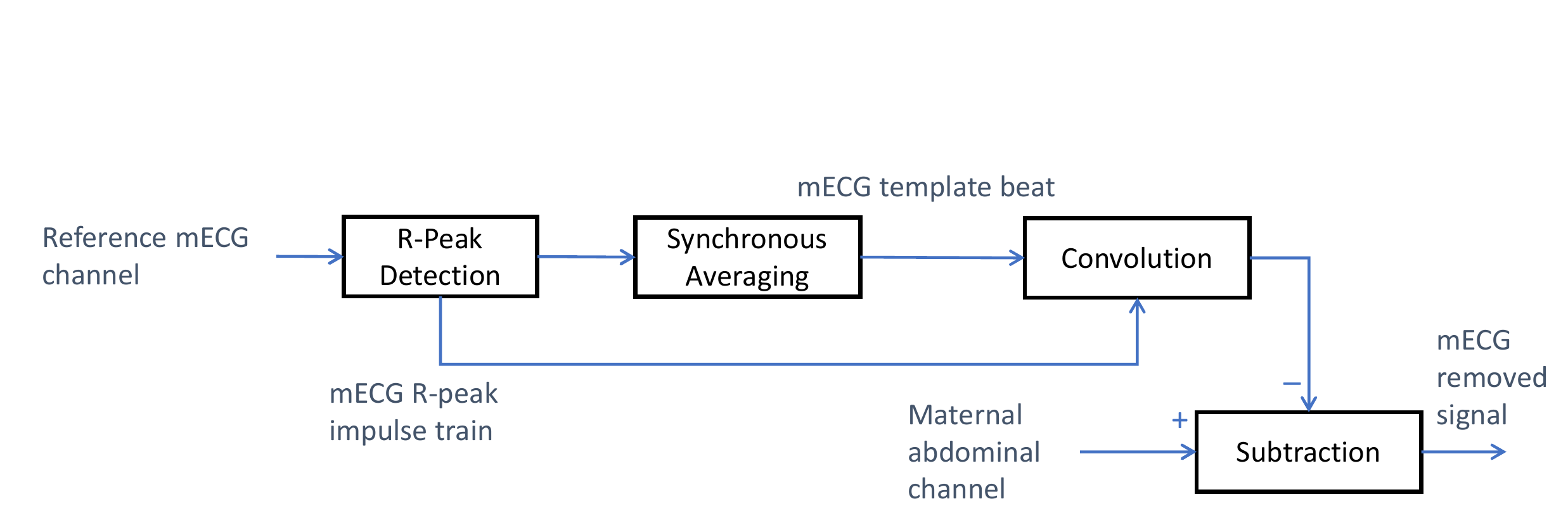}
\caption{Demonstration of the concept of optimal cyclostationary Wiener filtering for mECG cancellation.}
\label{fig:cyclostationaryWienerFilter}
\end{figure}

The above example is closely related to ECG denoising using a data model of the form (\ref{eq:ECGmodel}). Accordingly, if the inter-beat variations of the ECG were negligible, an ECG would be a \textit{wide-sense cyclostationary} process. In that case, one could \textit{optimally}--- in the Wiener filtering sense--- filter the ECG as demonstrated in Fig.~\ref{fig:cyclostationaryWienerFilter}: 1) detect the R-peaks, 2) perform synchronous averaging (or \textit{robust weighted averaging} \cite{Leski2004}) to find the average ECG beat, 3) reconstruct the denoised ECG by repeating the average beat at the R-peak locations \cite{jamshidian2018fetal}. Now suppose that $z(t) = x(t) + \eta(t)$ is a signal acquired from a maternal abdominal lead, $x(t)$ is the mECG and we are interested in the background signal $\eta(t)$, which is the fECG plus other noises. In this case, the above algorithm simply reduces to template subtraction: \textit{``construct a maternal ECG template and subtract this template by aligning it under the maternal R-peaks of the abdominal leads.''} However, since in reality the ECG has RR-interval deviations and morphological variations, instead of simple template subtraction that does not account for beat-wise heart-rate and morphological variations, it is better to make the procedure beat-wise adaptive, to compensate the beat-wise variations of the ECG (parametrized by $\gamma_n$ in the data model (\ref{eq:ECGmodel})). 

For example, the \textit{cardiac phase signal} introduced in Section (\ref{sec:morphologicmodel}) can be used to compensate the RR-interval deviations by time-warping \cite{Sameni2008a}. The minor beat-wise variations can further be compensated using classical beat alignment techniques \cite{sornmo1998,aastrom2000}. The template subtraction may also be made beat-wise adaptive, using Kalman filtering schemes as detailed in Section \ref{sec:kalmanfilter}. In fact, by applying such beat alignment techniques, the beat-wise deviations parametrized by $\gamma_n$ in (\ref{eq:ECGmodel}) are compensated and the resulting signal would become cyclostationary. As a result, the optimal cyclostationary Wiener filter for removing the mECG from maternal abdominal recordings is basically a template subtraction in the transformed domain (after compensating the beat-wise deviations of the mECG).

\subsection{Adaptive filters for fECG extraction}
\label{sec:adaptivefilters}
Adaptive filters are one of the popular methods used for mECG cancellation and fECG extraction. The procedure consists of training an adaptive filter for either removing the mECG using one or several maternal reference channels \cite{Widrow75,Outram95}, or directly training the filter for extracting the fetal QRS waves \cite{Farvet1968,Park92}. \textit{Ad hoc}, adaptive filters such as \textit{partition-based weighted sum filters} \cite{Shao2004}, and least square error fittings \cite{Martens2007}, have also been used for this purpose. A comparative study of template subtraction and several adaptive filters including the \textit{least mean squares} (LMS), \textit{recursive least squares} (RLS), and an \textit{ad hoc} filter coined \textit{echo state neural network} (ESN) was reported in \cite{behar2014comparison,joachimbehar2014}.

\begin{figure}
    \centering
    \includegraphics[trim=1.8in 2.3in 2.2in 2.8in,clip,width = 0.8\columnwidth]{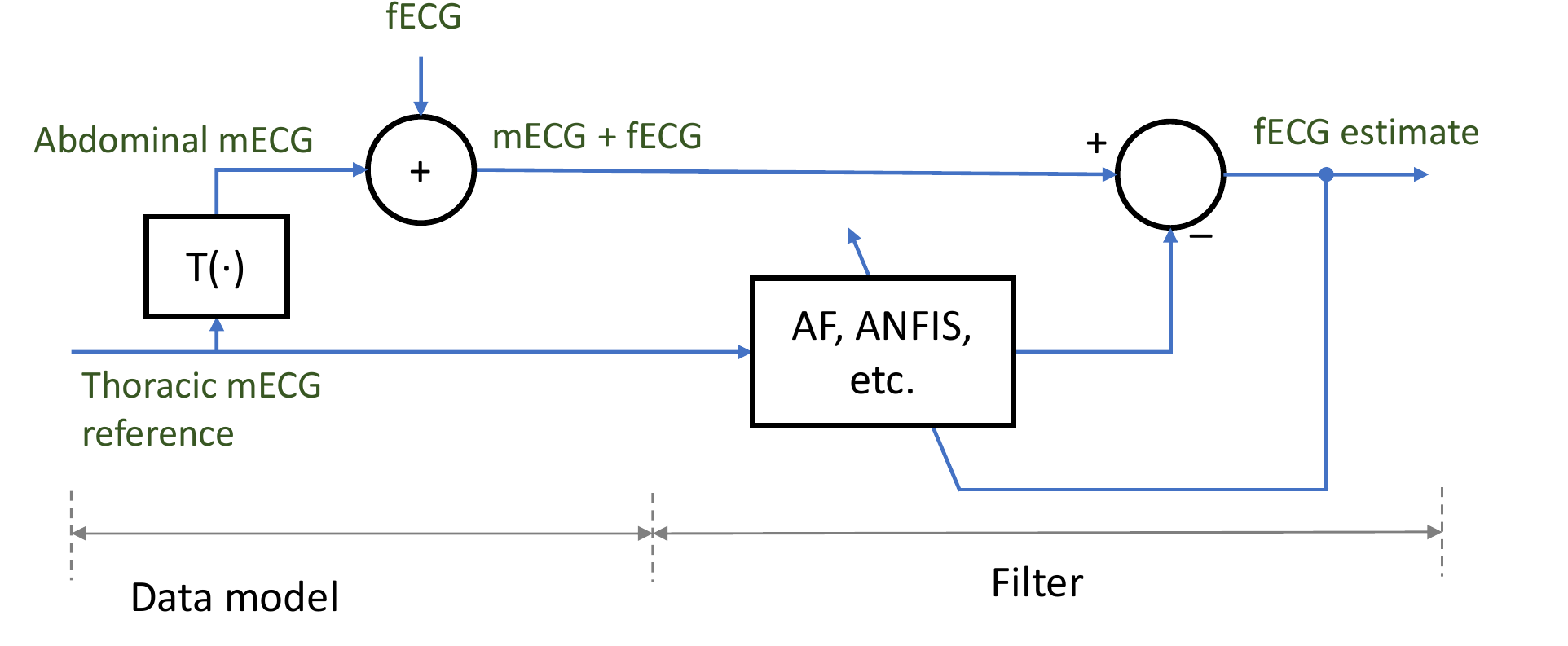}
    \caption{Adaptive filters for maternal ECG cancellation; concept adopted from \cite{assaleh2006extraction}}
    \label{fig:adaptivefilter}
\end{figure}

As demonstrated in Fig.~\ref{fig:adaptivefilter}, adaptive filtering methods for mECG removal, either require a reference mECG channel that is morphologically similar to the contaminating waveform, or require several channels to approximately reconstruct any morphological shape from the reference channels using adaptive \cite{Widrow75}, neural networks or neuro-fuzzy inference systems \cite{assaleh2006extraction}. Both of these approaches are practically inconvenient and with limiting performance, since the morphology of the mECG contaminants highly depends on the electrode locations and it is not always possible to reconstruct the complete mECG morphology from a (linear) combination of the reference electrodes, especially due to the limitations of finite dimensional dipole model of the heart, detailed in Section \ref{sec:volumeconductor}.

\subsection{Kalman filters for fECG extraction}
\label{sec:kalmanfilter}
Adaptive methods of mECG cancellation should ideally not rely on the electrode placement and the mECG morphology of the reference channel. This objective has motivated the development of Kalman filters for fECG extraction \cite{SSJC06,SSJ08,Sameni2008,HadiNarimaniMS2014,NarimaniSameni2015}. The Kalman filter and its extensions are adaptive in their nature and are therefore ideal for ECG signals with beat-wise morphological variations.

In \cite{SSJC06}, an extended Kalman filter (EKF) was suggested for denoising ECG signals recorded from noisy data. The \textit{process model} required for this EKF was based on an extension of the McSharry-Clifford synthetic ECG model \cite{McSharry2003,SSJB05}. The EKF formulation was later used in \cite{Sameni2008,SSJ08} for removing mECG artifacts from maternal abdominal recordings. Accordingly, following the volume conduction and dipolar data models (\ref{eq:datamodel}) and (\ref{eq:SyntheticDipole}), we can assume that the maternal abdominal signals consist of the mECG $s_m(t)$, the fECG $s_f(t)$ and background noise $\nu(t)$. Using the nonlinear state-space model proposed in \cite{SSJC06}, for mECG modeling the following set of process and observation equations can be written for the maternal body surface recorded signals $x(t)$:

\begin{itemize}
    \item \textit{Process equations}:
\begin{equation}
  \begin{array}{l}
  \theta(t+1)=[\theta(t) + \omega_m(t)]\mod(2\pi)\\
  s_m(t+1)=s_m(t)\displaystyle-\omega_m(t)\sum_{i=1}^k\frac{\alpha_i \tilde{\theta}_i(t)}{b_i^2} \exp(\frac{-\tilde{\theta}_i(t)^2}{2b_i^2})+w(t)
  \end{array}
\label{eq:ECGKFProcess}  
\end{equation}

\item \textit{Observation equations}:
\begin{equation}
	\begin{array}{l}
		\phi(t) = \theta(t) + \nu(t)\\
		x(t) = s_m(t) + s_f(t) + \eta(t)
	\end{array}
\label{eq:ECGKFObservation}  
\end{equation}
\end{itemize}
where $\tilde{\theta}_i(t)=[\theta(t)-\theta_i]\mod(2\pi)$, $\omega_m(t) = 2\pi f_m(t)/f_s$ is the maternal normalized angular velocity, $f_m(t)$ is the instantaneous maternal heart-rate in Hertz, $f_s$ is the sampling frequency in Hertz, $\alpha_i$, $b_i$, and $\theta_i$ are the amplitude, width and center parameters of the $i$th Gaussian kernel, and $k$ is the number of Gaussian kernels used for modeling the mECG morphology. In (\ref{eq:ECGKFProcess}) and (\ref{eq:ECGKFObservation}), $\theta(t)$ and $s_m(t)$ are the state variables; $\phi(t)$ is the cardiac phase measurement obtained by maternal RR-interval calculation and a linear phase map as demonstrated in Fig.~\ref{fig:Linearphase}; $x(t)$ is the maternal abdominal ECG measurement; $w(t)$ denotes the process noise; $\nu(t)$ is the phase measurement noise and $\eta(t)$ is the ECG measurement noise. According to the procedure detailed in \cite{SSJC06}, this model can be used in an EKF for estimating the mECG $\hat{s}_m(t)$. At the same time, the residual signal $x(t) - \hat{s}_m(t)$ (known as the \textit{innovation process} of the Kalman filter) is an estimate of $s_f(t) + \eta(t)$. The source codes required for implementing this method--- and the other methods detailed in this chapter--- are online available in the \textit{open-source electrophysiological toolbox} (OSET) \cite{OSET3.14}.

An advantage of the Kalman filtering framework is that besides signal estimation and denoising, it intrinsically provides confidence intervals for the estimations as well. By defining $\textbf{x}(t)=[\theta(t),s(t)]^T$ as the state vector at instant $t$ and $\hat{\textbf{x}}(t)$ as the \textit{posterior} estimate of $\textbf{x}(t)$, the posterior error of the estimation is defined as $\textbf{e}(t) = \textbf{x}(t)-\hat{\textbf{x}}(t)$ with a covariance matrix $\mathbf{P}(t) = \mathbb{E}\{(\textbf{e}(t) - \mathbb{E}\{\textbf{e}(t)\}) (\textbf{e}(t) - \mathbb{E}\{\textbf{e}(t))^T\}$. The matrix $\mathbf{P}(t)$ is an essential part of all the different variants of the Kalman filter and is calculated and updated as the filter propagates in time. The eigenvalues of this matrix can be used to form an \textit{error likelihood ellipsoid} (also known as \textit{concentration ellipsoid} \cite{VanTrees2001detection}) that represents the region of highest likelihood for the true state vector $\textbf{x}(t)$. This likelihood ellipsoid provides a confidence region for the estimated signals.

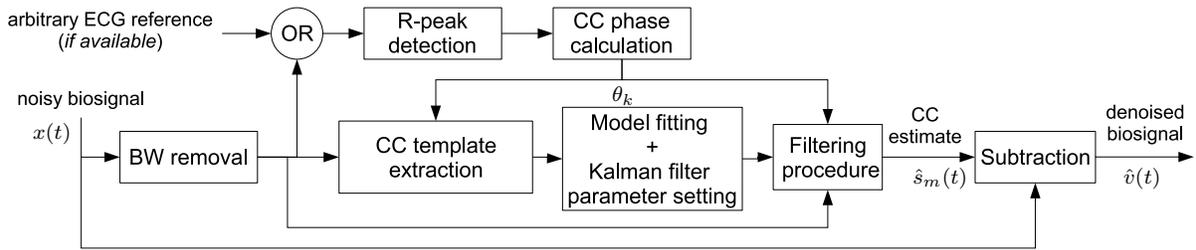
\begin{figure*}[tb]\centering
	\setlength{\unitlength}{1mm}
	\begin{picture}(160,40)			\put(0,0){\includegraphics[trim=0in 1cm 0in .8cm,clip,width=160mm]{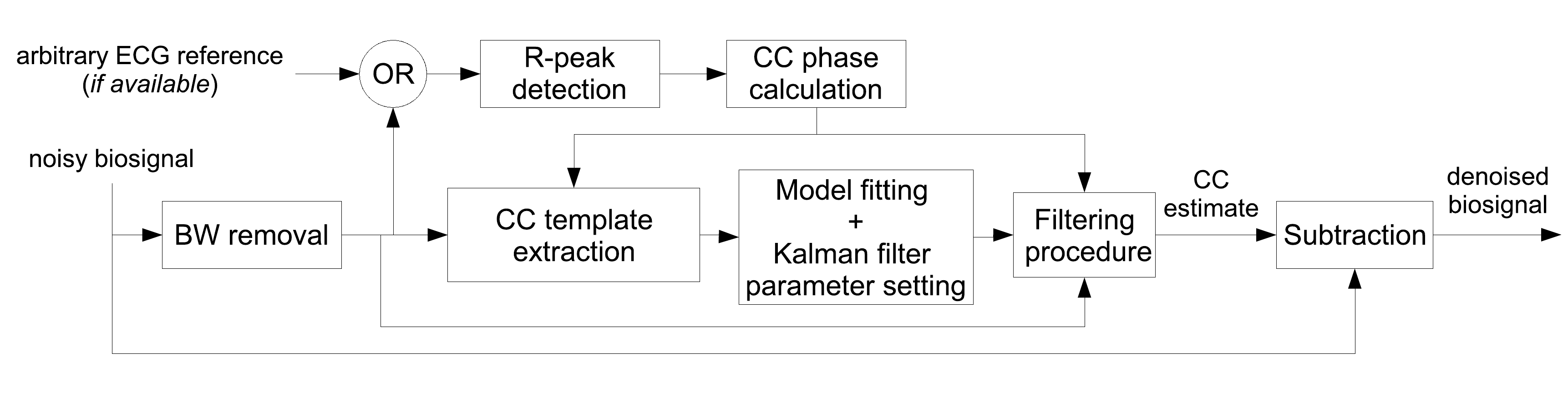}}
\put(5,15){\small{$x(t)$}}
\put(122,9){\small{$\hat{s}_m(t)$}}
\put(150,9){\small{$\hat{v}(t)$}}
\put(82,20){\small{$\theta_k$}}
	\end{picture}
\caption{The overall denoising scheme. As shown in this figure the R-peaks of the contaminating signals (CC) may either be detected from an arbitrary reference ECG or from the noisy biosignal after baseline wander (BW) removal; adopted from \cite{SSJ08}}
\label{fig:BlockDiagram}
\end{figure*}

The overall procedure for removing mECG signals by using the Kalman filtering framework is illustrated in Fig. \ref{fig:BlockDiagram} and may be summarized as follows:
\begin{enumerate}
{\setlength{\rightmargin}{.1cm}\setlength{\leftmargin}{.1cm}\setlength{\listparindent}{0cm}}

\item \textit{Baseline wander removal}. For the reliable extraction of the average mECG templates, the baseline wander of the noisy records should be removed beforehand.

\item \textit{mECG R-peak detection}. These peaks are required for constructing the phase signal $\theta(t)$, which is in turns needed for synchronizing the noisy ECG with the dynamic model in (\ref{eq:ECGKFProcess}). They are also used for extracting the mean mECG by synchronous averaging over the maternal heart beats. Depending on the power of the contaminating mECG, as compared with the background signals and noise, the maternal R-peaks may be detectable from the noisy recordings or from an arbitrary chest lead or abdominal channel synchronously recorded with the noisy dataset.

\item \textit{mECG template extraction}. Using the R-peaks, the \textit{ensemble average} (EA) and standard deviation of the mECG are extracted through synchronous averaging. Several methods have been proposed in the literature for synchronous averaging. One of the most effective approaches is the \textit{robust weighted averaging} method \cite{Leski2002}, which outperforms conventional EA extraction methods and is useful for noisy nonstationary mixtures.

\item \textit{Model fitting}. As proposed in \cite{SSJC06,CSMJ05b}, by using a nonlinear least square estimation, the parameters of the Gaussian kernel defined in (\ref{eq:ECGKFProcess}) are found, such that the model will best fit the mean mECG waveform.

\item \textit{Covariance matrix calculations}. The standard deviation of the average mECG is used to find the entries of the process and observation noise covariance matrices, as required for (extended) Kalman filtering.

\item \textit{Filtering}. Having the required model parameters, the mECG may be estimated by the EKF framework and the desired background signal (fECG plus noise) is found from $\hat{v}(t) = x(t) - \hat{s}_m(t)$.
\item \textit{fECG Post-processing}. The residual signals containing fECG and noise is post-processed for improving the fECG signal quality. Various methods such as an adaptive filter, a wavelet denoiser, or even a secondary EKF stage (this time customized for fECG denoising) can be used in this stage.
\end{enumerate}

Note further that for online applications or denoising long nonstationary datasets, all the dynamic model parameters and the covariance matrices can be updated over time, by recalculating them from the most recent cardiac beats. Further details regarding the Kalman filter based approach and its extensions such as the extended Kalman smoother (EKS), unscenting Kalman filter (UKF) and H-infinity filter can be followed from \cite{SSJC06,SSJ08,HadiNarimaniMS2014,jamshidian2019temporally}.

\begin{figure}[tb]
\centering
\begin{subfigure}{0.9\columnwidth}{\centering\includegraphics[trim=0in 0in 0in 0in,clip,width=4.2in]{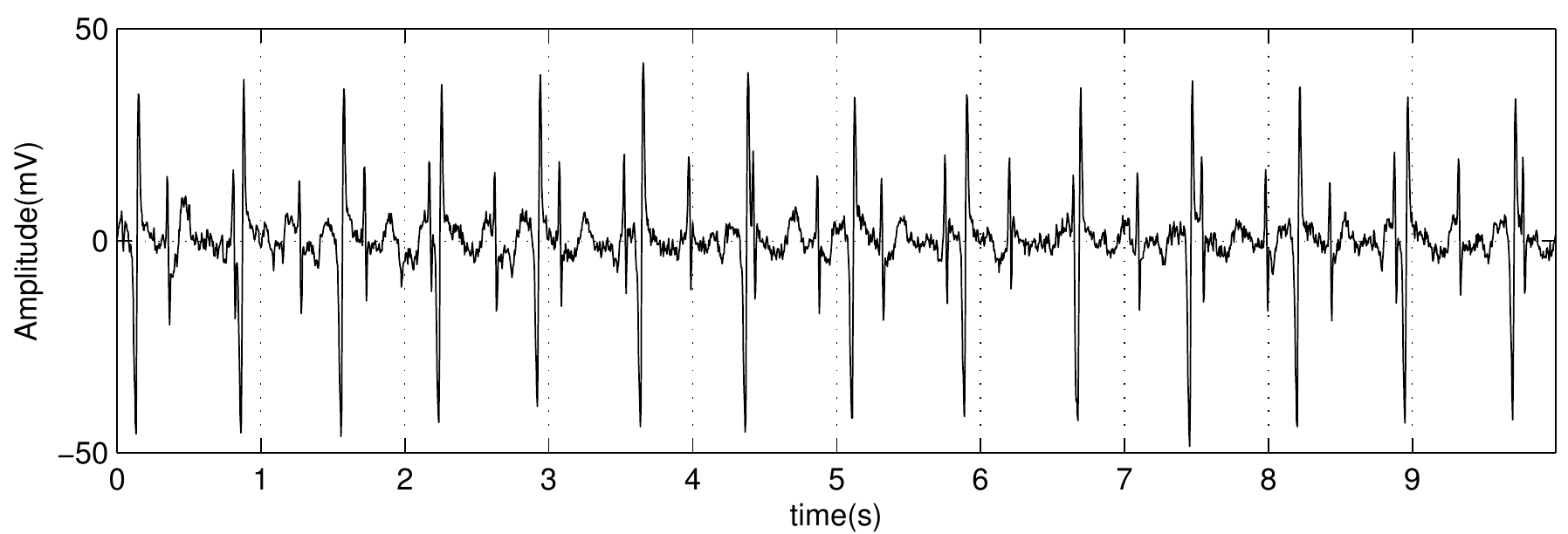}\subcaption{Original}\label{fig:DaISyoriginal}}\end{subfigure}
\begin{subfigure}{0.9\columnwidth}{\centering\includegraphics[trim=0in 0in 0in 0in,clip,width=4.2in]{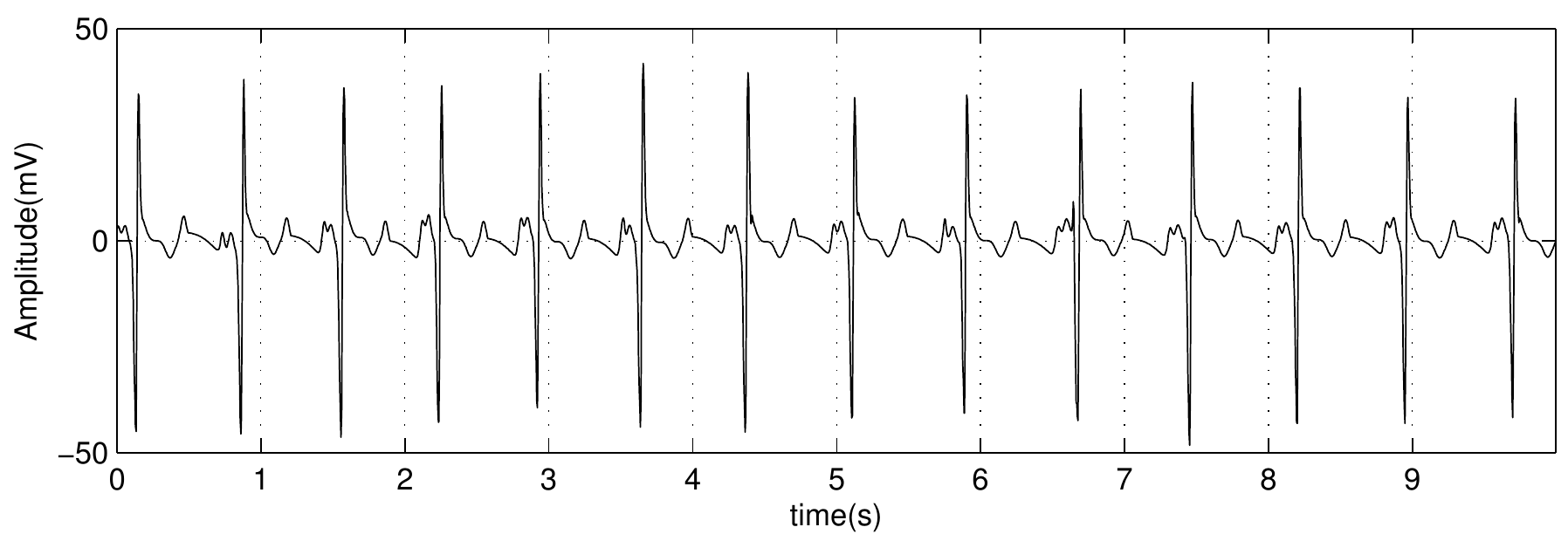}\subcaption{EKF of the maternal ECG}\label{fig:DaISyEKS}}
\end{subfigure}
\begin{subfigure}{0.9\columnwidth}{\centering\includegraphics[trim=0in 0in 0in 0in,clip,width=4.2in]{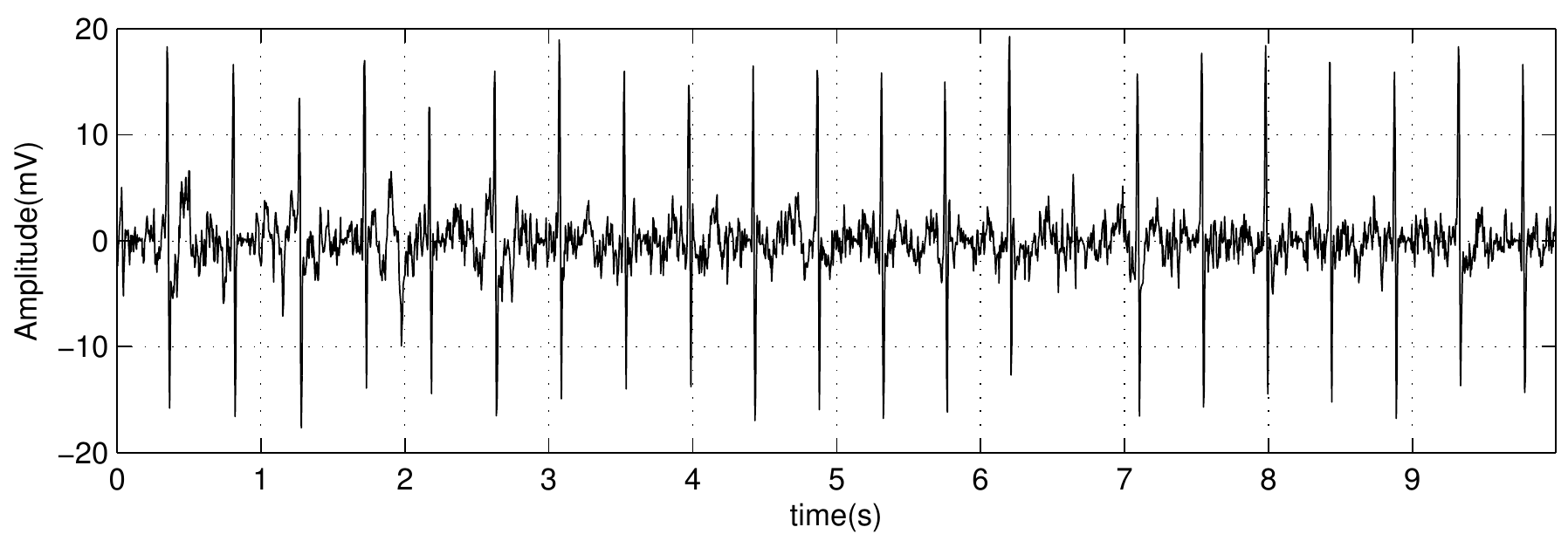}\subcaption{Residual fetal signal}\label{fig:DaISyEKSFetal}}
\end{subfigure}
\begin{subfigure}{0.9\columnwidth}{\centering\includegraphics[trim=0in 0in 0in 0in,clip,width=4.2in]{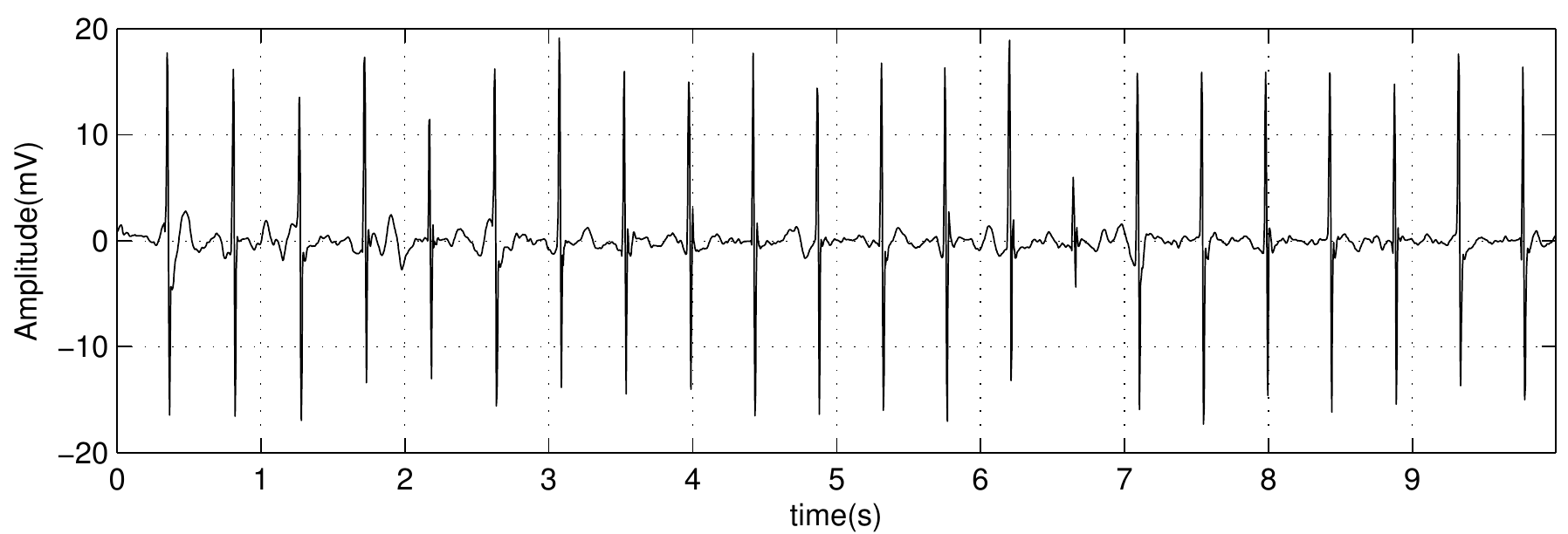}\subcaption{Fetal signal after post-processing}\label{fig:DaISyEKSFetalPostProcessed}}
\end{subfigure}
\caption{The first channel of the DaISy dataset \cite{DaISy}, recorded from a maternal abdominal lead before and after the EKF procedure, adopted from \cite{Sameni2008}.}
\label{fig:realeval}
\end{figure}

In Fig.~\ref{fig:DaISyoriginal}, the first channel of the DaISy fECG dataset is used for illustration \cite{DaISy}. The mECG estimate and the fetal residual components are depicted in Fig.~\ref{fig:DaISyEKS} and Fig.~\ref{fig:DaISyEKSFetal}. As a post-processing step, the extended Kalman filtering algorithm is applied to the residual fetal components, this time by training the filter parameters over the fECG. The post-processed fECG are depicted in Fig.~\ref{fig:DaISyEKSFetalPostProcessed}. From these results, it is seen that the Kalman filter is very effective for the extraction of fECG components from noisy maternal abdominal mixtures, even from as few as a single channel. However, as noticed from Fig.~\ref{fig:DaISyEKSFetalPostProcessed}, between t=6~s and t=7~s, the filter has failed to discriminate between the maternal and fetal components when the ECG waves of the mother and fetus have fully overlapped in time. The reason is that when the maternal and fetal components coincide in time, there are no other \textit{a priori} information for separating the maternal and fetal components. This is in fact an intrinsic limitation of single-channel methods, which motivates the application of multichannel recordings.

\begin{figure}[tb]\sidecaption
\centering
\includegraphics[trim=0in 0in 0in 0in,clip,width=4.1in]{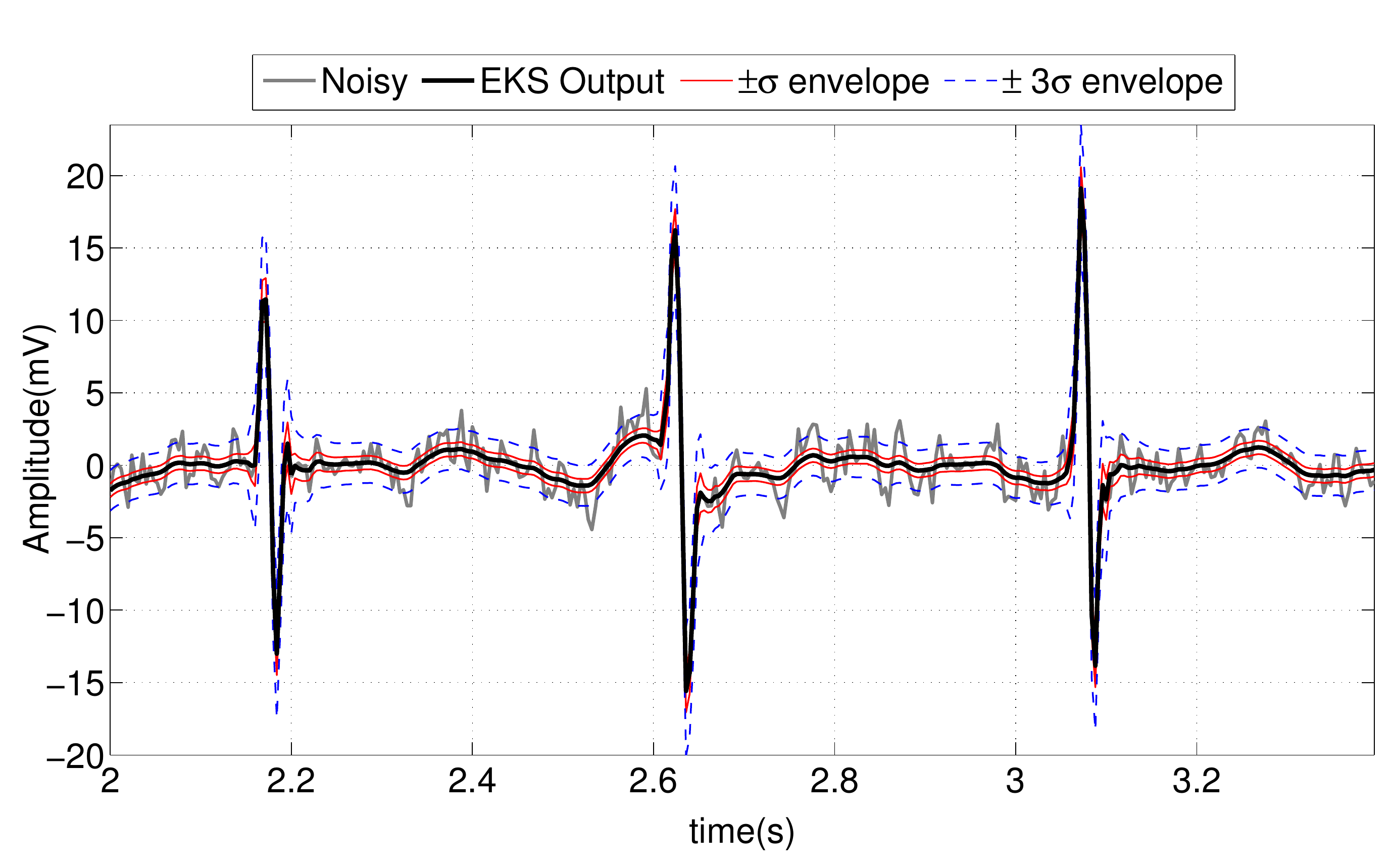}
\caption{Several fetal ECG beats adopted from Fig. \ref{fig:realeval}, before and after the post-processing EKF, together with the $\pm\sigma(t)$ and $\pm 3\sigma(t)$ confidence envelopes, adopted from~\cite{Sameni2008}.}
\label{fig:ConfidenceReg}
\end{figure}

As noted before, an important feature of Bayesian filtering is the ability of predicting the accuracy of the estimates. For the Kalman filter, this is readily achieved through the calculation of the error covariance matrix $\mathbf{P}(t)$. Suppose that the entry of the covariance matrix $\mathbf{P}(t)$ corresponding to the ECG estimate is denoted $\sigma(t)^2$ and the ECG estimation error is Gaussian, then the estimated ECG is bounded within the $\pm\sigma(t)$ envelope in 68\% of the sample points. This is due to the fact that approximately 68\% of the values drawn from a Gaussian distribution are within one standard deviation away from the mean, about 95\% of its values are within two standard deviations, and about 99.7\% lie within three standard deviations. These probabilities are different for non-Gaussian errors obtained by a nonlinear estimator such as the EKF. However, the $\pm\sigma(t)$ envelope can still be used as an approximate measure of error spread \cite[p. 79]{VanTrees2001detection}. In Fig.~ \ref{fig:ConfidenceReg}, several beats of the fECG before and after post-processing by an extended Kalman filter, together with their corresponding $\pm\sigma(t)$ and $\pm 3\sigma(t)$ envelopes are plotted. It is seen that the error envelopes provide the confidence region of the denoised fECG. 

\section{Multi-channel fetal electrocardiogram extraction}
\label{sec:multichannel}
Due to the limitations of single-channel fECG analysis detailed in the previous section, advanced fECG extraction algorithms are commonly multichannel. Some of the advantages of multichannel fECG acquisition and analysis include:
\begin{itemize}
    \item Improved SNR due to spatial filtering and joint analysis of multiple channels
    \item Robustness to fetal position and displacement, due to the spatial diversity of the leads
    \item Robustness to the possible detachment of a few of the electrodes
    \item Ability of extracting the fECG even during overlapping ECG waves of the mother and fetus
    \item Obtaining multiple perspectives of the fetal heart
\end{itemize}
Reconsidering the maternal abdominal recordings data model (\ref{eq:datamodel}),
in the multichannel case, it can be represented in the following matrix form:
\begin{equation}
\mathbf{x}(t) = [\mathbf{H}_m \quad \mathbf{H}_f \quad \mathbf{H}_v] \left[\begin{array}{c} \mathbf{s}_m(t) \\ \mathbf{s}_f(t) \\ \mathbf{v}(t) \end{array}\right] + \mathbf{n}(t) = \mathbf{A}\mathbf{s}(t) + \mathbf{n}(t) = \sum_{k = 1}^{p} \mathbf{a}_k s_k(t) + \mathbf{n}(t)
\label{eq:datamodelcompact}
\end{equation}
where $p \stackrel{\Delta}{=}
m+l+k$ is the total effective number of sources due to the maternal and fetal ECG and structured noise, $\mathbf{A} = [\mathbf{a}_1, \ldots, \mathbf{a}_p]\in \mathbb{R}^{n\times p}$ is the overall source-sensor mixing matrix (or the lead-field matrix) and $\mathbf{s}(t) = [s_1(t), \ldots, s_p(t)] \in \mathbb{R}^p$ contains all the cardiac sources and structured noise components.

The objective of multichannel analysis is to recover an estimate of $\mathbf{s}(t)$ (or more specifically $\mathbf{s}_f(t)$), from $\mathbf{x}(t)$, using the available assumptions regarding the mECG, fECG, and noises. A classical approach to solving this problem is to estimate the matrix $\mathbf{B}\in \mathbb{R}^{p\times n}$, such that $\mathbf{B}\mathbf{A} = \mathbf{I}$. Therefore,
\begin{equation}
\mathbf{y}(t) = \mathbf{B}\mathbf{x}(t) = \mathbf{s}(t) + \mathbf{B}\mathbf{n}(t)
\label{eq:datamodelinverse}
\end{equation}
which is a noisy estimate of the source vector $\mathbf{s}(t)$. Since both the source vector $\mathbf{s}(t)$ and the mixing matrix $\mathbf{A}$ are unknown, the problem is categorized as a \textit{blind or semi-blind source separation} (BSS) problem \cite{ComonJutten2010handbook}. In this problem, if the number of observed channels is equal to or greater than the number of effective number of sources, i.e., $n \geq p$, and $\mathbf{A}$ is non-singular, the observed mixture is \textit{determined} or \textit{over-determined}. Therefore, noting that $\mathbf{s}_m(t)$, $\mathbf{s}_f(t)$ and  $\mathbf{v}(t)$ can be considered as groups of statistically independent sources with inter independence and intra dependencies, BSS algorithms such as (noisy) independent component analysis (ICA) \cite{Car98,Lathauwer2000,Zarzoso01}, semi-blind source separation algorithms such as periodic component analysis ($\pi$CA) \cite{Sameni2008a}, and more recently nonstationary component analysis (NSCA) \cite{jamshidian2019temporally} have been effectively used to solve this problem. The general challenges of this problem are:
\begin{enumerate}
\item \textit{Amplitude and sign ambiguity}: An intrinsic ambiguity of the multichannel data-model (\ref{eq:datamodelcompact}) is that the source vector amplitude and sign may not be retrieved merely from the measurements $\mathbf{x}(t)$. This can be explained by the fact that exchanging an arbitrary non-zero scaling factor $\alpha$ and $1/\alpha$ between the $k$th column of the matrix $\mathbf{A}$ and the source $s_k(t)$ does not change the measurements. Therefore, there is no way to retrieve the source amplitudes and sign, from the measurements alone. 

\item \textit{Estimated source order:} Retrieving the order of sources is another limitation that may not be resolved from the measurements alone (without other priors or constraints). The reason is that taking an arbitrary permutation matrix $\mathbf{P}$, $\mathbf{A}\mathbf{s}(t)$ and $\mathbf{A}\mathbf{P}\mathbf{P}^{T}\mathbf{s}(t)$ are identical.

\item \textit{Noisy mixtures:} It is clear from the right hand side of (\ref{eq:datamodelinverse}), that even if the separation matrix $\mathbf{B}$ is perfectly estimated, i.e. $\mathbf{B}\mathbf{A}=\mathbf{I}$, due to the noise term $\mathbf{B}\mathbf{n}(t)$, the resulting mixture can remain noisy, except for the non-probable special case that the observation noise lies in the \textit{null-space} of the separation matrix $\mathbf{B}$, resulting in $\mathbf{B}\mathbf{n}(t)=\mathbf{0}$. Otherwise, the noise can even be amplified and the desired components, such as the fECG, may in cases be totally obscured by noise. In fact, the problem due to full-rank observation noise is twofold. On the one hand, the noise hampers the estimation of the separation matrix. On the other hand, it remains or is even amplified during source separation. Therefore, whenever possible, it is better to minimize or remove the channel-wise full-rank noise before source separation. In the latter case (channel-wise noise removal), any processing of the multichannel data should be performed by using filters that approximately have a \textit{linear phase} (constant \textit{group delay}) over the bandwidth of interest. Moreover, the difference between the group delays of the filters applied to different channels should be negligible, as compared with the sampling time of the data, to avoid the displacement of the components of different channels during preprocessing. This is a fundamental requirement for synchronous multichannel analysis, which has been underemphasized in the literature.

\item \textit{Non-punctual sources:} The heart is not a punctual source. This fact has several implications on fECG extraction, including: 1) the fECG morphology can change as the fetus moves with respect to the maternal body surface leads; 2) during source separation, depending on the heart-sensor distance and the SNR of the measurements, more than one source is associated to the mother and the fetus. The notion of effective number of sources detailed in Section \ref{sec:volumeconductor}, corresponds to this fact. It has been previously shown that even though among the extracted sources, only a few might visually resemble the fECG, when one applies synchronous averaging to the different channels extracted by BSS algorithms (by aligning the R-peak positions and averaging over several beats), the fECG emerges from all channels. This point was first illustrated in \cite{SJS06} and justified in \cite{Sameni2008} using multi-pole expansion of body surface potentials. An example of adult and fetal ECG obtained by synchronous averaging after applying a typical ICA algorithm is shown in Fig.~\ref{fig:synchavg}, for illustration. This implies that for non-punctual sources, perfect separation of the sources (maternal and fetal ECG) is not fully achieved. Although in practice, the number of cardiac source signals extracted from multichannel ECG--- including maternal abdominal recordings--- is limited by the number of channels, distance to the heart and the SNR of the recordings.
\begin{figure}
    \centering
    \begin{subfigure}{2.7in}
    {\centering\includegraphics[width = 2.7in]{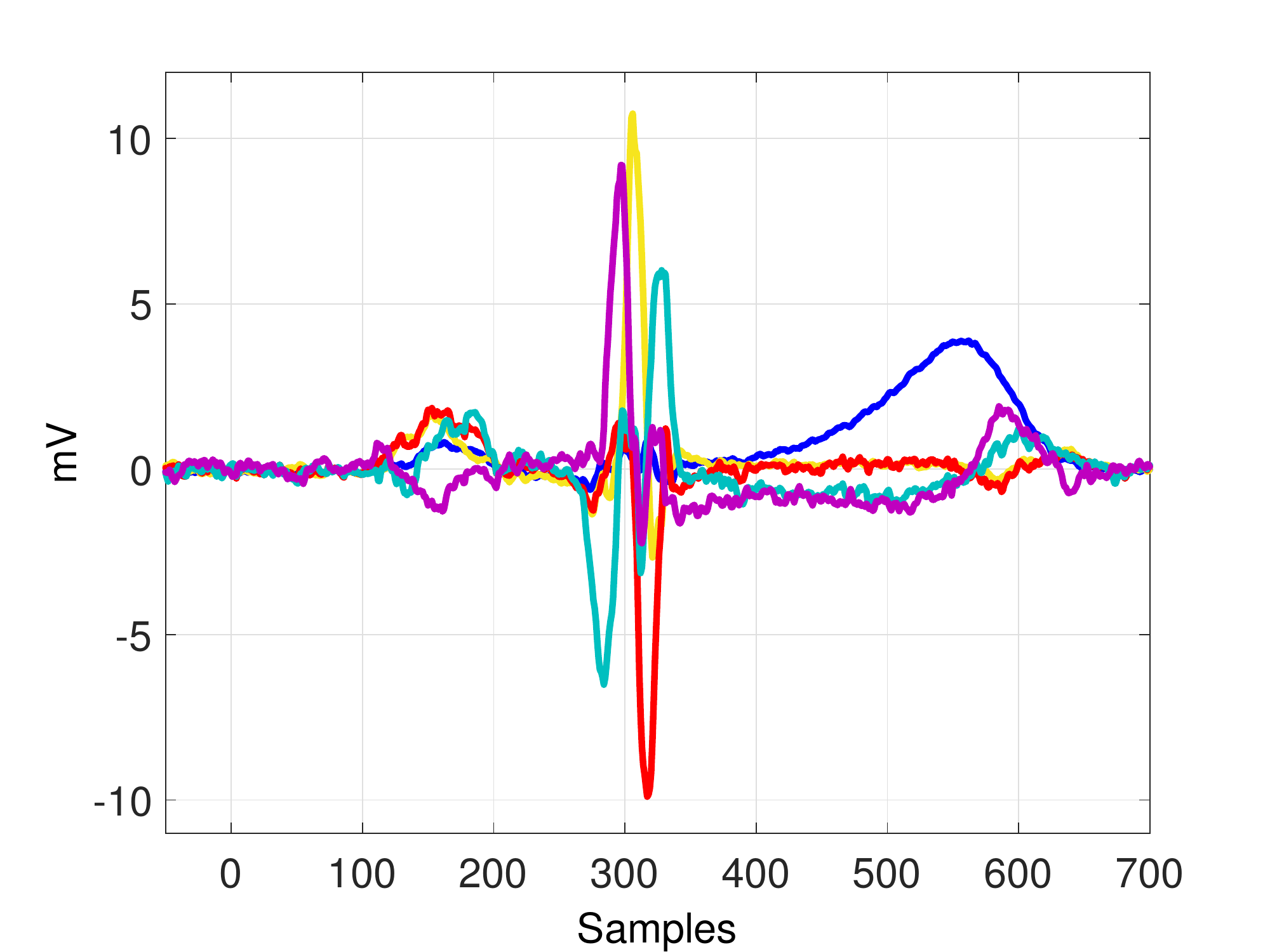}\subcaption{Adult ECG}}    
    \end{subfigure}
    \begin{subfigure}{2.7in}
    {\centering\includegraphics[width = 2.7in]{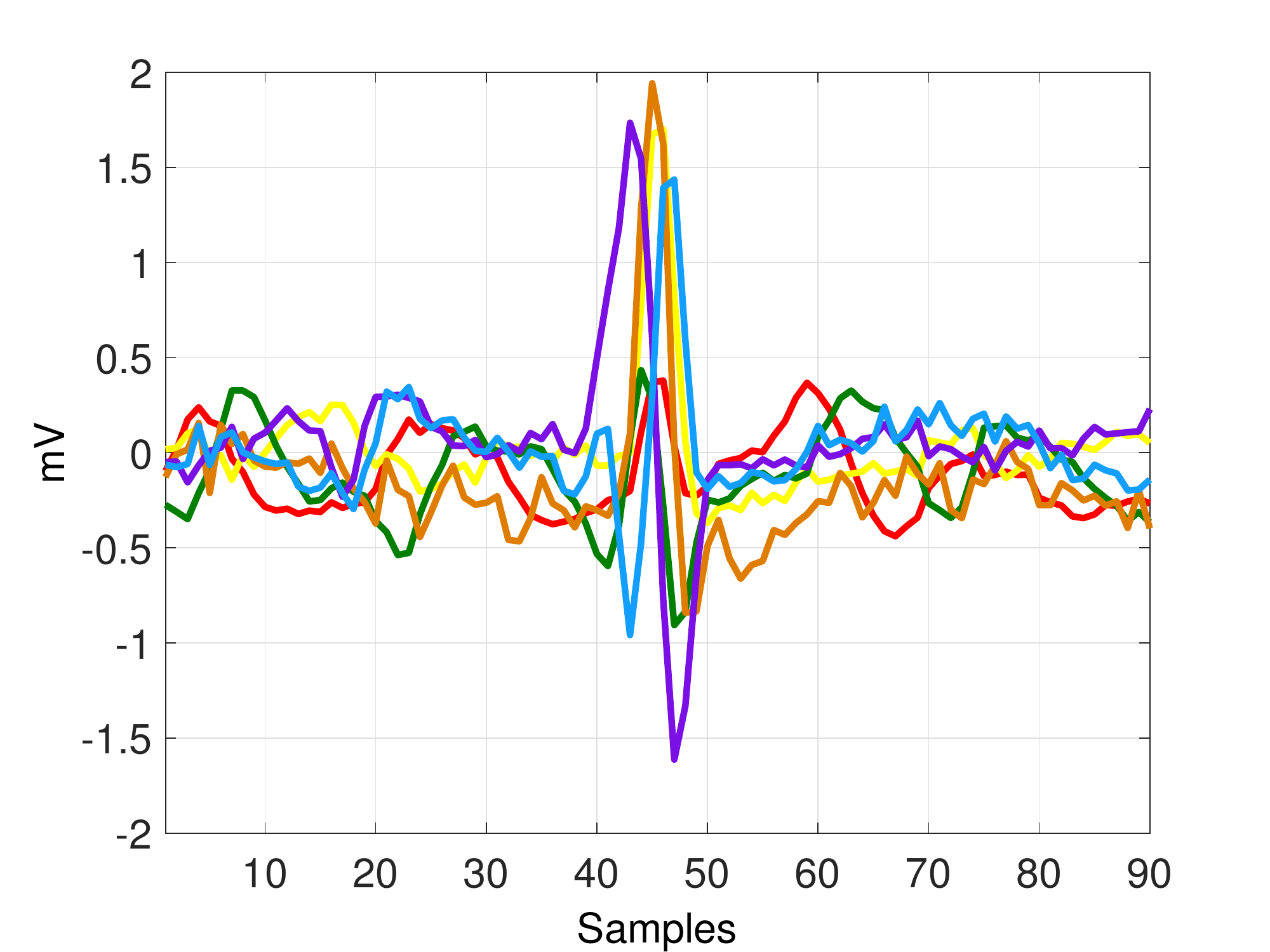}\subcaption{Fetal ECG}}\end{subfigure}\caption{An illustration of the concept of non-punctuality of the cardiac sources resulting in multidimensionality of the components extracted from adult and fetal ECG. Synchronous averaging has been performed over the different channels extracted by independent component analysis to demonstrate the existence of the ECG components in all channels.}
    \label{fig:synchavg}
\end{figure}

\item \textit{Low-rank measurements:} If the number of abdominal channels are insufficient ($n < p$) or when the maternal-fetal mixture is singular (e.g., due to the closeness of the sensors or special fetal positioning in the womb), the signal mixture is \textit{under-determined}. In this case, due to the rank-deficiency of the mixture, linear transforms are unable to separate the maternal and fetal subspaces \cite{SJS2010,jamshidian2018fetal}.

\item \textit{Time-variant mixtures:} When the mixing matrices $\mathbf{H}_m$ and $\mathbf{H}_f$ are functions of time, e.g., due to fetal movement during signal acquisition, the sources may no longer be retrieved by stationary source separation algorithms. In this context, adaptive source separation algorithms are required \cite{Cardoso1996}. These methods have also been specifically used for online fECG extraction \cite{fatemi2017online,jamshidian2018fetal}.
\end{enumerate}
In the sequel, some of the different approaches of fECG extraction from multichannel recordings are reviewed.

\subsection{Independent component analysis}
\label{sec:ICA}
Independent component analysis (ICA) is the most common class of algorithms for solving blind and semi-blind source separation (BSS) problems such as (\ref{eq:datamodelcompact}), where both the mixing matrix $\mathbf{A}$ and the source vector $\mathbf{s}(t)$ are unknown (with or without noise) \cite{ComonJutten2010handbook}. The problem of retrieving the sources and mixing matrix at the same time is clearly ill-posed. Therefore, additional assumptions and priors about the source and/or mixture are required. In ICA, one seeks linear mixtures of the form $\mathbf{y}(t)=\mathbf{B}\mathbf{x}(t)$, which maximize some measure of statistical independence between the estimated sources, also known as a \textit{contrast function}.

Many ICA algorithms attempt to solve the problem in several phases, for example by first pre-whitening and sphering the data by principal component analysis (PCA) (\ref{fig:PCAICA}). Pre-whitening acts as an intermediate step for achieving independence and only leaves the estimation of a rotation matrix to achieve independence.
\begin{figure}[tb]
\centering
\begin{tabular}{ccccccc}
	$\textbf{x}(t)$ & $\CapRightArrow{Whitening}{2.5mm}$ & $\textbf{v}(t)$ & $\CapRightArrow{Sphering}{1mm}$ & $\textbf{z}(t)$ & $\CapRightArrow{ICA}{2.5mm}$ & $\textbf{y}(t)$\\
 & (\textit{rotation}) & & (\textit{scaling}) & & (\textit{rotation}) &
\end{tabular}
\caption{General scheme of ICA algorithms with spatial pre-whitening}
\label{fig:PCAICA}
\end{figure}

An algebraic approach to ICA is to seek the separation matrix $\mathbf{B}$, such that it diagonalizes a set of matrices containing second or higher order statistics derived from the multichannel recordings \cite{ComonJutten2010handbook}. For signals with temporal structure, there are various algorithms that use this algebraic approach. Considering that no more than two matrices can be simultaneously diagonalized by using a single linear transform, many algebraic algorithms have been developed for the approximate joint diagonalization of such matrices. The first and most widely used algorithm in this context, is known as \textit{joint approximate diagonalization of eigenmatrices} (JADE) \cite{Cardoso1993,CardosoSourceCodes}. To date, fECG extraction has been one of the classical biomedical applications for testing and comparing various ICA algorithms. Some of the pioneer contributions in this area include: \cite{Car98,Lathauwer2000,Zarzoso01}.

\subsection{Independent subspace analysis}
\label{sec:ISA}
Independent subspace analysis (ISA) has been introduced as a variant of ICA, for problems in which one deals with groups of signals having inter-group independence and intra-group dependencies. ISA was first introduced in \cite{Comon95} and mathematically developed in \cite{Car98}, where the notion of ICA was generalized to the notion of \textit{multidimensional ICA}. Accordingly, ISA relies on the idea of \textit{vector-valued} components rather than \textit{scalar} source signals. The first--- and most commonly studied--- application of ISA has been for fECG extraction. Throughout the chapter, we have learned that the cardiac signals of either the mother or the fetus are generally multidimensional. Therefore, the maternal and fetal ECG components form signal subspaces with internal dependencies, while the components of the maternal and fetal subspaces are independent from each other.

ISA may be realized by applying an initial ICA step on mutichannel observations and then empirically regrouping the independent components that belong to the same subspace from prior knowledge of the subspace structures to achieve a \textit{canonical representation} of each subspace. In fact, there is an intrinsic ambiguity in retrieving the components inside the subspaces, which may not be resolved with the same measure of independence used for subspace separation. In other words, from the source separation viewpoint, no representation of the extracted mECG and fECG components inside their signal subspaces can be considered to be better than the another. Therefore, the components that belong to the same subspace are regrouped after the initial ICA step. However, the challenges of ISA are:
\begin{enumerate}
    \item To find the dimensions of each subspace \cite{Car98}.
    \item Automatic regrouping of the components \cite{weiss99segmentation,bach03beyond,stogbauer-2004-70}.
    \item The impact of subspace distances and noise on the stability of the extracted subspaces \cite{Meinecke2002,HMM03}.
\end{enumerate}
For fECG extraction, previous studies have focused on the feasibility of extracting the independent subspaces \cite{Car98,Lathauwer2000} and regrouping strategies \cite{bach03beyond}.

\subsection{Generalized eigenvalue decomposition}
\label{sec:gevd}
Although ICA and ISA are very effective for fECG extraction, they do not make explicit use of the pseudo-periodicity of the maternal and fetal ECG and the fact that multiple sources may correspond to the mECG and the fECG (due to the non-punctuality of the cardiac sources detailed before). In order to be used in fully automated algorithms, it is also convenient to be able to rank the extracted sources corresponding to the mECG and/or fECG automatically. These requirements resulted in the development of source separation algorithms, which are specifically customized for cardiac signals. Algorithms such as periodic component analysis ($\pi$CA) \cite{Sameni2008a}, and nonstationary component analysis \cite{jamshidian2019temporally} were developed for this purpose. These methods are based on an algebraic transform known as generalized eigenvalue decomposition, which was previously used in one of the basic source separation algorithms known as AMUSE \cite{Tong1991}.

For real symmetric matrices $\mathbf{A},\mathbf{B} \in \mathbb{R}^{n \times n}$, \textit{generalized eigenvalue decomposition} (GEVD) of the matrix pair $(\mathbf{A},\mathbf{B})$ consists of finding $\mathbf{W} \in \mathbb{R}^{n \times n}$ and $\bm{\Lambda}\in \mathbb{R}^{n \times n}$, such that
\begin{equation}\label{eq:EVD}
\begin{array}{l}
     \mathbf{W}^T \mathbf{A} \mathbf{W} = \bm{\Lambda}\\
     \mathbf{W}^T \mathbf{B} \mathbf{W} = \mathbf{I}_n
\end{array}
\end{equation}
where $\bm{\Lambda} = \Diag(\lambda_1, \ldots, \lambda_n)$ contains the generalized eigenvalues corresponding to the eigenmatrix $\mathbf{W}=[\textbf{w}_1,\ldots,\textbf{w}_n]$, with real eigenvalues sorted in descending order on its diagonal. Symmetric positive definite matrix pairs have real positive eigenvalues and the first eigenvector $\textbf{w}=\textbf{w}_1$ maximizes the \textit{Rayleigh quotient} \cite{Strang1988}:
\begin{equation}\label{eq:Rayleigh}
J(\textbf{w}) = \frac{\textbf{w}^T \mathbf{A} \textbf{w}}{\textbf{w}^T \mathbf{B} \textbf{w}}
\end{equation}
It can be shown that all ICA methods based on pre-whitening can be eventually converted into a GEVD problem of two (problem-specific) matrices \cite{Sameni2008a}. Therefore, in semi-blind source separation problems, in which prior knowledge regarding the underlying components exists, the problem of source separation can be considered as a \textit{matrix design problem}. The performance of GEVD-based source separation and generic methods for choosing the proper matrix pair have been addressed in previous research \cite{yeredor2011performance,yeredor2009optimal}.

GEVD can for example be used for the separation of temporally correlated (or periodic) sources from other signals. For example, for a zero-mean wide-sense stationary or cyclostationary real observation vector $\textbf{x}(t)$, the covariance matrix is:
\begin{equation}
\mathbf{C}_x(\tau)=\mathbb{E}_t\{\textbf{x}(t+\tau)\textbf{x}(t)^T\}
\label{eq:cov}
\end{equation}
where $\mathbb{E}_t\{\cdot\}$ indicates averaging over $t$. The AMUSE algorithm is a source separation algorithm that jointly whitens the data and diagonalizes $\mathbf{C}_x(\tau)$ for some arbitrary $\tau$, i.e. the solution of the GEVD problem of the matrix pair $(\mathbf{C}_x(\tau),\mathbf{C}_x(0))$ \cite{Tong1991,Parra2003}. What hampers the performance of GEVD for source separation, is the fact that real world sources are rarely fully periodic. Therefore, more advanced source separation algorithms use (approximate) joint diagonalization of more than two matrices, which are more robust to data outliers and computational errors as compared with AMUSE \cite{Belouchrani1997,Cardoso99}. In this context, the second-order blind identification (SOBI) algorithm is an example of a time-domain algorithm that whitens the data and approximately diagonalizes $\mathbf{C}_x(\tau)$ for several time-lags $\tau$ \cite{Belouchrani1997}. Similar time-domain methods have also been proposed for cyclostationary sources, in which the data is again pre-whitened and matrices corresponding to cyclostationary statistics of the dataset are (approximately) diagonalized \cite{Ferreol2000}. An alternative approach is to use signal priors such as the pseudo-periodicity and ``bumpy'' shape of the ECG, as detailed below.

\subsection{Periodic component analysis}
\label{sec:PiCA}
In (pseudo-)periodic component analysis ($\pi$CA)\footnote{The term $\pi$CA was originally coined in \cite{SaulA00}, for extracting periodic signals, which resulted in GEVD of a pair matrices as in AMUSE \cite{Tong1991}.}, the matrix pair $(\mathbf{C}_1,\mathbf{C}_0)$ are jointly diagonalized by GEVD, where $\mathbf{C}_0=\mathbf{C}_x$ is the covariance matrix of $\mathbf{x}(t)$ and $\mathbf{C}_1$ is a variable-period version of the lagged-covariance matrix (\ref{eq:cov}), using the time-varying period of the ECG defined in (\ref{eq:tau}): 
\begin{equation}
\mathbf{C}_1=\mathbb{E}_t\{\textbf{x}(t+\tau_t)\textbf{x}(t)^T\}
\label{eq:covtvar}
\end{equation}
In order to assure the symmetry of $\mathbf{C}_1$ and the realness of its eigenvalues, the following step is applied before GEVD:
\begin{equation}\label{eq:equalization}
\mathbf{C}_1 \gets \frac{(\mathbf{C}_1+\mathbf{C}_1^T)}{2}
\end{equation}
Next, considering $\mathbf{W}$ as the joint diagonalizer of the matrix pair $(\mathbf{C}_1,\mathbf{C}_0)$, the linear transform
\begin{equation}
\textbf{y}(t) = \textbf{W}^T \textbf{x}(t)
\label{eq:lintrans}
\end{equation}
extracts uncorrelated sources $\textbf{y}(t)=[y_1(t),...,y_n(t)]^T$  with maximal correlation at time-variant periods $\tau_t$, which is the heart-rate of interest. Therefore, $\textbf{y}(t)$ ranks the sources in order of similarity with the desired heart-rate. In other words, $y_1(t)$ is the most periodic component and $y_n(t)$ is the least periodic with respect to the R-peaks of the ECG. This method is flexible in the cardiac period used for source separation. For instance for fECG extraction, let $\theta_m(t)$ and $\theta_f(t)$ be the maternal and fetal ECG phases found from the maternal and fetal R-peaks (as defined in Section \ref{sec:cardiacphase}) and $\mathbf{C}_m$ and $\mathbf{C}_f$ represent the lagged covariance matrices of the maternal and fetal heart-rates, found by averaging (\ref{eq:covtvar}) over the maternal and fetal periods $\tau_t^m$ and $\tau_t^f$, respectively. Then different variants of GEVD is obtained, if the matrix $\mathbf{C}_1$ used in GEVD be set to any of the following matrices \cite{Sameni2008a}:
\begin{subequations}
    \begin{align}
	(\mathbf{C}_1,\mathbf{C}_0) = (\mathbf{C}_m, \mathbf{C}_x)\\
	(\mathbf{C}_1,\mathbf{C}_0) = (\mathbf{C}_f, \mathbf{C}_x)\\
	(\mathbf{C}_1,\mathbf{C}_0) = (\mathbf{C}_m, \mathbf{C}_f)	\end{align}
\label{eq:C}	
\end{subequations}


If we assume the data to be pre-whitened, the diagonalization of the matrices defined in (\ref{eq:C}) is respectively equivalent to finding (a) the most periodic components with respect to the mECG, (b) the most periodic components with respect to the fECG, and (c) the most periodic components with respect to the mECG while being the least periodic components with respect to the fECG. In this latter case the extracted components should gradually change from the mECG to the fECG, from the first to the last component, but the components are not necessarily uncorrelated. It should of course be noted that the last two cases are difficult to implement in practice, as they require the prior extraction of the fetal R-peaks to form the $\mathbf{C}_f$ matrix. Another reservation is for abnormal maternal cardiac signals, which a measure of pseudo-periodicity can fail for mECG and fECG source separation.

\subsection{Nonstationary component analysis}
\label{sec:NSCA}
The reservations regarding possible abnormal mECG and the difficulty of fECG R-peak identification in noise have motivated source separation algorithms that are merely based on rather regular spiky or bumpy shapes of the maternal and fetal ECG. The theory is based on source separation algorithms for variance-nonstationary source mixtures, which is a special case of methods known as \textit{nonstationary component analysis} (NSCA) 
\cite{pham2001blind,yeredor2010second,jamshidian2019temporally}. Accordingly, let us consider multivariate signals $\mathbf{x}(t) \in \mathbb{R}^n$ ($t\in \mathcal{T}$), where $\mathcal{T}$ denotes the set of available discrete-time samples and $\mathcal{P} \subset \mathcal{T}$ is a subset of these samples, which are considered as being \textit{nonstationary} or \textit{odd events} that do not follow the (average) background model in certain aspects. For our application, they can correspond to the maternal or fetal QRS complexes. In this case, a sample-wise \textit{hypothesis test} can be performed for the identification of the temporally nonstationary events:
\begin{equation}
\begin{array}{rl}
         \mathcal{H}_0: t \notin \mathcal{P} \\
         \mathcal{H}_1: t \in \mathcal{P}
    \end{array}
\label{eq:hypothesistestGeneral}
\end{equation}
Denoting the subset of samples that satisfy the alternative hypothesis $\mathcal{H}_1$ with $\mathcal{P}$, a special case of GEVD is obtained by finding the matrix $\mathbf{W}$, which satisfies (\ref{eq:EVD}) for $\mathbf{A} = \mathbb{E}_u\{\mathbf{x}(u)\mathbf{x}(u)^T\}$ and $\mathbf{B} = \mathbb{E}_t\{\mathbf{x}(t) \mathbf{x}(t)^T\}$, where $\mathbb{E}_t\{\cdot\}$ and $\mathbb{E}_u\{\cdot\}$ denote averaging over all time samples $t\in \mathcal{T}$ and $u\in \mathcal{P}$, respectively. Using this matrix, the linear transform $\mathbf{y}(t) = \mathbf{W}^T \mathbf{x}(t)$ extracts $n$ uncorrelated channels with maximal energy over the subset of time samples $u \in \mathcal{P}$. Applying this method for ECG extraction, $\mathbf{W}$ retrieves uncorrelated linear mixtures of $\mathbf{x}(t)$ with maximal energy during the QRS complex.

As detailed in \cite{jamshidian2019temporally}, in the simplest case, the nonstationary sample set $\mathcal{P}$ can be identified by thresholding the time-varying power of an arbitrary reference channel $r(t)$ (which can even be one of the channels of $\mathbf{x}(t)$, or a mixture of them), over a sliding window of length $w$:
\begin{equation}
P_w(t) = \displaystyle \frac{1}{w}\sum_{a = -\frac{w}{2}}^{\frac{w}{2}} |r(t-a)|^2
\label{eq:TVEnergy}
\end{equation}
The ratio of $P_w(t)$ for two windows of lengths $w= w_1$ and $w=w_2$ ($w_2 \gg w_1$) can be used as a measure for detecting fast local nonstationary epochs within a slowly varying (or stationary) background activity:
\begin{equation}
\rho(t) = \displaystyle \frac{P_{w_1}(t)}{P_{w_2}(t)}
\label{eq:TVEnergyRatio}
\end{equation}
which is the \textit{local power envelope} (LPE) of the reference channel. For a global measure, the denominator {$P_{w_2}(t)$} can be replaced with the average signal power $P_\infty$. The values of $\rho(t)$ significantly smaller or larger than one correspond to time epochs that are different (nonstationary) from the background activity. The rationale behind the above definition is that a stationary signal, such as the non-ECG background signals and noises, have a consistent energy profile over time and notable deviations of the LPE from unity (with appropriate window lengths $w_1$ and $w_2$) are indicators of nonstationary epochs such as the maternal and fetal QRS complexes. Therefore, the LPE can be used to extract the time epochs of the maternal or fetal QRS as follows:
\begin{equation}
\theta_{\text{LPE}} = \{t\quad\!\!\!|\quad\!\!\!  {\rho(t) \geq \zeta_u \text{ or } \rho(t) \leq \zeta_l}, t\in \mathcal{T}\}
\label{eq:localenergyindex}
\end{equation}
where $\zeta_u$ and $\zeta_l$ are predefined upper and lower thresholds satisfying $\zeta_u > 1 > \zeta_l \geq 0$. In \cite{jamshidian2019temporally}, other indexes based on the \textit{innovation process} of an extended Kalman filter trained over the mECG were proposed for the identification and extraction of the fECG.

\subsection{Approximate joint diagonalization using ECG-specific priors}
\label{sec:AJD}
Maternal and fetal ECG source separation from background noise can benefit from the advantages of methods such as $\pi$CA and NSCA at the same time. Suppose that the matrices $\mathbf{C}_{i}$ ($i = 1, \ldots, K$) are positive semi-definite matrices containing second or higher order statistics regarding the maternal and fetal ECG. For example, the matrices can be the lagged-covariance matrices corresponding to the maternal or fetal heart, or the covariance matrices obtained by energy thresholding, as in NSCA. We may now seek the joint approximate diagonalizer $\mathbf{W} \in \mathbb{R}^{n \times n}$, such that the matrices
\begin{equation}
	\mathbf{W}^T \mathbf{C}_i \mathbf{W} = \bm{\Lambda}_i, \quad i = 1, \ldots, K
\label{eq:AJD}		
\end{equation}
are ``as diagonal as possible.'' It is known that for $K>2$, the diagonalization is only achieved approximately by using different variants of approximate joint diagonalization (AJD). Depending on the application and diagonalization algorithm, in order to achieve uncorrelated sources, the total covariance matrix $\mathbf{C}_x = \mathbb{E}\{(\mathbf{x}(t) - \mathbf{m}_x)(\mathbf{x}(t) - \mathbf{m}_x)^T\}$ ($\mathbf{m}_x = \mathbb{E}\{\mathbf{x}(t)\}$), may also be among the set of matrices to be diagonalized\footnote{Enforcing the diagonalization of $\mathbf{C}_x$ guarantees decorrelation of the extracted sources, at a cost of consuming $n(n-1)/2$ degrees of freedom of the matrix $\mathbf{W}$. This is why some BSS algorithms do not enforce whitening or sphering, but rather include the covariance matrix among the approximately diagonalized set of matrices, at a cost of reduced performance \cite{laheld1994adaptive}.}. The approach based on AJD is more robust as compared with $\pi$CA and NSCA, which only work with two matrices. It is also more effective than JADE and other generic ICA algorithms, as it uses specific features of the ECG of the mother and the fetus. However, the order of sources is no longer guaranteed in AJD.

\subsection{Illustration}
\label{sec:results}
\begin{figure}[tb]\sidecaption
\centering
\includegraphics[trim=.9in .15in .85in .4in,clip,width=4.5in]{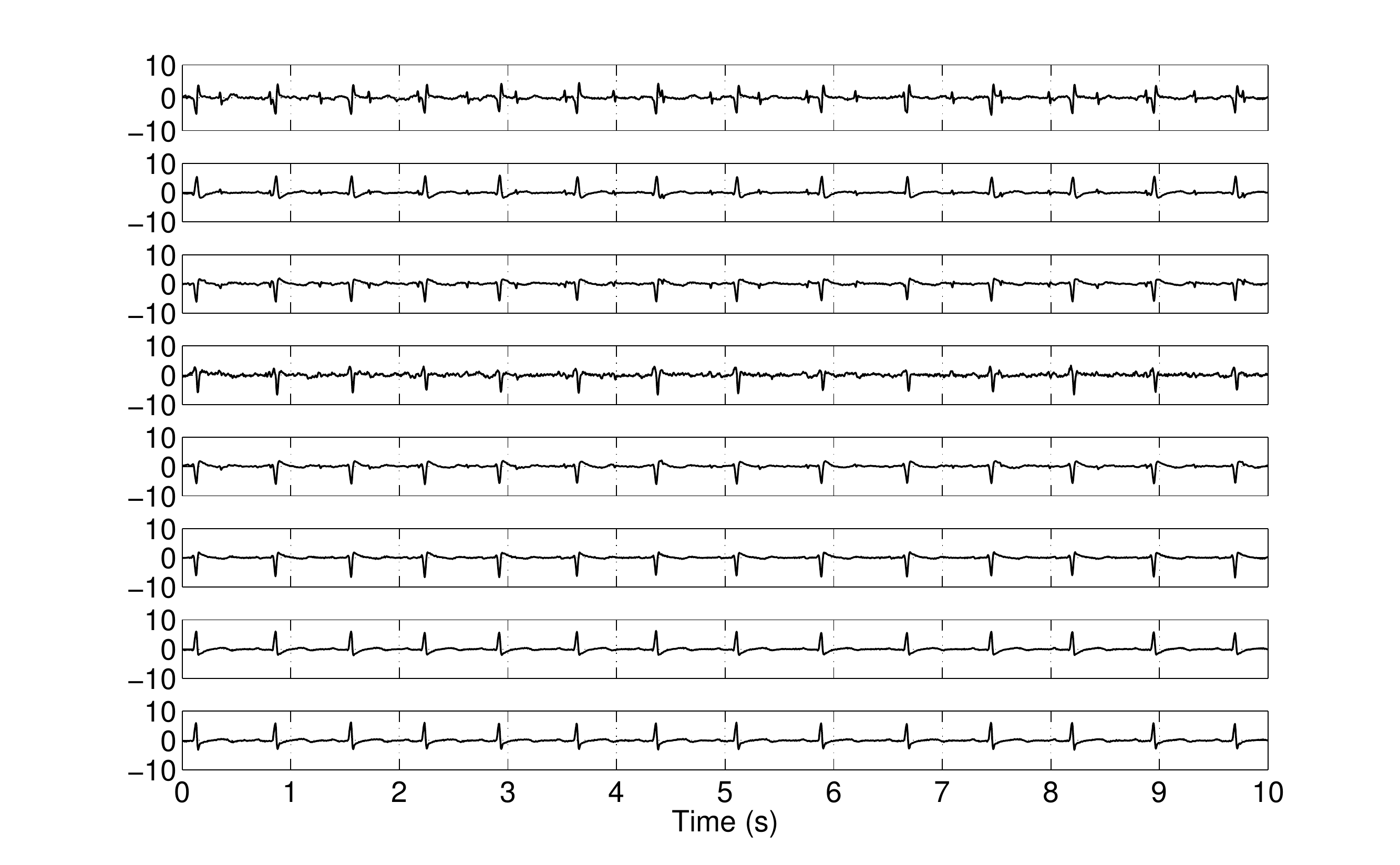}
\caption{The DaISy dataset consisting of five maternal abdominal and three thoracic channels \cite{DaISy}.}
\label{fig:original}
\end{figure}

The DaISy fECG dataset is used for illustration \cite{DaISy}. This sample data consists of five abdominal and three thoracic channels recorded from the abdomen and chest of a pregnant woman at a sampling rate of 250~Hz. The eight channels of the dataset are depicted in Fig.~\ref{fig:original}.

\begin{figure}[tb]\sidecaption
\centering
\includegraphics[trim=.9in .15in .85in .4in,clip,width=4.5in]{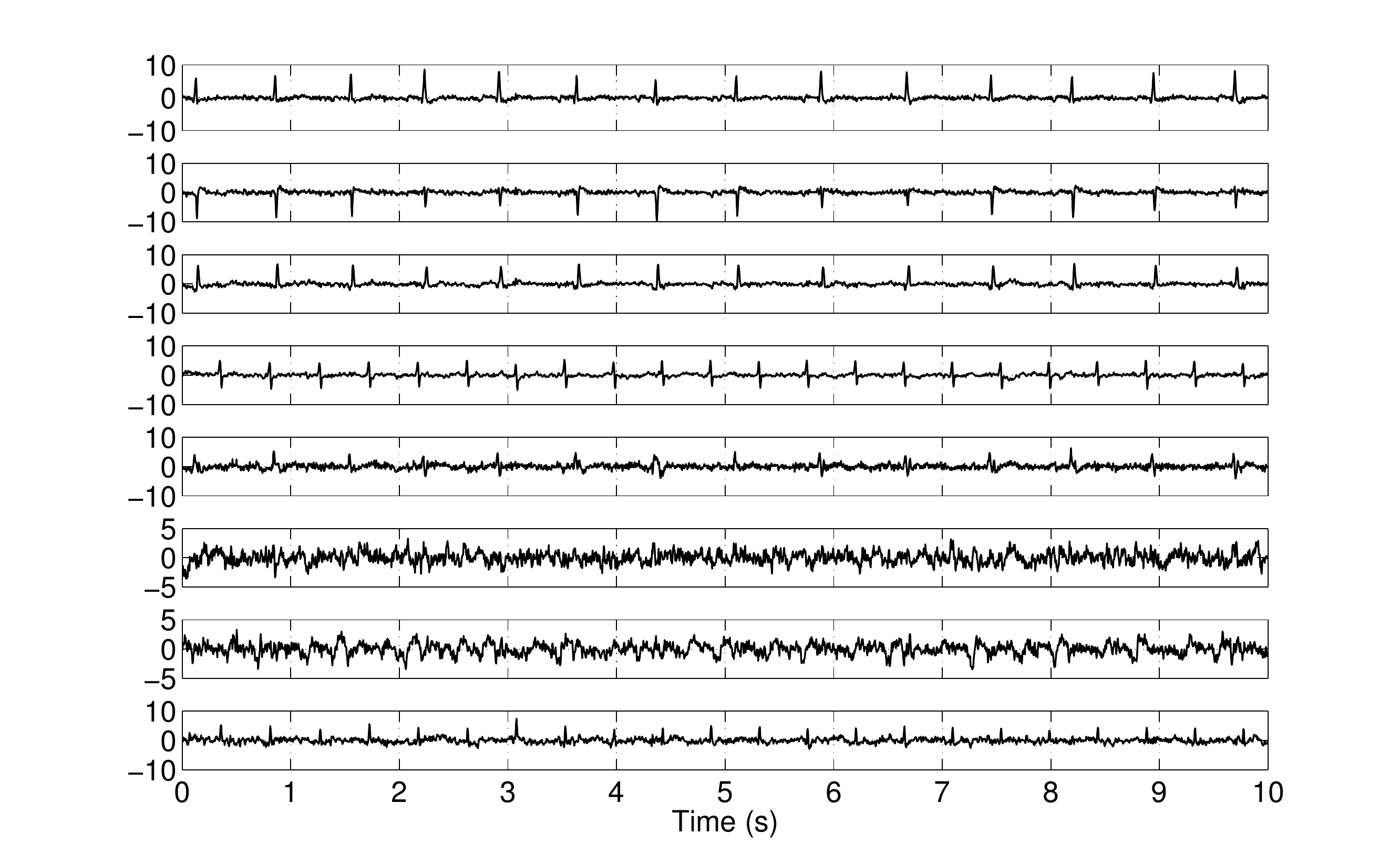}
\caption{Independent components extracted from the dataset of Fig.~\ref{fig:original}, using the JADE algorithm. Notice that components 1, 2, 3, and 5 correspond to the maternal subspace and components 4 and 8 to the fetal subspace.}
\label{fig:ICA}
\end{figure}

The result of applying independent subspace decomposition \cite{Car98}, using the JADE algorithm \cite{Cardoso1993,CardosoSourceCodes} is depicted in Fig.~\ref{fig:ICA}. Accordingly, the first, second, third and fifth components correspond to the mECG subspace, the fourth and eighth components correspond to the fECG, and the sixth and seventh components are noise.

\begin{figure}[tb]\sidecaption
\centering
\includegraphics[trim=.9in .15in .85in .4in,clip,width=4.5in]{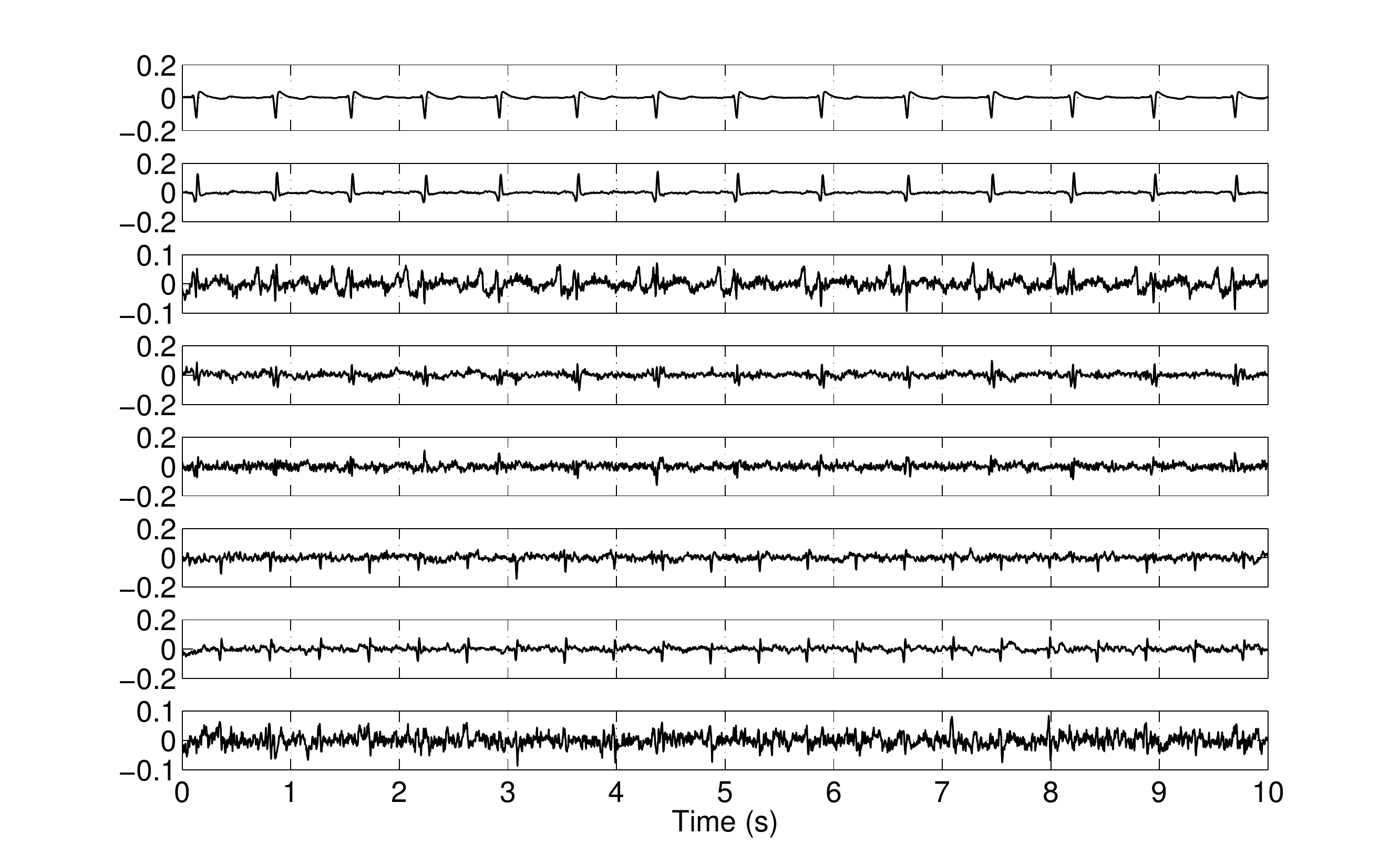}
\caption{Periodic components extracted by $\pi$CA, from the dataset of Fig.~\ref{fig:original}, with \textbf{maternal} ECG beat synchronization. The maternal ECG contribution has reduced from top to bottom.}
\label{fig:maternal}
\end{figure}

By performing R-wave detection on the last maternal thoracic channels of Fig.~\ref{fig:original} (channel eight), the mECG phase $\theta_m(t)$ is calculated as detailed in Section \ref{sec:cardiacphase}. Next, the time-varying mECG period $\tau_t^m$ is calculated, from which the matrix $\mathbf{C}_m$ and the generalized eigenmatrix $\mathbf{W}$ (the joint diagonalizer) of the $(\mathbf{C}_m,\mathbf{C}_x)$ pair is found and its columns are sorted in descending order of the corresponding eigenvalues. The result periodic components calculated from (\ref{eq:lintrans}) are depicted in Fig.~ \ref{fig:maternal}. Accordingly, the first component, which corresponds to the largest eigenvalue, has the most resemblance with the mECG, while as the eigenvalues decrease, the signals become less similar to the mECG. Although two of the extracted components (components six and seven) are the fetal components, the extraction of the fECG has not been explicitly enforced by the algorithm. This can be explained by considering that $\pi$CA is ranking the extracted components according to their resemblance with the mECG period, while the fetal components do not resemble the maternal ECG when they are averaged synchronously with respect to the maternal R-peaks. The fetal components are therefore extracted among the last components, merely as components that are uncorrelated with the mECG and the other signals.
 
\begin{figure}[tb]\sidecaption
\centering
\includegraphics[trim=.9in .15in .85in .4in,clip,width=4.5in]{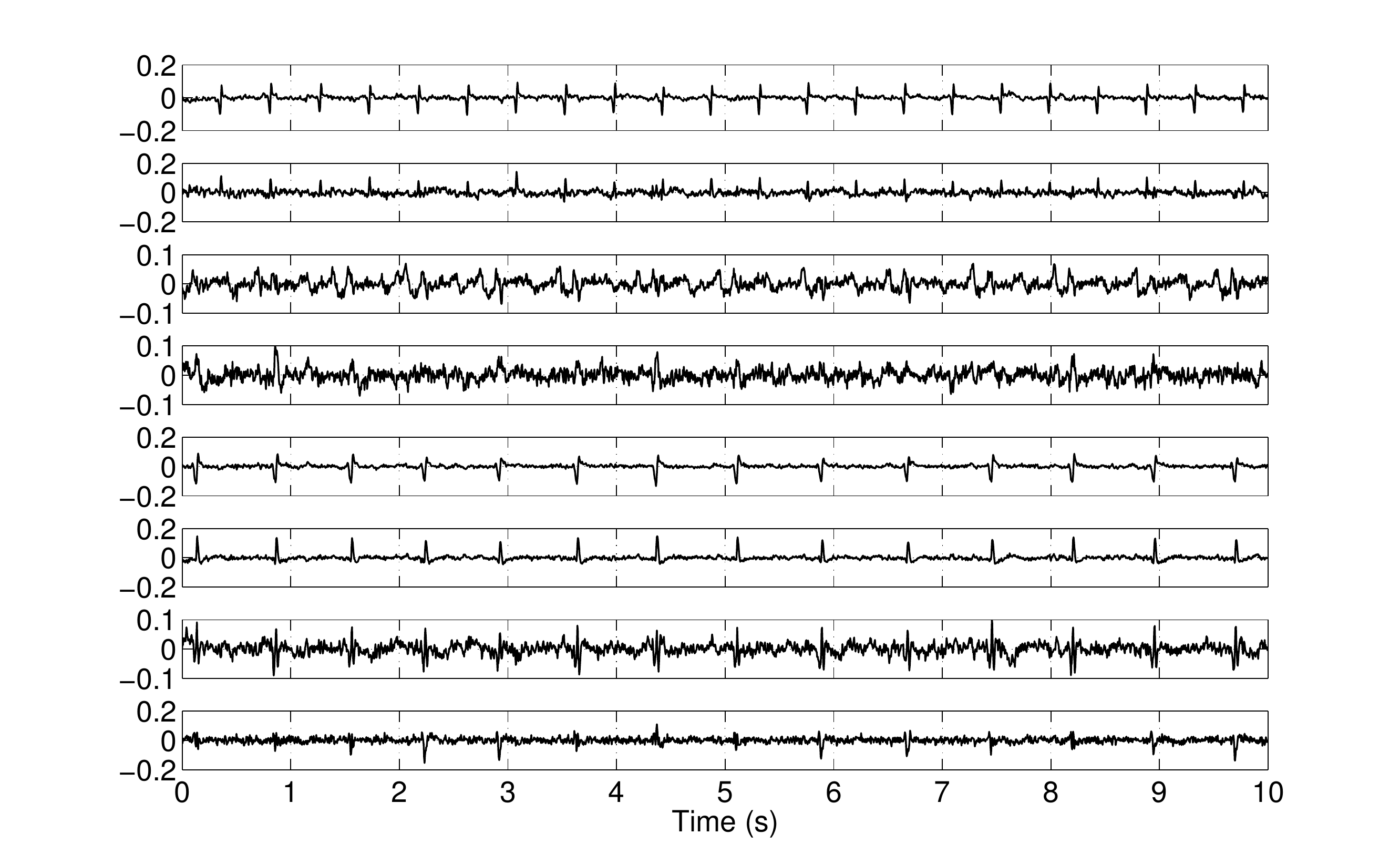}
\caption{Periodic components extracted by $\pi$CA, from the dataset of Fig.~\ref{fig:original}, with \textbf{fetal} ECG beat synchronization. It is observed that the fetal ECG contribution reduces from top to bottom.}
\label{fig:fetal}
\end{figure}

As explained in Section \ref{sec:PiCA}, it is also possible to consider the fECG periodicity in the matrix $\mathbf{C}_f$, which requires the fetal R-peaks for extracting the time-varying fetal period $\tau_t^f$. To illustrate this case, the fECG component extracted by JADE in the fourth channel of Fig.~\ref{fig:ICA} is used for fetal R-peak detection and phase calculation. Having calculated the fECG phase $\theta_f(t)$, GEVD is applied to $(\mathbf{C}_f,\mathbf{C}_x)$ to extract the periodic components of the fECG. The resultant periodic components are depicted in Fig.~\ref{fig:fetal}. In this case, it is observed that the extracted components are ranked according to their resemblance with the fECG.

\begin{figure}[tb]\sidecaption
\centering
\includegraphics[trim=.9in .15in .85in .4in,clip,width=4.5in]{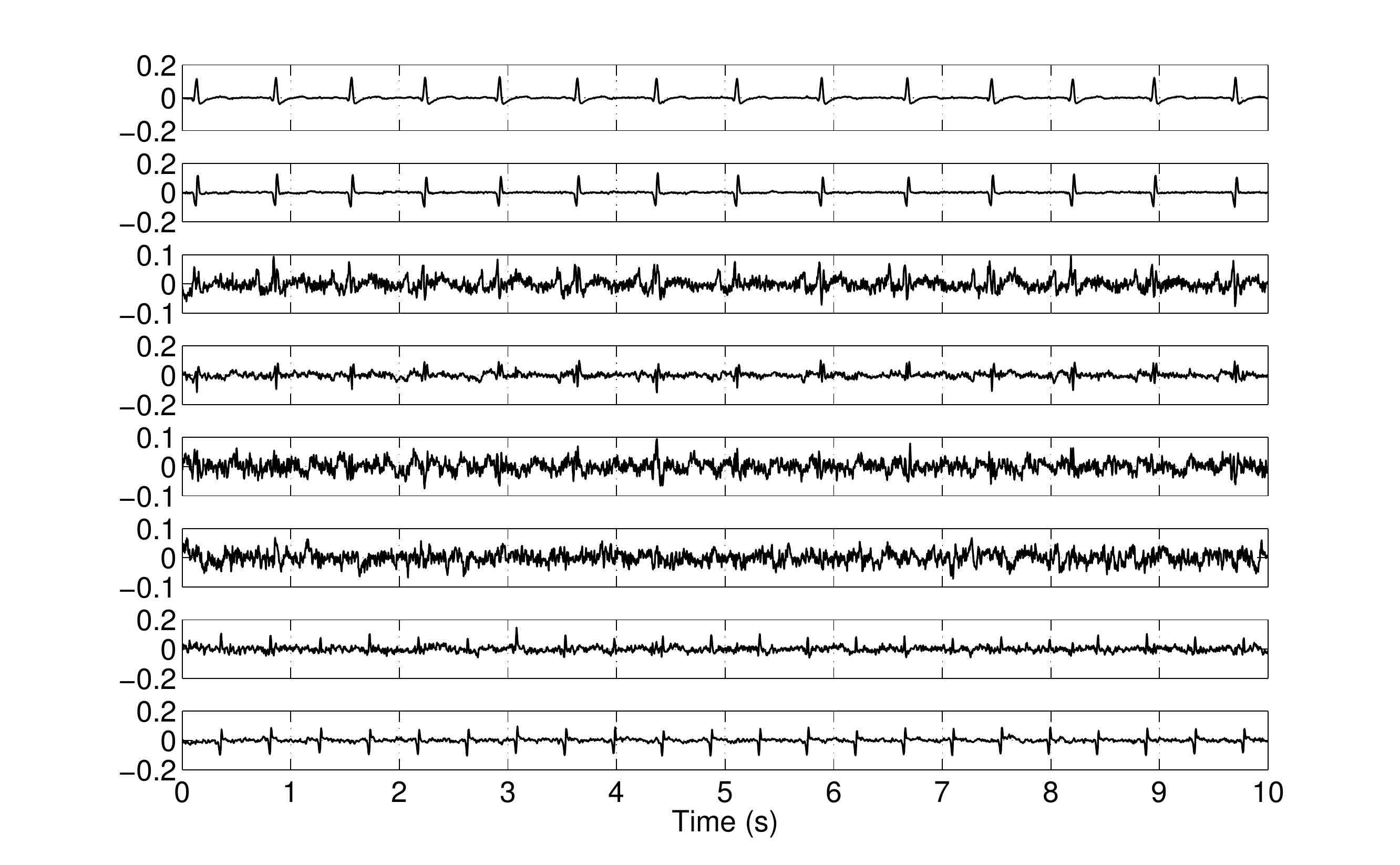}
\caption{Periodic components extracted by $\pi$CA from the dataset of Fig.~\ref{fig:original}, with \textbf{maternal and fetal} ECG beat synchronization. The maternal (fetal) ECG contribution reduces (increases) from top to bottom.}
\label{fig:maternalfetal}
\end{figure}

The next results correspond to the last type of covariance matrix defined in (\ref{eq:C}) by performing GEVD over the matrix pair $(\mathbf{C}_m, \mathbf{C}_f)$ and calculating the periodic components from (\ref{eq:lintrans}). The resulting components are depicted in Fig.~\ref{fig:maternalfetal}. As expected, the first component has the most resemblance with the mECG, while the last component mostly resembles the fECG and the intermediate components are blended from the maternal to fetal ECG plus noise. Note that in this case, the extracted components are no longer uncorrelated, since the covariance matrix of the data has not been diagonalized.

\begin{figure}[tb]
\begin{subfigure}{\textwidth}
\centering\includegraphics[trim=.21in 0in 0in 0in,clip,width=3in, width=4.5in]{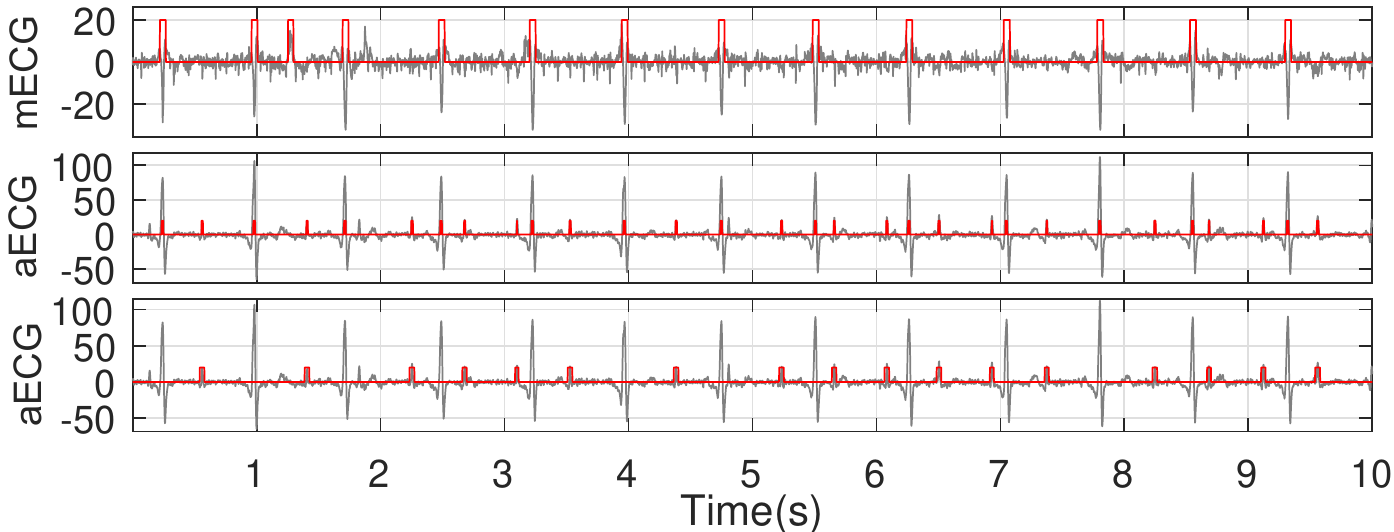}
\subcaption{Reference channels and nonstationary time epochs}\label{fig:caseNo1NSTE}
\end{subfigure}
\begin{subfigure}{\textwidth}
\centering\includegraphics[width=4.5in]{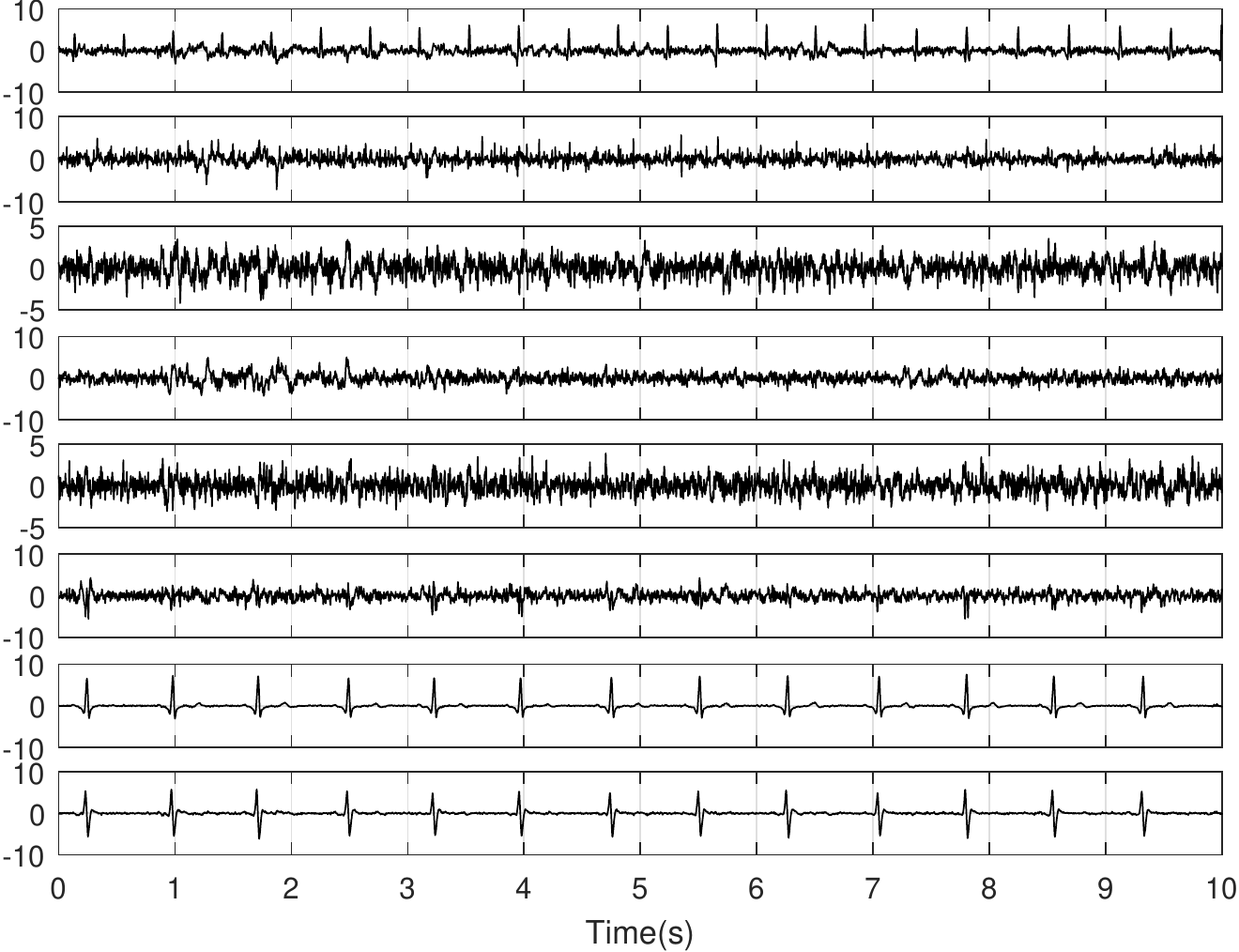}
\subcaption{NSCA result}\label{fig:caseNo1NSCA1}
\end{subfigure}
\caption{The result of NSCA for the sample data of Fig.~\ref{fig:original}. a) the reference mECG local power envelope time epochs (top panel), an abdominal channel local power envelope time epochs (middle panel), and the merged local power envelope time epochs after excluding the mECG time epochs (bottom panel). The nonstationary epochs are shown as red pulses. b) The NSCA result. Adopted from \cite{jamshidian2019temporally}}
\label{fig:NSCAGEVDEnergy}
\end{figure}

The next illustration corresponds to the NSCA algorithm. In this case, the local power envelope index detailed in Section~\ref{sec:NSCA} is used to detect the local power envelope from the first channel of Fig.~\ref{fig:original}. Considering a typical fetal QRS length of approximately 50~ms, the sliding window lengths of the nonstationarity detector in (\ref{eq:TVEnergyRatio}) were set to $w_1$=10~ms and $w_2$=200~ms. The local power envelopes detected by these window lengths can belong to either the mECG or fECG. Therefore, the local peak envelopes of the mECG were independently detected from the last maternal thoracic channel (as a channel which does not have any dominant fetal R-peak due to the electrode location). For this channel, the sliding window lengths were set to $w_1$=20~ms and $w_2$=400~ms, which are adapted for detecting the mECG segments by thresholding. Next, according to the fusion technique explained in \cite{jamshidian2019temporally}, the temporally nonstationary epochs of channel one were excluded from the nonstationary epochs of channel eight, resulting in time instants, which mainly correspond to the fECG and not the mECG. The resulting nonstationary time epochs were used to calculate the required NSCA covariance matrix according to the hypothesis test (\ref{eq:hypothesistestGeneral}). Finally, GEVD was performed on the covariance matrices and the sources were obtained from (\ref{eq:lintrans}). The result of this method together with the detected nonstationary time epochs are shown in Fig.~\ref{fig:NSCAGEVDEnergy}, where it is observed that the fECG is successfully extracted and the components are ranked from top to bottom according to their similarity to the fECG. Furthermore, it is seen that the method has been able to extract the fECG even during the temporal overlaps of the mECG and fECG, despite the fact that some of the fetal QRS peaks have not been considered among the temporally nonstationary epochs (notice the missed fetal R-peaks at t = 1.0, 1.8, 4.0 and 4.8 seconds in the nonstationary epochs of Fig.~\ref{fig:caseNo1NSTE}). Further details regarding this example can be found in \cite{jamshidian2019temporally}.

\section{Advanced methods for fetal ECG extraction}
\label{sec:advances}
In this section, some of the advanced methods, which have been developed in the literature for fECG extraction under special circumstances such as low-rank and time-variant mixtures are reviewed.

\subsection{Low-rank measurements and non linearly-separable fetal and maternal ECG}
As noted throughout the chapter, due to the number and placement of the electrodes, and also the fetal positioning, the maternal abdominal recordings can become rank deficient. As a result, it may happen that the fetal and maternal ECG are not separable using any of the aforementioned linear transforms. In this case, nonlinear methods can be used to separate the maternal and fetal subspaces, or additional synthetic channels can be added to compensate the rank deficiency of the mixtures. 

In order to solve the non-separability of the mECG, it has been proposed to synthetically generate $q$ excess ``clean'' mECG channels, i.e., synthetic channels that resemble the mECG, but do not have any fECG, and to augment the excess channels as auxiliary channel(s) $\mathbf{x}_a(t) \in \mathbb{R}^q$ with the original measured signals \cite{jamshidian2018fetal}:
\begin{equation}
\tilde{\mathbf{x}}(t) = \left[\begin{array}{c} \mathbf{x}(t) \\ \mathbf{x}_a(t) \end{array}\right]
\label{eq:augmenteddatamodel}
\end{equation}
where $\tilde{\mathbf{x}}(t) \in \mathbb{R}^{n+q}$. It was shown in \cite{jamshidian2018fetal} that the $q$ additional synthetic channels amend the rank-deficiency of the problem and help in obtaining a determined or over-determined mixture, from which the fECG could be extracted using conventional ICA, $\pi$CA or NSCA algorithms. Apparently, the auxiliary channel generation and augmentation is a nonlinear procedure, which utilizes the maternal signals' null-space. To implement this method, a channel that resembles the maternal abdominal leads, but is not exactly the same as the other abdominal recorded channels is needed, which at the same time prevents the multichannel data from becoming singular and does not contain any traces of the fECG.

\begin{figure}[tb]
    \centering
    \begin{subfigure}[b]{0.48\textwidth}
        \includegraphics[width=0.9\textwidth]{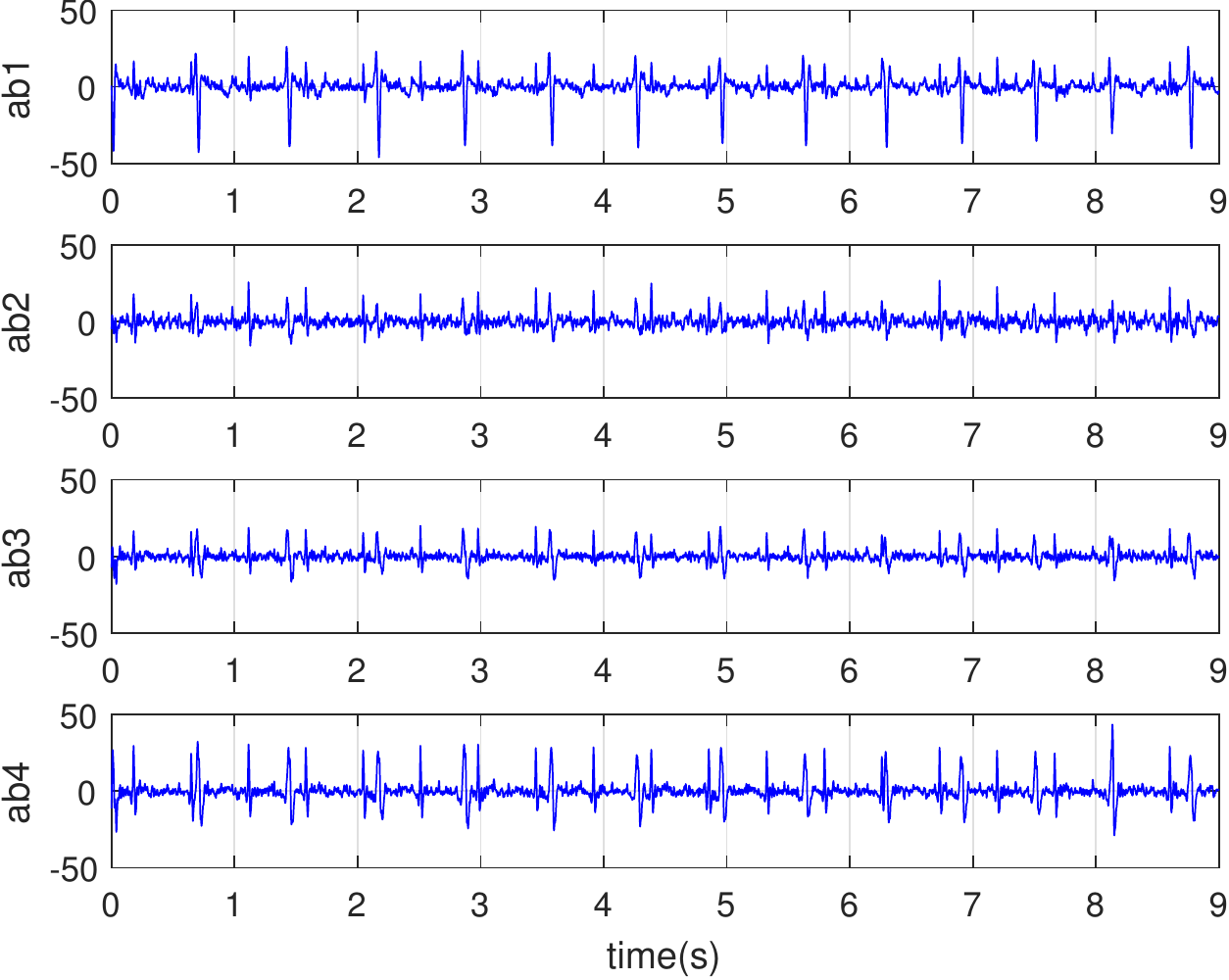}
        \subcaption{Raw abdominal segment}
        \label{fig:ADFECG}
    \end{subfigure}
    ~ 
    \begin{subfigure}[b]{0.48\textwidth}
        \includegraphics[width=0.9\textwidth]{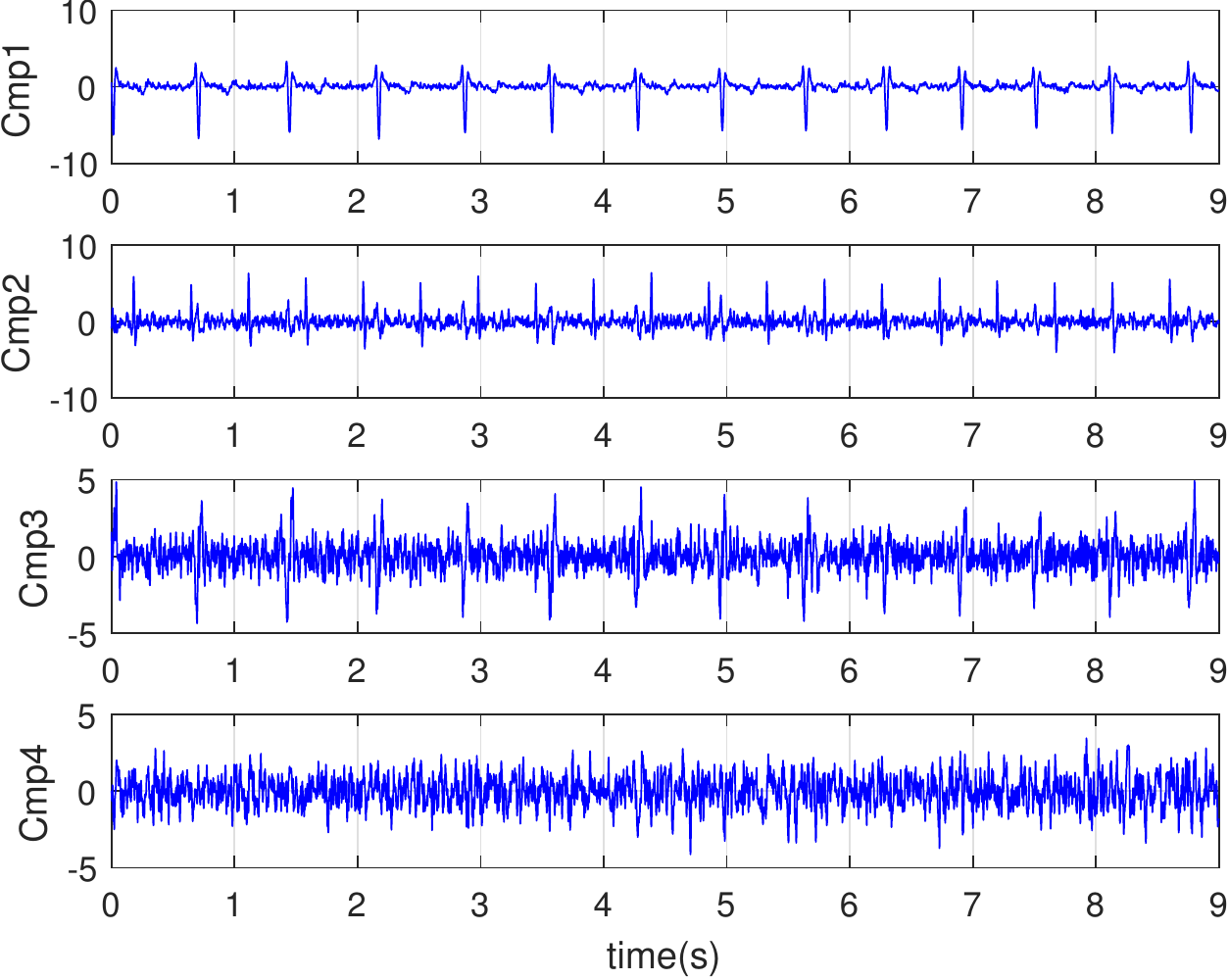}
        \subcaption{JADE on raw segment}
        \label{fig:ADFECGRawJade}
    \end{subfigure}
    \begin{subfigure}[b]{0.48\textwidth}
    \includegraphics[width=0.9\textwidth]{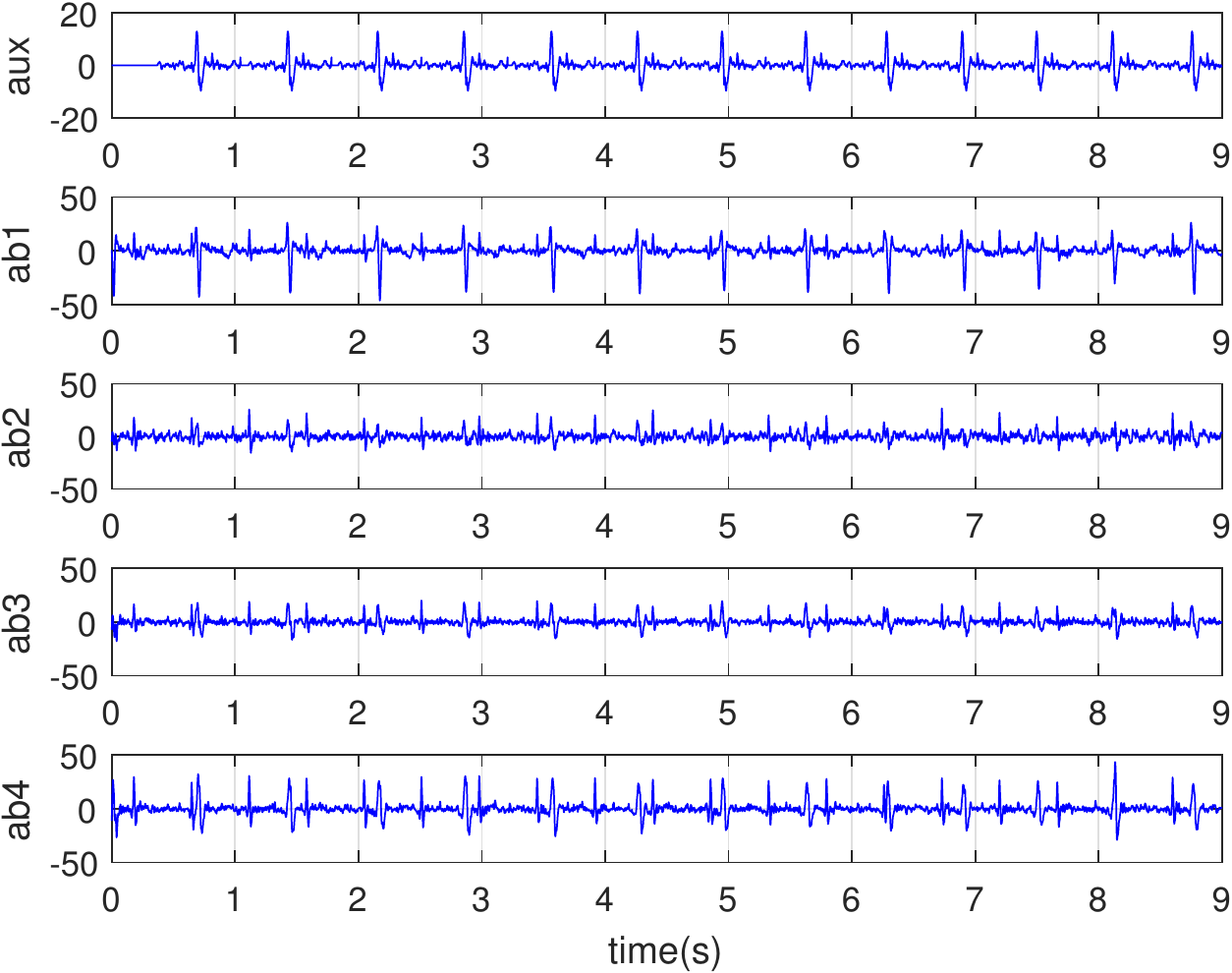}
        \subcaption{Augmented with auxiliary mECG channel}
        \label{fig:ADFECGAugmented}
    \end{subfigure} 
    ~
    \begin{subfigure}[b]{0.48\textwidth}
        \includegraphics[width=0.9\textwidth]{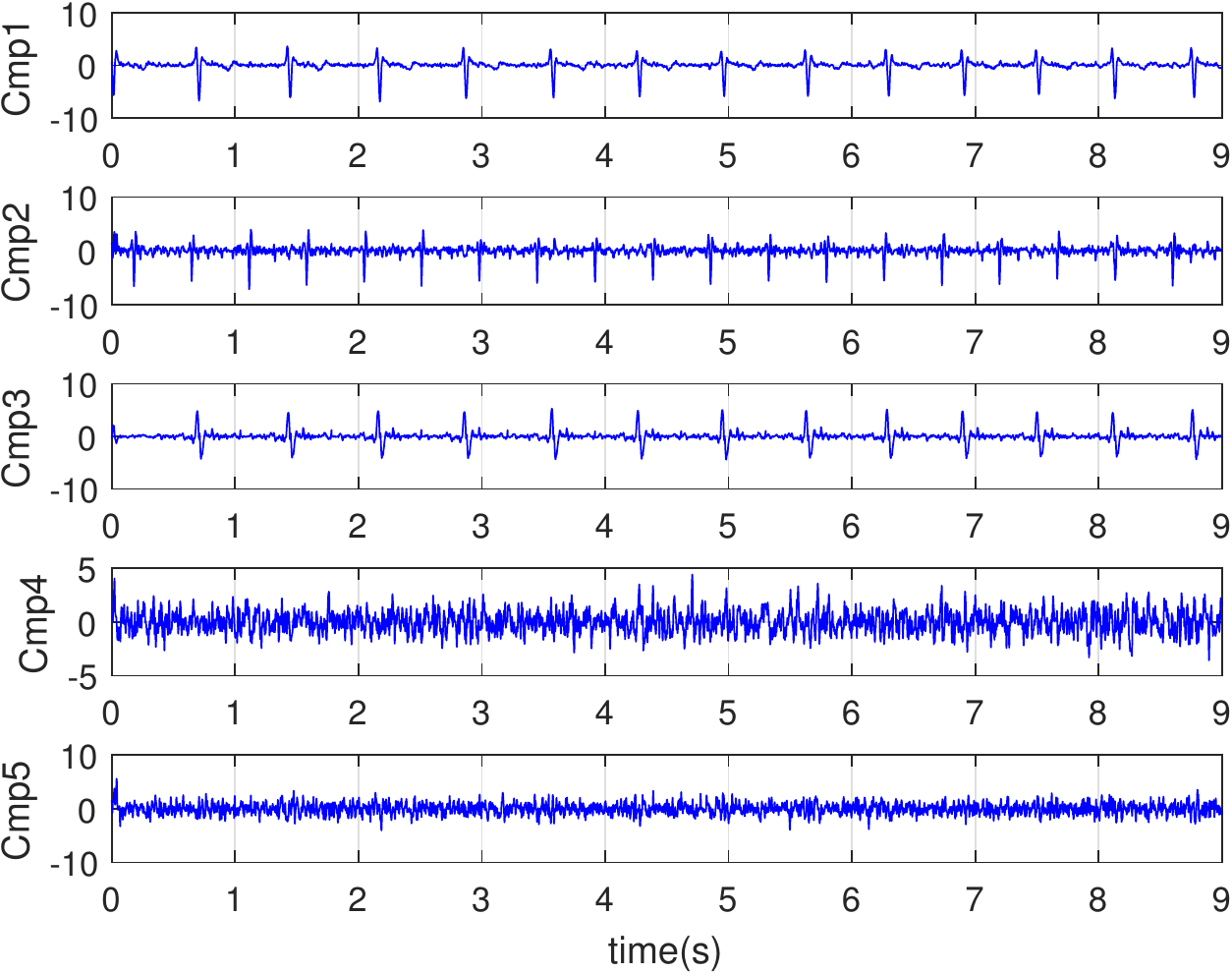}
        \subcaption{JADE on augmented channels}
        \label{fig:ADFECGAugmentedJADE}
    \end{subfigure}
    \caption{(a) A segment of four abdominal channels of the ADFECG database; (b) the result of JADE on the original data segment; (c) the data segment augmented with an auxiliary mECG channel added as the first channel; (d) the result of JADE on the augmented data segment. Adopted from \cite{jamshidian2018fetal,FahimehJamshidianTehraniPHD2015}.}
	\label{fig:JADEOC}
\end{figure}

The ECG cyclostationarity detailed in Section \ref{sec:templatesubtraction}, together with the realistic ECG generator described in  \ref{sec:morphologicmodel} provide the means of constructing the required synthetic maternal abdominal ECG. For this, a set of reference channels are selected. Next, the average mECG morphology is calculated by weighted averaging \cite{Leski2004}. The average morphology is either repeated directly at the positions of the maternal R-peaks to construct a synthetic auxiliary channel (according to \ref{eq:ECGmodel}), or the mECG is extracted by single-channel adaptive or extended Kalman filtering, as detailed in Sections \ref{sec:adaptivefilters} and \ref{sec:kalmanfilter}. The resulting mECG channels are next augmented with the original channels according to (\ref{eq:augmenteddatamodel}). The augmented data is finally given to multichannel source separation algorithms to recover the maternal and fetal ECG components. Note that this technique may not generally be proved to resolve the problem of rank deficiency, as it is data-dependent. However, as demonstrated in  \cite{jamshidian2018fetal}, it has been shown to resolve the rank deficiency of some of the most popular online available fECG datasets, which have few number of channels and other multichannel BSS algorithms have failed \cite{MIT-BIH-ADFECG,MIT-BIH-NIFECG}. For illustration, a sample data adopted from the \textit{abdominal and direct fetal electrocardiogram} (ADFECG) database \cite{MIT-BIH-ADFECG} is shown in Fig.~\ref{fig:JADEOC}. As shown in this figure, the maternal and fetal ECG were not fully separable by applying JADE on the original four channels, since traces of the mECG exist in the fECG component. However, by adding an auxiliary channel according to the procedure detailed in \cite{jamshidian2018fetal}, JADE has achieved in fully separating the mECG and fECG.

\subsection{Maternal-fetal subspace decomposition by deflation}
\begin{figure}[tb]\centering
\includegraphics[trim=0in 0in 0in 0in,clip,width=\textwidth]{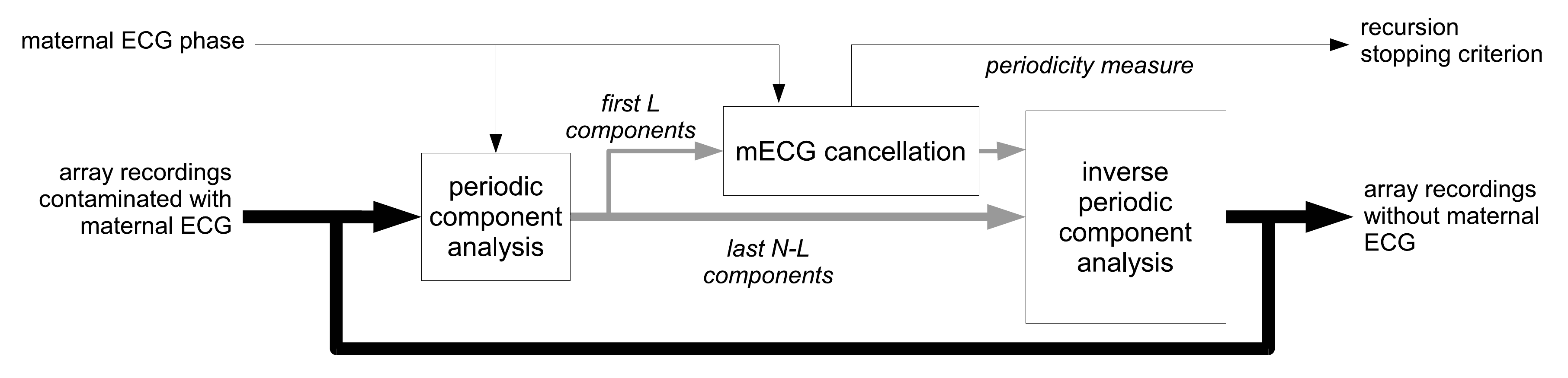}
\caption{The DEFL algorithm for separating the mECG from abdominal recordings, in highly noisy and rank-deficient scenarios \cite{SJS2010,patentSameni1_Published}}
\label{fig:ECGIteration}
\end{figure}

In \cite{Sameni2008,SJS2010,patentSameni1_Published}, a deflation-based procedure, knwon as \textit{denoising by deflation} (DEFL), was proposed for the general problem of rank-deficient and noisy source separation, with special interest in noninvasive fECG extraction. DEFL is a subspace denoising algorithm, which separates the undesired signals of multichannel noisy data using a sequence of \textit{linear decomposition}, \textit{denoising} and \textit{linear re-composition}, in successive iterations. The overall block diagram of DEFL for mECG cancellation is shown in Fig.~\ref{fig:ECGIteration}. Accordingly, the linear decomposition unit is generally a GEVD procedure such as $\pi$CA (or NSCA), using the R-peaks of the mother. The outputs of this unit are ranked in descending (ascending) order of resemblance with the signal (noise) subspace. This block concentrates the components of the maternal subspace in the first few components of its output. The unit is followed by a linear or nonlinear monotonic denoising filter that is applied to the first $L$ components ($1\leq L < N$) of the previous block. This filter can be any of the single-channel filters detailed in Section \ref{sec:singlechannel}, applied to each channel separately, or a multichannel filter applied to the first $L$ components together. Although, such denoising could have been directly applied to the original data $\textbf{x}(t)$, but by applying it after the linear decomposition step, we benefit from the improved signal quality of the first few components extracted by the linear decomposition block. This improvement is the direct consequence of maximizing the $\pi$CA or NSCA cost functions during the GEVD procedure. Finally, the residual signals of the $L$ denoised components and the other $N-L$ unchanged components are transformed back to the observation space, using the inverse of the linear decomposition matrix. In each iteration of the algorithm, portions of the mECG, fECG and noise subspaces are separated, and the procedure is repeated until the output signals satisfy some predefined measure of signal/noise separability.

According to Fig.~\ref{fig:ECGIteration}, each iteration of DEFL can be summarized as follows:
\begin{equation}
\mathbf{y}_i(t) = \mathbf{W}_i^{-T}\mathbf{G}\big(\mathbf{W}_i^T\mathbf{x}_i(t);L\big)
\label{eq:offline}
\end{equation}
where $\mathbf{x}_i(t)$ is the input data of the $i$th iteration ($\mathbf{x}_1(t) = \mathbf{x}(t)$), $\mathbf{y}_i$ is output of the iteration, $\mathbf{G}(\cdot;L)$ is the denoising operator applied to the first $L$ channels of the input, and $\mathbf{W}_i$ is the spatial filter ($\pi$CA or NSCA).

The concept behind (\ref{eq:offline}) is analogical to \textit{wavelet shrinkage} used for single-channel denoising. An important property of the DEFL algorithm is that unlike most ICA-based denoising schemes, the data dimensionality is preserved. Moreover, due to the denoising block between the linear projection stages, it overall performs as a nonlinear filtering scheme, which can deal with full-rank and even non-additive noise mixtures. An adaptive version of this algorithm has also been developed for online fECG extraction \cite{fatemi2017online}.

\subsection{Block-wise and online fetal ECG extraction}
For long multichannel data records, the batch processing requires a huge amount of memory and processing time. Moreover, during long recording sessions, it often happens that the fetus moves, which means that the fetal position changes with respect to the fixed maternal abdominal sensors. Therefore, stationary source separation algorithms, which as in (\ref{eq:datamodelcompact}) assume a constant mixing matrix $\mathbf{A}$, will fail or result in partial fECG source separation. To resolve this issue, the data is either partitioned and processed block-wise, or algorithms specific for online processing are used. 

\subsubsection{Block-wise analysis}
Depending on the application, the maternal abdominal data can be partitioned into overlapping or non-overlapping blocks and any of the fECG extraction schemes detailed in previous sections is applied to each block. This is the most popular method, as it does not require any change in the algorithms used for fECG extraction. However, the challenge is how to automatically identify and recombine the extracted fECG of successive blocks. Especially, as noted in Section \ref{sec:multichannel}, ICA algorithms which are one of the most popular methods for fECG extraction, do not guarantee to preserve the order and amplitude of the sources over different data blocks. As a result, for non-supervised algorithms, a post fECG extraction unit is required, which automatically detects, normalizes and aligns the fECG of successive blocks. Automatic signal quality indexes have been proposed in the literature for adult ECG signal quality assessment \cite{li2007robust,clifford2011signal,Andreotti2017}. In \cite{jamshidian2018fetal}, several signal quality indexes were specifically proposed for the fECG and successfully tested over several available datasets.

\subsubsection{Online source separation}
An alternative solution for processing long fECG data records is to use sample-wise online source separation algorithms. Adaptive source separation algorithms are well-known in the blind source separation literature. One of the most popular algorithms in this context is known as \textit{equivariant adaptive source separation via independence} (EASI) \cite{Cardoso1996}. In this method, the separating matrix at time instant $t$, denoted by $\mathbf{B}(t)$, is adaptively updated using an equivariant serial update:
\begin{equation}
\mathbf{B}(t+1) = \mathbf{B}(t) - \lambda(t) \mathbf{H}(\mathbf{y}(t)) \mathbf{B}(t)
\label{eq:EASI}
\end{equation}
where $\lambda(t)$ is an update factor, $\mathbf{y}(t) = \mathbf{B}(t) \mathbf{x}(t)$ are adaptive estimates of the independent sources and $\mathbf{H}(\cdot)$ is a nonlinear function of the estimated sources \textit{cumulants} \cite{Cardoso1996}. For time-varying mixtures, the mixing matrix $\mathbf{A}$ defined in (\ref{eq:datamodelcompact}) becomes time-variant and the algorithm seeks the demixing matrix such that $\mathbf{B}(t)\mathbf{A}(t)$ approaches identity, that is where $\|\mathbf{H}(\mathbf{y}(t))\|\rightarrow 0$. This approach also works for the cases, in which the variations are due to the sources rather than the mixture. For instance, suppose that the mixing matrix $\mathbf{A}(t)=\mathbf{A}$ is constant, but the sources are nonstationary. As a result, the function $\mathbf{H}(\cdot)$ deviates from zero as the recursion reaches the nonstationary epochs of the signals. Various source separation algorithms, which use sample-wise updates (instead of global averaging) can be used for online fECG extraction \cite[Ch. 3.2]{Hyvarinen2001}, \cite[Ch. 4.5]{ComonJutten2010handbook}.

Finally note that for an online implementation of GEVD-based algorithms (such as $\pi$CA and NSCA), the covariance matrices $\mathbf{C}_{x}$ and $\mathbf{C}_m$ are both updated in time, i.e.,
\begin{equation}
\begin{array}{l}
\mathbf{C}_x(t) = \alpha \mathbf{C}_x(t-1) + \mathbf{x}(t) \mathbf{x}(t)^T\\
\mathbf{C}_m(t) = \beta \mathbf{C}_m(t-1) + \mathbf{x}(t) \mathbf{x}(t-\tau_t)^T
\end{array}
\end{equation}
where $\alpha, \beta \leq 1$ are forgetting factors and $\tau_t$ is the time-variant heart-rate period defined in (\ref{eq:tau}). Further details regarding the various online variants of fECG extraction algorithms can be followed from \cite{fatemi2017online,BiglariSameni2016,jamshidian2018fetal}.

\section{Fetal ECG post-processing}

\subsection{Fetal R-peak detection}
After extracting the fECG from maternal abdominal recordings, the next step is to extract clinical parameters from the fECG. The fetal heart-rate (fHR) is the first and most commonly used parameter, which in turns requires the detection of the fECG R-peaks. In this context, classical R-peak detectors such as local peak search over sliding windows or the well-known Pan-Tompkins method used in adult ECG \cite{Pan1985}, are the most common. However, considering the relatively low SNR of the fECG and its limited morphological shapes, specific fetal R-peak detectors have been developed that are robust to noise \cite{BiglariSameni2016,jamshidian2018fetal}. These methods are based on a \textit{matched filter} using fixed or adaptive QRS-like templates. A wide range of these techniques were studied and compared with one another during the Physionet/Computing in Cardiology Challenge 2013 \cite{silva2013noninvasive}. 

After fetal R-peaks, the fHR time series is commonly post-processed to refine the calculated heart-rate time-series and to correct the excess and missing R-peaks. These corrections have been commonly performed by rule-based methods, which correct the outlier R-peaks (and the corresponding heart-beats), while keeping the normal beats unchanged \cite{silva2013noninvasive,dessi2014advanced,FahimehJamshidianTehraniPHD2015}.

\subsection{Fetal ECG enhancement}
\label{sec:post-processing}
Depending on the signal quality, after mECG cancellation, the fECG might be directly detectable from one or more of the residual channels, or additional stages may be required for extracting the fECG from the residual background noise. As detailed in Section \ref{sec:singlechannel}, numerous techniques have been proposed for ECG denoising, including Kalman filters \cite{SSJC06,SSJB05,SSJ05}, wavelet denoisers \cite{kabir2012denoising,sameni2017online}, filtering using piecewise smoothness priors \cite{sameni2017online}, etc. An example of such post-processing for fECG enhancement was demonstrated in Fig.~\ref{fig:realeval}.

For morphological analysis due to the relatively low SNR of fECG signals--- even after mECG and background noise cancellation--- the SNR improvement obtained by post-processing filters can still be insufficient for reliable fECG parameter extraction. In this case, an effective approach is to use synchronous weighted averaging of successive beats \cite{Leski2004}. This procedure is known to improve the SNR by a factor of $K$, where $K$ is the number of averaged beats.  

\subsection{Fetal ECG morphological parameter extraction}
To date, the morphological parameters of the fECG and their relationship with the well-being of the fetus are still under study. Researchers have extracted parameters such as the QT-interval \cite{AJOG2011,BeharAJOG2013,joachimbehar2014} and the ST-segment \cite{CliffordAJOG2009,McDonnellAJOG2011}. The typical benchmark for these studies is commonly the invasive fECG obtained from the fetal scalp electrodes acquired during labor. However, it is currently difficult to evaluate the fECG parameters independently, since there are very few open-access fECG databases with expert annotations. Considering that the technology of fECG acquisition and processing is emerging as a standard procedure, it is foreseen that fetal ECG-based parameter extraction will be the main focus of research in future studies.

\section{Conclusion}
In this chapter some of the major technologies and algorithms used for the acquisition and noninvasive processing of fetal electrocardiogram signals from maternal abdominal recordings were reviewed. The recent advances in this domain, especially during the past decade, demonstrate that the technology is emerging as a stable and reliable alternative for invasive methods. A promising future trend is to combine this technology with other low-cost fetal cardiac monitoring modalities such as the phonocardiogram (PCG) and the Doppler technology. The extension of these technologies to multiple pregnancies, pathological cases and its combination with other vital aspects such as the development of the fetal central nervous system (CNS) and cerebral growth are among the future challenges of this domain. The availability of open-access data with clinical annotations and open-source devices and algorithms are among the requirements that can significantly accelerate the development of this technology.

\bibliographystyle{spmpsci}
\bibliography{refbib} 

\begin{thebibliography}{100}
\providecommand{\url}[1]{{#1}}
\providecommand{\urlprefix}{URL }
\expandafter\ifx\csname urlstyle\endcsname\relax
  \providecommand{\doi}[1]{DOI~\discretionary{}{}{}#1}\else
  \providecommand{\doi}{DOI~\discretionary{}{}{}\begingroup
  \urlstyle{rm}\Url}\fi

\bibitem{Andreotti2016}
Andreotti, F., Behar, J., Zaunseder, S., Oster, J., Clifford, G.D.: {Fetal ECG
  Synthetic Database (FECGSYNDB)} (2016).
\newblock \doi{10.13026/C21P4T}.
\newblock \urlprefix\url{https://physionet.org/content/fecgsyndb/}

\bibitem{Andreotti2017}
Andreotti, F., Gr{\"a}{\ss}er, F., Malberg, H., Zaunseder, S.: Non-invasive
  fetal {ECG} signal quality assessment for multichannel heart rate estimation.
\newblock {IEEE} Transactions on Biomedical Engineering \textbf{64}(12),
  2793--2802 (2017).
\newblock \doi{10.1109/tbme.2017.2675543}

\bibitem{andreotti2014robust}
Andreotti, F., Riedl, M., Himmelsbach, T., Wedekind, D., Wessel, N., Stepan,
  H., Schmieder, C., Jank, A., Malberg, H., Zaunseder, S.: {Robust fetal ECG
  extraction and detection from abdominal leads}.
\newblock Physiological measurement \textbf{35}(8), 1551 (2014)

\bibitem{assaleh2006extraction}
Assaleh, K.: Extraction of fetal electrocardiogram using adaptive neuro-fuzzy
  inference systems.
\newblock IEEE Transactions on Biomedical Engineering \textbf{54}(1), 59--68
  (2006)

\bibitem{aastrom2000}
{\AA}str{\"o}m, M., Santos, E.C., S{\"o}rnmo, L., Laguna, P., Wohlfart, B.:
  Vectorcardiographic loop alignment and the measurement of morphologic
  beat-to-beat variability in noisy signals.
\newblock Biomedical Engineering, IEEE Transactions on \textbf{47}(4), 497--506
  (2000)

\bibitem{DaISy}
{B. De Moor}: {Database for the Identification of Systems (DaISy)}.
\newblock \urlprefix\url{http://homes.esat.kuleuven.be/~smc/daisy/}

\bibitem{bach03beyond}
Bach, F., Jordan, M.: Beyond independent components: trees and clusters.
\newblock Journal of Machine Learning Research \textbf{4}, 1205--1233 (2003).
\newblock \urlprefix\url{http://cmm.ensmp.fr/~bach/bach03a.pdf}

\bibitem{joachimbehar2014}
Behar, J.: Extraction of clinical information from the non-invasive fetal
  electrocardiogram.
\newblock Ph.D. thesis, Oxford University, UK (2014)

\bibitem{behar2014comparison}
Behar, J., Johnson, A., Clifford, G.D., Oster, J.: {A comparison of single
  channel fetal ECG extraction methods}.
\newblock Annals of biomedical engineering \textbf{42}(6), 1340--1353 (2014)

\bibitem{BeharAJOG2013}
Behar, J., Wolfberg, A., Zhu, T., Oster, J., Niksch, A., Mah, D., Chun, T.,
  Greenberg, J., Tanner, C., Harrop, J., et~al.: {Evaluation of the fetal QT
  interval using non-invasive fetal ECG technology}.
\newblock American Journal of Obstetrics \& Gynecology \textbf{210}(1),
  S283--S284 (2014)

\bibitem{Belouchrani1997}
Belouchrani, A., Abed-Meraim, K., Cardoso, J.F., Moulines, E.: {A Blind Source
  Separation Technique Using Second-Order Statistics}.
\newblock {IEEE} Trans. Signal Processing \textbf{45}, 434--444 (1997)

\bibitem{bb16291}
Ben-Arie, J., Rao, K.: Nonorthogonal representation of signals by {G}aussians
  and {G}abor functions.
\newblock {IEEE} Trans. Circuits Syst. {II} \textbf{42}(6), 402--413 (1995)

\bibitem{BiglariSameni2016}
Biglari, H., Sameni, R.: Fetal motion estimation from noninvasive cardiac
  signal recordings.
\newblock Physiological Measurement \textbf{37}(11), 2003--2023 (2016).
\newblock \urlprefix\url{http://stacks.iop.org/0967-3334/37/i=11/a=2003}

\bibitem{CardosoSourceCodes}
Cardoso, J.F.: Source Codes for Blind Source Separation and Independent
  Component Analysis.
\newblock \urlprefix\url{http://www2.iap.fr/users/cardoso/}

\bibitem{Car98}
Cardoso, J.F.: {Multidimensional independent component analysis}.
\newblock In: Proceedings of the IEEE International Conference on Acoustics,
  Speech, and Signal Processing (ICASSP'98), vol.~4, pp. 1941--1944 (1998)

\bibitem{Cardoso99}
Cardoso, J.F.: High-order contrasts for independent component analysis.
\newblock Neural Comput. \textbf{11}(1), 157--192 (1999).
\newblock \doi{http://dx.doi.org/10.1162/089976699300016863}

\bibitem{Cardoso1996}
Cardoso, J.F., Laheld, B.: Equivariant adaptive source separation.
\newblock Signal Processing, IEEE Transactions on \textbf{44}(12), 3017 --3030
  (1996).
\newblock \doi{10.1109/78.553476}

\bibitem{Cardoso1993}
Cardoso, J.F., Souloumiac, A.: {Blind beamforming for non Gaussian signals}.
\newblock {IEE - Proceedings -F} \textbf{140}, 362--370 (1993)

\bibitem{GDC06}
Clifford, G.: A novel framework for signal representation and source
  separation.
\newblock Journal of Biological Systems \textbf{14}(2), 169--183 (2006)

\bibitem{clifford2011signal}
Clifford, G., Lopez, D., Li, Q., Rezek, I.: Signal quality indices and data
  fusion for determining acceptability of electrocardiograms collected in noisy
  ambulatory environments.
\newblock In: Computing in Cardiology, 2011, pp. 285--288. IEEE (2011)

\bibitem{CliffordAJOG2009}
Clifford, G., Sameni, R., Ward, J., Robertson, J., Pettigrew, C., Wolfberg, A.:
  {Comparing the fetal ST-segment acquired using a FSE and abdominal sensors}.
\newblock American Journal of Obstetrics \& Gynecology \textbf{201}(6), S242
  (2009)

\bibitem{AJOG2011}
Clifford, G., Sameni, R., Ward, J., Robinson, J., Wolfberg, A.J.: {Clinically
  accurate fetal ECG parameters acquired from maternal abdominal sensors}.
\newblock {American Journal of Obstetrics and Gynecology} \textbf{205}(1),
  47.e1--47.e5 (2011)

\bibitem{cliffordSPIE04}
Clifford, G.D., McSharry, P.E.: A realistic coupled nonlinear artificial
  {{ECG}}, {BP}, and respiratory signal generator for assessing noise
  performance of biomedical signal processing algorithms.
\newblock Proc of SPIE International Symposium on Fluctuations and Noise
  \textbf{5467}(34), 290--301 (2004)

\bibitem{CSMJ05b}
Clifford, G.D., Shoeb, A., McSharry, P.E., Janz, B.A.: {Model-based Filtering,
  Compression and Classification of the ECG}.
\newblock International Journal of Bioelectromagnetism \textbf{7}(1), 158--161
  (2005)

\bibitem{IEC60601-2-25:2011}
Commission, I.E.: {Medical electrical equipment - Part 2-25: Particular
  requirements for the basic safety and essential performance of
  electrocardiographs}.
\newblock Standard, International Standard (2011)

\bibitem{Comon95}
Comon, P.: Supervised classification, a probabilistic approach.
\newblock In: Verleysen (ed.) ESANN-European Symposium on Artificial Neural
  Networks, pp. 111--128. D~facto Publ., Brussels (1995).
\newblock [invited paper]

\bibitem{ComonJutten2010handbook}
Comon, P., Jutten, C. (eds.): Handbook of Blind Source Separation: Independent
  Component Analysis and Applications.
\newblock Independent Component Analysis and Applications Series. Elsevier
  Science (2010)

\bibitem{Cremer1906}
Cremer, M.: {\"Uber die Direkte Ableitung der Aktionstrome des Menschlichen
  Herzens vom Oesophagus und \"Uber das Elektrokardiogramm des Fetus}.
\newblock {M\"unchener Medizinische Wochenschrift} \textbf{53}, 811--813 (1906)

\bibitem{dessi2014advanced}
Dess{\`\i}, A., Pani, D., Raffo, L.: An advanced algorithm for fetal heart rate
  estimation from non-invasive low electrode density recordings.
\newblock Physiological measurement \textbf{35}(8), 1621 (2014)

\bibitem{Devedeux1993}
Devedeux, D., Marque, C., Mansour, S., Germain, G., Duch{\^e}ne, J.: Uterine
  electromyography: a critical review.
\newblock American journal of obstetrics and gynecology \textbf{169}(6),
  1636--1653 (1993)

\bibitem{Farvet1968}
Farvet, A.G.: {Computer Matched Filter Location of Fetal R-Waves}.
\newblock Medical \& Biological Engineering \textbf{6}(5), 467--475 (1968)

\bibitem{fatemi2017online}
Fatemi, M., Sameni, R.: {An online subspace denoising algorithm for maternal
  ECG removal from fetal ECG signals}.
\newblock Iranian Journal of Science and Technology, Transactions of Electrical
  Engineering \textbf{41}(1), 65--79 (2017)

\bibitem{Ferreol2000}
Ferreol, A., Chevalier, P.: {On the Behavior of Current Second and Higher Order
  Blind Source Separation Methods for Cyclostationary Sources}.
\newblock {IEEE} Trans. Signal Processing \textbf{48}, 1712--1725 (2000)

\bibitem{gardCyclo}
Gardner, W.A.: Cyclostationarity in communications and signal processing.
\newblock Tech. rep., DTIC Document (1994)

\bibitem{hall2006}
Hall, J., Hall, J., Guyton, A.: Guyton \& Hall physiology review.
\newblock Guyton Physiology Series. Elsevier Saunders (2006)

\bibitem{HMM03}
Hamerling, S., Meinecke, F., {M\"{u}ller}, K.R.: {Analysing ICA components by
  injecting noise}.
\newblock In: Proceedings of the 4th Int. Symp. on Independent Component
  Analysis and Blind Source Separation (ICA2003), pp. 149--154. Nara, Japan
  (2003).
\newblock
  \urlprefix\url{http://www.lis.inpg.fr/pages_perso/bliss/deliverables/d19.html}

\bibitem{Hyvarinen2001}
Hyvarinen, A., Karhunen, J., Oja, E.: Independent Component Analysis.
\newblock Wiley-Interscience (2001)

\bibitem{jafarnia2007modified}
Jafarnia-Dabanloo, N., McLernon, D., Zhang, H., Ayatollahi, A., Johari-Majd,
  V.: {A modified Zeeman model for producing HRV signals and its application to
  ECG signal generation}.
\newblock Journal of theoretical biology \textbf{244}(2), 180--189 (2007)

\bibitem{FahimehJamshidianTehraniPHD2015}
Jamshidian-Tehrani, F.: {Online Noninvasive Fetal Cardiac Signal Extraction}.
\newblock Ph.D. thesis, Artificial Intelligence, School of Electrical \&
  Computer Engineering, Shiraz University (2019)

\bibitem{jamshidian2018fetal}
Jamshidian-Tehrani, F., Sameni, R.: {Fetal ECG extraction from time-varying and
  low-rank noninvasive maternal abdominal recordings}.
\newblock Physiological measurement \textbf{39}(12), 125008 (2018)

\bibitem{jamshidian2019temporally}
Jamshidian-Tehrani, F., Sameni, R., Jutten, C.: Temporally nonstationary
  component analysis; application to noninvasive fetal electrocardiogram
  extraction.
\newblock IEEE Transactions on Biomedical Engineering  (2019).
\newblock \urlprefix\url{https://doi.org/10.1109/TBME.2019.2936943}

\bibitem{webster2009medical}
John W.~Clark, J.: The origin of biopotentials.
\newblock In: J.G. Webster (ed.) Medical instrumentation: application and
  design, chap.~4. John Wiley \& Sons (2009)

\bibitem{kabir2012denoising}
Kabir, M.A., Shahnaz, C.: {Denoising of ECG signals based on noise reduction
  algorithms in EMD and wavelet domains}.
\newblock Biomedical Signal Processing and Control \textbf{7}(5), 481--489
  (2012)

\bibitem{Kanjilal1997}
Kanjilal, P., Kanjilal, P., Palit, S., Saha, G.: {Fetal ECG extraction from
  single-channel maternal ECG using singular value decomposition}.
\newblock Biomedical Engineering, IEEE Transactions on \textbf{44}(1), 51--59
  (1997).
\newblock \doi{10.1109/10.553712}

\bibitem{kester2005data}
Kester, W., Engineeri, A.D.I.: Data conversion handbook.
\newblock Newnes (2005)

\bibitem{Khamene2000}
Khamene, A., Negahdaripour, S.: {A new method for the extraction of fetal ECG
  from the composite abdominal signal}.
\newblock Biomedical Engineering, IEEE Transactions on \textbf{47}(4), 507--516
  (2000).
\newblock \doi{10.1109/10.828150}

\bibitem{laguna1996adaptive}
Laguna, P., Jan{\'e}, R., Olmos, S., Thakor, N.V., Rix, H., Caminal, P.:
  {Adaptive estimation of QRS complex wave features of ECG signal by the
  Hermite model}.
\newblock Medical and Biological Engineering and computing \textbf{34}(1),
  58--68 (1996)

\bibitem{laheld1994adaptive}
Laheld, B., Cardoso, J.F.: Adaptive source separation without pre-whitening.
\newblock In: {EUSIPCO-94 - The 7th European Signal Processing Conf.}, pp.
  183--186. Edinburgh, UK (1994)

\bibitem{Larks1962}
Larks, S.D.: Present status of fetal electrocardiography.
\newblock Bio-Medical Electronics, IRE Transactions on \textbf{9}(3), 176--180
  (1962).
\newblock \doi{10.1109/TBMEL.1962.4322994}

\bibitem{Lathauwer2000}
de~Lathauwer, L., de~Moor, B., Vandewalle, J.: Fetal electrocardiogram
  extraction by blind source subspace separation.
\newblock Biomedical Engineering, IEEE Transactions on \textbf{47}(5), 567
  --572 (2000).
\newblock \doi{10.1109/10.841326}

\bibitem{Leski2002}
Leski, J.: Robust weighted averaging [of biomedical signals].
\newblock Biomedical Engineering, IEEE Transactions on \textbf{49}(8), 796--804
  (2002).
\newblock \doi{10.1109/TBME.2002.800757}

\bibitem{Leski2004}
Leski, J., Gacek, A.: {Computationally effective algorithm for robust weighted
  averaging}.
\newblock Biomedical Engineering, IEEE Transactions on \textbf{51}(7),
  1280--1284 (2004).
\newblock \doi{10.1109/TBME.2004.827953}

\bibitem{li2007robust}
Li, Q., Mark, R.G., Clifford, G.D.: {Robust heart rate estimation from multiple
  asynchronous noisy sources using signal quality indices and a Kalman filter}.
\newblock Physiological measurement \textbf{29}(1), 15 (2007)

\bibitem{li2017efficient}
Li, R., Frasch, M.G., Wu, H.T.: {Efficient fetal-maternal ECG signal separation
  from two channel maternal abdominal ECG via diffusion-based channel
  selection}.
\newblock Frontiers in physiology \textbf{8}, 277 (2017)

\bibitem{Liu2015}
Liu, G., Luan, Y.: An adaptive integrated algorithm for noninvasive fetal {ECG}
  separation and noise reduction based on {ICA}-{EEMD}-{WS}.
\newblock Medical {\&} Biological Engineering {\&} Computing \textbf{53}(11),
  1113--1127 (2015).
\newblock \doi{10.1007/s11517-015-1389-1}

\bibitem{Ma2015}
Ma, Y., Xiao, Y., Wei, G., Sun, J.: A multichannel nonlinear adaptive noise
  canceller based on generalized {FLANN} for fetal {ECG} extraction.
\newblock Measurement Science and Technology \textbf{27}(1), 015703 (2015).
\newblock \doi{10.1088/0957-0233/27/1/015703}

\bibitem{Malmivuo2000}
Malmivuo, J.: Biomagnetism.
\newblock In: J.D. Bronzino (ed.) The Biomedical Engineering Handbook, 2nd
  edn., chap.~16. Boca Raton: CRC Press LLC (2000)

\bibitem{MP95}
Malmivuo, J.A., Plonsey, R.: {Bioelectromagnetism, Principles and Applications
  of Bioelectric and Biomagnetic Fields}.
\newblock Oxford University Press (1995).
\newblock \urlprefix\url{http://butler.cc.tut.fi/~malmivuo/bem/bembook}

\bibitem{Martens2007}
Martens, S.M., Rabotti, C., Mischi, M., Sluijter, R.J.: {A robust fetal ECG
  detection method for abdominal recordings}.
\newblock Physiological measurement \textbf{28}(4), 373 (2007).
\newblock \doi{10.1088/0967-3334/28/4/004}.
\newblock \urlprefix\url{http://dx.doi.org/10.1088/0967-3334/28/4/004}

\bibitem{McDonnellAJOG2011}
McDonnell, C., Clifford, G., Sameni, R., Ward, J., Robertson, J., Wolfberg, A.:
  {Comparison of abdominal sensors to a fetal scalp electrode for fetal ST
  analysis during labor}.
\newblock American Journal of Obstetrics \& Gynecology \textbf{204}(1), S256
  (2011)

\bibitem{McSharry2003}
McSharry, P.E., Clifford, G.D., Tarassenko, L., Smith, L.A.: {A Dynamic Model
  for Generating Synthetic Electrocardiogram Signals}.
\newblock Biomedical Engineering, IEEE Transactions on \textbf{50}, 289--294
  (2003)

\bibitem{Meinecke2002}
Meinecke, F., Ziehe, A., Kawanabe, M., {M\"{u}ller}, K.R.: A resampling
  approach to estimate the stability of one-dimensional or multidimensional
  independent components.
\newblock Biomedical Engineering, IEEE Transactions on \textbf{49}(12 Pt 2),
  1514--25 (2002)

\bibitem{HadiNarimaniMS2014}
Narimani, H.: {Application of Kalman and H-$\infty$ filters in
  Electrocardiogram Denoising}.
\newblock Master's thesis, Biomedical Engineering, School of Electrical \&
  Computer Engineering, Shiraz University (2014)

\bibitem{NarimaniSameni2015}
Narimani, H., Sameni, R.: {Electrocardiogram Denoising Using H-Infinity
  Filters}.
\newblock In: Electrical Engineering (ICEE), 2015 23rd Iranian Conference on
  (2015)

\bibitem{niknazar2013fetal}
Niknazar, M., Rivet, B., Jutten, C.: {Fetal ECG extraction by extended state
  Kalman filtering based on single-channel recordings}.
\newblock Biomedical Engineering, IEEE Transactions on \textbf{60}(5),
  1345--1352 (2013)

\bibitem{Oostendorp1989}
Oostendorp, T.: Modeling the Fetal {ECG}.
\newblock Ph.D. dissertation, K. U. Nijmegen, The Netherlands (1989).
\newblock \urlprefix\url{http://hdl.handle.net/2066/113606}

\bibitem{Outram95}
Outram, N.J., Ifeachor, E.C., Eetvelt, P.W.J.V., Curnow, J.S.H.: Techniques for
  optimal enhancement and feature extraction of fetal electrocardiogram.
\newblock In: IEE Proc.-Sci. Meas. Technol., vol. 142, pp. 482--489 (1995)

\bibitem{Pan1985}
Pan, J., Tompkins, W.J.: {A Real-Time QRS Detection Algorithm}.
\newblock Biomedical Engineering, IEEE Transactions on \textbf{BME-32}(3),
  230--236 (1985).
\newblock \doi{10.1109/TBME.1985.325532}

\bibitem{Park92}
Park, Y., Lee, K., Youn, D., Kim, N., Kim, W., Park, S.: {On detecting the
  presence of fetal R-wave using the moving averaged magnitude difference
  algorithm}.
\newblock Biomedical Engineering, IEEE Transactions on \textbf{39}(8), 868--871
  (1992).
\newblock \doi{10.1109/10.148396}

\bibitem{Parra2003}
Parra, L., Sajda, P.: {Blind Source Separation via Generalized Eigenvalue
  Decomposition}.
\newblock {Journal of Machine Learning Research} \textbf{4}, 1261--1269 (2003)

\bibitem{pham2001blind}
Pham, D.T., Cardoso, J.F.: Blind separation of instantaneous mixtures of
  nonstationary sources.
\newblock IEEE Transactions on signal processing \textbf{49}(9), 1837--1848
  (2001)

\bibitem{MIT-BIH-ADFECG}
PhysioNet: {Abdominal and Direct Fetal Electrocardiogram Database}.
\newblock National Institutes of Health.
\newblock \urlprefix\url{https://physionet.org/physiobank/database/adfecgdb/}

\bibitem{MIT-BIH-NIFECG}
PhysioNet: {Noninvasive Fetal ECG Database}.
\newblock National Institutes of Health.
\newblock \urlprefix\url{physionet.org/pn3/nifecgdb/}

\bibitem{Sameni2008}
Sameni, R.: {Extraction of Fetal Cardiac Signals from an Array of Maternal
  Abdominal Recordings}.
\newblock Ph.D. thesis, Sharif University of Technology -- Institut National
  Polytechnique de Grenoble (2008).
\newblock
  \urlprefix\url{http://www.sameni.info/Publications/Thesis/PhDThesis.pdf}

\bibitem{sameni2017online}
Sameni, R.: Online filtering using piecewise smoothness priors: application to
  normal and abnormal electrocardiogram denoising.
\newblock Signal Processing \textbf{133}(4), 52--63 (2017).
\newblock \urlprefix\url{https://doi.org/10.1016/j.sigpro.2016.10.019}

\bibitem{OSET3.14}
Sameni, R.: {The Open-Source Electrophysiological Toolbox (OSET), version 3.14}
  (2018).
\newblock \urlprefix\url{https://gitlab.com/rsameni/OSET/}

\bibitem{SameniClifford2010}
Sameni, R., Clifford, G.D.: {A Review of Fetal ECG Signal Processing; Issues
  and Promising Directions}.
\newblock {The Open Pacing, Electrophysiology \& Therapy Journal (TOPETJ)}
  \textbf{3}, 4--20 (2010).
\newblock \doi{10.2174/1876536X01003010004}

\bibitem{SCJS06}
Sameni, R., Clifford, G.D., Jutten, C., Shamsollahi, M.B.: {Multichannel ECG
  and Noise Modeling: Application to Maternal and Fetal ECG Signals}.
\newblock {EURASIP Journal on Advances in Signal Processing} \textbf{2007},
  {Article ID 43407, 14 pages} (2007).
\newblock \urlprefix\url{https://doi.org/10.1155/2007/43407}

\bibitem{patentSameni1_Published}
Sameni, R., Jutten, C., Shamsollahi, M., Clifford, G.: {Extraction of Fetal
  Cardiac Signals} (2010).
\newblock US Patent

\bibitem{SJS06}
Sameni, R., Jutten, C., Shamsollahi, M.B.: {What ICA Provides for ECG
  Processing: Application to Noninvasive Fetal ECG Extraction}.
\newblock In: {Proc. of the International Symposium on Signal Processing and
  Information Technology (ISSPIT'06)}, pp. 656--661. {Vancouver, Canada} (2006)

\bibitem{Sameni2008a}
Sameni, R., Jutten, C., Shamsollahi, M.B.: {Multichannel Electrocardiogram
  Decomposition using Periodic Component Analysis}.
\newblock Biomedical Engineering, IEEE Transactions on \textbf{55}(8),
  1935--1940 (2008).
\newblock \urlprefix\url{https://doi.org/10.1109/TBME.2008.919714}

\bibitem{SJS2010}
Sameni, R., Jutten, C., Shamsollahi, M.B.: {A Deflation Procedure for Subspace
  Decomposition}.
\newblock IEEE Transactions on Signal Processing \textbf{58}(4), 2363--2374
  (2010)

\bibitem{SSJ05}
Sameni, R., Shamsollahi, M.B., Jutten, C.: {Filtering Electrocardiogram Signals
  Using the Extended Kalman Filter}.
\newblock In: Proceedings of the 27th Annual International Conference of the
  IEEE Engineering in Medicine and Biology Society (EMBS), pp. 5639--5642.
  Shanghai, China (2005).
\newblock \urlprefix\url{https://doi.org/10.1109/IEMBS.2005.1615765}

\bibitem{SSJ08}
Sameni, R., Shamsollahi, M.B., Jutten, C.: {Model-based Bayesian filtering of
  cardiac contaminants from biomedical recordings}.
\newblock Physiological Measurement \textbf{29}(5), 595--613 (2008).
\newblock \doi{10.1088/0967-3334/29/5/006}

\bibitem{SSJB05}
Sameni, R., Shamsollahi, M.B., Jutten, C., Babaie-Zadeh, M.: {Filtering Noisy
  {ECG} Signals Using the Extended {K}alman Filter Based on a Modified Dynamic
  {ECG} Model}.
\newblock In: Proceedings of the 32nd Annual International Conference on
  Computers in Cardiology, pp. 1017--1020. Lyon, France (2005)

\bibitem{SSJC06}
Sameni, R., Shamsollahi, M.B., Jutten, C., Clifford, G.D.: {A Nonlinear
  Bayesian Filtering Framework for {ECG} Denoising}.
\newblock Biomedical Engineering, IEEE Transactions on \textbf{54}(12),
  2172--2185 (2007).
\newblock \urlprefix\url{https://doi.org/10.1109/TBME.2007.897817}

\bibitem{SaulA00}
Saul, L.K., Allen, J.B.: {Periodic Component Analysis: An Eigenvalue Method for
  Representing Periodic Structure in Speech}.
\newblock In: NIPS, pp. 807--813 (2000).
\newblock
  \urlprefix\url{http://www.cs.cmu.edu/Groups/NIPS/00papers-pub-on-web/SaulAllen.pdf}

\bibitem{scher1960factor}
Scher, A.M., Young, A., Meredith, W.M.: Factor analysis of the
  electrocardiogram: test of electrocardiographic theory: normal hearts.
\newblock Circulation Research \textbf{8}(3), 519--526 (1960)

\bibitem{Shao2004}
Shao, M., Barner, K., Goodman, M.: {An interference cancellation algorithm for
  noninvasive extraction of transabdominal fetal electroencephalogram
  (TaFEEG)}.
\newblock Biomedical Engineering, IEEE Transactions on \textbf{51}(3), 471--483
  (2004).
\newblock \doi{10.1109/TBME.2003.821011}

\bibitem{silva2013noninvasive}
Silva, I., Behar, J., Sameni, R., Zhu, T., Oster, J., Clifford, G.D., Moody,
  G.B.: {Noninvasive fetal ECG: the physionet/computing in cardiology challenge
  2013}.
\newblock In: Computing in Cardiology Conference (CinC), 2013, pp. 149--152.
  IEEE (2013)

\bibitem{Snowden2001}
Snowden, S., Simpson, N.A., Walker, J.J.: A digital system for recording the
  electrical activity of the uterus.
\newblock Physiol Meas \textbf{22}(4), 673--679 (2001)

\bibitem{sornmo1998}
S{\"o}rnmo, L.: Vectorcardiographic loop alignment and morphologic beat-to-beat
  variability.
\newblock Biomedical Engineering, IEEE Transactions on \textbf{45}(12),
  1401--1413 (1998)

\bibitem{sornmo1981method}
S\"ornmo, L., Borjesson, P.O., Nygards, M.E., Pahlm, O.: {A Method for
  Evaluation of QRS Shape Features Using a Mathematical Model for the ECG}.
\newblock IEEE Transactions on Biomedical Engineering \textbf{BME-28}(10),
  713--717 (1981)

\bibitem{Stinstra2001}
Stinstra, J.: Reliability of the fetal magnetocardiogram.
\newblock Ph.D. thesis, University of Twente, Enschede, The Netherlands (2001).
\newblock \urlprefix\url{http://doc.utwente.nl/35964/}

\bibitem{stogbauer-2004-70}
Stogbauer, H., Kraskov, A., Astakhov, S.A., Grassberger, P.: Least dependent
  component analysis based on mutual information.
\newblock Physical Review E \textbf{70}, 066123 (2004).
\newblock \urlprefix\url{doi:10.1103/PhysRevE.70.066123}

\bibitem{Strang1988}
Strang, G.: Linear Algebra and Its Applications, 3 edn.
\newblock Brooks/Cole (1988)

\bibitem{swarnalatha2010novel}
Swarnalatha, R., Prasad, D.: {A novel technique for extraction of FECG using
  multi stage adaptive filtering}.
\newblock Journal of Applied Sciences \textbf{10}(4), 319--324 (2010)

\bibitem{TI:SBAA160A}
Texas Instruments: {Analog front-end design for ECG systems using delta-sigma
  ADCs} (2010)

\bibitem{Tong1991}
Tong, L., Liu, R.W., Soon, V., Huang, Y.F.: {Indeterminacy and identifiability
  of blind identification}.
\newblock {IEEE} Trans. Circuits Syst. \textbf{38}, 499--509 (1991)

\bibitem{VanTrees2001detection}
van Trees, H.: {Detection, Estimation, and Modulation Theory. Part I}.
\newblock John Wiley \& Sons (2001)

\bibitem{Vigneron2003}
Vigneron, V., Paraschiv-Ionescu, A., Azancot, A., Sibony, O., Jutten, C.:
  {Fetal electrocardiogram extraction based on non-stationary ICA and wavelet
  denoising}.
\newblock Signal Processing and Its Applications, 2003. Proceedings. Seventh
  International Symposium on \textbf{2}, 69--72 vol.2 (1-4 July 2003).
\newblock \doi{10.1109/ISSPA.2003.1224817}

\bibitem{Webster1998}
Webster, J.G. (ed.): {Medical Instrumentation: Application and Design}, third
  edn.
\newblock John Wiley \& Sons (1998)

\bibitem{weiss99segmentation}
Weiss, Y.: Segmentation using eigenvectors: A unifying view.
\newblock In: Proc. IEEE Int. Conf. Computer Vision (2), pp. 975--982 (1999).
\newblock \urlprefix\url{citeseer.ist.psu.edu/weiss99segmentation.html}

\bibitem{Widrow75}
Widrow, B., Glover, J., McCool, J., Kaunitz, J., Williams, C., Hearn, H.,
  Zeidler, J., Dong, E., Goodlin, R.: {Adaptive Noise Cancelling: Principles
  and Applications}.
\newblock Proc. {IEEE} \textbf{63}(12), 1692--1716 (1975)

\bibitem{yeredor2009optimal}
Yeredor, A.: On optimal selection of correlation matrices for
  matrix-pencil-based separation.
\newblock In: International Conference on Independent Component Analysis and
  Signal Separation, pp. 187--194. Springer, Berlin, Heidelberg (2009)

\bibitem{yeredor2010second}
Yeredor, A.: Second-order methods based on color.
\newblock In: Handbook of Blind Source Separation, pp. 227--279 (Ch. 7).
  Elsevier (2010)

\bibitem{yeredor2011performance}
Yeredor, A.: {Performance analysis of GEVD-based source separation with
  second-order statistics}.
\newblock IEEE Transactions on Signal Processing \textbf{59}(10), 5077--5082
  (2011)

\bibitem{Zarzoso01}
Zarzoso, V., Nandi, A.: Noninvasive fetal electrocardiogram extraction: blind
  separation versus adaptive noise cancellation.
\newblock Biomedical Engineering, IEEE Transactions on \textbf{48}(1), 12--18
  (2001).
\newblock \doi{10.1109/10.900244}

\end{thebibliography}

\end{document}